\documentclass[journal,12pt,draftclsnofoot,onecolumn,twoside]{IEEEtran}
\usepackage{times,amsmath,epsfig,amssymb,cite,bm,float,epstopdf,graphicx,graphics,amsthm}
\usepackage{dblfloatfix}
\usepackage{balance}
\usepackage[stable]{footmisc}
\usepackage{tikz,pgfplots}
\usepackage[stable]{footmisc}
\usepackage{lettrine}
\usepackage[font={footnotesize},labelfont={footnotesize},belowskip=-3pt]{caption}
\usepackage{algorithm,algpseudocode}
\usepackage{tabularx,multirow,booktabs}
\usepackage{rotating,afterpage,lscape}
\usepackage{subfigure,float}
\usepackage{adjustbox}
\usepackage{enumitem}
\usepackage{afterpage}

\tikzset{>=latex}
\newcolumntype{C}{>{\centering\arraybackslash}X} 
\title{Unified Statistical Channel Model for Turbulence-Induced Fading in Underwater Wireless Optical Communication Systems}

\small{\author{Emna~Zedini,~\IEEEmembership{Member,~IEEE,}~Hassan~M.~Oubei,~\IEEEmembership{Member,~IEEE,} Abla~Kammoun,~\IEEEmembership{Member,~IEEE,}~Mounir~Hamdi,~\IEEEmembership{Fellow,~IEEE,}
~Boon~S.~Ooi,~\IEEEmembership{Senior~Member,~IEEE,}~and~Mohamed-Slim~Alouini,~\IEEEmembership{Fellow,~IEEE}
\thanks{E. Zedini and M. Hamdi are with the College of Science and Engineering, Hamad Bin
Khalifa University (HBKU), Doha, Qatar (e-mails:\{ezedini, mhamdi\}@hbku.edu.qa).}
\thanks{H.M. Oubei, A. Kammoun, B.S. Ooi and M.-S. Alouini are with the Computer, Electrical, and Mathematical Science and Engineering (CEMSE) Division, King Abdullah University of Science and Technology (KAUST) Thuwal, Makkah Province, Saudi Arabia (e-mails:\{hassan.oubei, abla.kammoun, boon.ooi, slim.alouini\}@kaust.edu.sa).}}}

\begin{document}
\pgfkeys{/pgf/fpu}
\pgfmathparse{16383+1}
\edef\tmp{\pgfmathresult}
\pgfkeys{/pgf/fpu=false}

\newcommand{\boundellipse}[3]
{(#1) ellipse (#2 and #3)
}
\maketitle
\begin{abstract}
A unified statistical model is proposed to characterize turbulence-induced fading in underwater wireless optical communication (UWOC) channels in the presence of air bubbles and temperature gradient for fresh and salty waters, based on experimental data. In this model, the channel irradiance fluctuations are characterized by the mixture Exponential-Generalized Gamma (EGG) distribution. We use the expectation maximization (EM) algorithm to obtain the maximum likelihood parameter estimation of the new model.
Interestingly, the proposed model is shown to provide a perfect fit with the measured data under all channel conditions for both types of water. The major advantage of the new model is that it has a simple mathematical form making it attractive from a performance analysis point of view. Indeed, we show that the application of the EGG model leads to
closed-form and analytically tractable expressions for key UWOC system performance metrics such as the outage probability, the average bit-error rate, and the ergodic capacity. To the best of our knowledge, this is the first-ever comprehensive channel model addressing the statistics of optical beam irradiance fluctuations in underwater wireless optical channels due to both air bubbles and temperature gradient.\\\\
\end{abstract}

\section{Introduction}
Underwater wireless optical communication (UWOC) systems have recently attracted considerable research attention as an appropriate and efficient transmission solution for a variety of underwater applications including offshore oil field exploration, oceanographic data collection, maritime archaeology, environmental monitoring, disaster prevention, and port security among others \cite{SurveyZeng}.
This rapidly growing interest stems from the recent advances in signal processing, digital communication, and low-cost visible light-emitting diodes (LEDs) and laser diodes (LD) that have the lowest attenuation in seawater \cite{LEDadvances, HassanHighDR, OubeiQAMOFDM, Oubei20m}. UWOC systems, operating in the blue/green portion of the spectrum in the 400-550 nm wavelength band, promise high data rates, low-latency, high transmission security, and reduced energy
consumption, compared with their acoustic counterparts \cite{SurveyZeng, Simodetection, HighBdW}.

Nevertheless, the reliability of such systems is highly affected by absorption and scattering effects \cite{SurveyZeng} as well as underwater optical turbulence (UOT). The identification of an accurate description for the absorption and scattering effects in UWOC channels has been extensively addressed in several recent works \cite{Jaruwatanadilok,Gabriel2,KihongWCL}. UOT results from rapid changes in the refractive index of the water caused by temperature fluctuations, salinity variations as well as the presence of air bubbles in seawater that affects the propagation of optical signals \cite{UOT,Nikishov2000,Korotkova2012}.
In oceans, air bubbles  are produced by breaking waves \cite{blanchard} and are found to significantly enhance the scattering process therein \cite{Zhang}. The presence of air bubbles in underwater and their effect on propagating optical signals are well established \cite{MOP:MOP26664,Woolf:01,Farmer:84}. In addition, the variations (gradient) in temperature and salinity in world water bodies are very common \cite{Boyle:87}. UOT distorts the intensity and phase of the propagating optical signal, which may degrade the performance of the UWOC system
\cite{UOT,Wang2017}. To mitigate these effects, various techniques have been presented.
The performance of UWOC systems using optical pre-amplification and multiple receivers has been investigated in \cite{Boucouvalas}. \cite{JamaliMIMO} studied the performance of multiple-input multiple-output (MIMO) UWOC systems with on-off keying (OOK). In \cite{PeppasIET}, the performance of UWOC systems employing spatial diversity and multi-pulse position modulation techniques is presented. The bit-error rate (BER) of multi-hop UWOC systems is evaluated in \cite{JamaliMultihop}.

 To design robust and reliable UWOC systems, it is important to investigate and understand the statistical distribution of optical signal fluctuations due to UOT. Early studies on UOT had mostly focused on theoretical investigations based on the formulation of free-space atmospheric turbulence models such as the Lognormal distribution to describe the irradiance fluctuations in the underwater environment \cite{laserbeamscin,Gercekcioglu,lognmunderwater}. However, the spectrum of refractive-index variations caused by temperature or pressure inhomogeneities in the atmosphere is much different from the refractive-index spectrum of temperature or salinity in  water. This makes the Lognormal distribution not appropriate for modeling the irradiance fluctuations in turbulent water. Therefore, there is a need for further investigation of new accurate statistical models to better characterize the turbulence-induced fading in UWOC.

 The influence of air bubbles has been characterized in several previous works based on Mie scattering theory \cite{Zhang,Davis}. It was only recently that the impact of bubbles on the distribution of the irradiance has been investigated through a set of lab experiments \cite{ExpLN,ConfUWOC}. These  works essentially show
that in the presence of air bubbles the distribution of the irradiance is accurately modeled by a mixture of the Exponential distribution and the Log-normal distribution which can also be replaced by the Gamma distribution. The presence of the Log-normal distribution or equivalently the Gamma distribution agrees with previous studies suggesting its use to model  underwater optical channels. The Exponential distribution, is however, less common. As shown in \cite{ConfUWOC}, it is used to model the loss in the received energy caused by air bubbles. Therefore, typical single-lobe distributions cannot appropriately fit the measured data in the presence of air bubbles, and a two-lobe statistical model is
required to predict the statistical behavior of UWOC turbulence-induced fading in all regions of the scintillation index.

In \cite{ExpLN}, the mixture Exponential-Lognormal model has been proposed to describe the irradiance fluctuations due to air bubbles in both fresh and salty waters in UWOC channels. However, the model does not take into account temperature or salinity gradient in the water channel and is shown not to accurately fit the measured data in all turbulence regimes where the scintillation index varies between 0.1 to 1
Moreover, the mathematical form of Lognormal-based distributions is not convenient for analytic calculations.
Furthermore, the design and the performance analysis of such systems is much more challenging. Indeed, the application of the Exponential-Lognormal in UWOC channels makes it very hard to obtain closed-form and easy-to-use expressions for important performance metrics such as the outage probability and the average BER.
The mathematical intractability of the Lognormal-based model becomes more evident when we know that the assessment of BER is based on numerical methods, as closed-form analytical expressions are not available for this model.
In \cite{Oubei:17sg}, Weibull distribution was used to characterize fluctuations of laser beam intensity in underwater caused by salinity gradient. The Weibull model showed an excellent agreement with measured data under all channel conditions. Statistical channel model for weak temperature-induced turbulence in UWOC systems was investigated in \cite{Oubei:17}. The Generalized Gamma distribution (GGD) was proposed to accurately describe both non turbulent thermally uniform and gradient based underwater wireless optical channels.
  In \cite{ConfUWOC}, we have proposed the mixture EG distribution to characterize optical signal irradiance fluctuations in underwater channel. The model provided a perfect fit with the measured data under all the channel conditions for both fresh and salty waters.

However, the aforementioned studies have investigated and modeled the statistics of laser beam irradiance fluctuations due either to air bubbles in thermally uniform channel or underwater channels with temperature or salinity gradient. To the best of the authors' knowledge, there is no comprehensive study that statistically describes optical beam irradiance fluctuations taking into account both air bubbles and temperature gradient.
 In this paper, we present a unified UWOC turbulence model that efficiently and statistically describe air bubbles and temperature-induced irradiance fluctuations from weak to strong turbulence condition using fresh as well as salty waters. Based on measured data, we propose the mixture Exponential-Generalized Gamma (EGG) distribution model that gives excellent goodness of fit under all channel conditions.
We use the expectation maximization (EM) algorithm to obtain the maximum likelihood (ML) estimates of the new model parameters.
When the water temperature is uniform throughout the tank, the received intensity of
the laser beam is best described by the simple Exponential-Gamma (EG) distribution
which is a special case of the EGG distribution.

In addition, we present a unified performance analysis of UWOC systems operating under intensity modulation/direct detection (IM/DD) as well as heterodyne techniques. As we will see in the sequel, the new model not only gives excellent agreement with the real measured data under all channel conditions but also can efficiently be used to obtain closed-form and mathematically tractable expressions for the system performance metrics such as the outage probability, the average bit-error rate (BER) for a variety of modulation schemes, and the ergodic capacity. We also derive new asymptotic expressions for all the performance metrics in the high SNR regime in terms of elementary functions.

The remainder of this paper is organized as follows. Section II illustrates the experimental setup for intensity fluctuations measurements. In Section III, we introduce the mixture EGG model and we provide its statistical parameters. The EM algorithm is also presented in detail. A comparison between our proposed EGG model, the EG model, and the Exponential-Lognormal model is also established by means of statistical goodness of fit tests, and presented in Section IV.
Section V is devoted to the experimental results and discussion. We then show how the application of the new model results in
closed-form expressions for fundamental system performance metrics along with the asymptotic analysis at high SNR regime in Section VI.
Section VII presents some numerical and simulation results to illustrate the mathematical formalism presented in this work. Finally, some concluding remarks are drawn in Section VIII.

\section{Experimental Setup}
\subsection{Turbulent UWOC Channels with Gradient Temperature}
\begin{figure}[h]
    \centering
\includegraphics[scale=0.45]{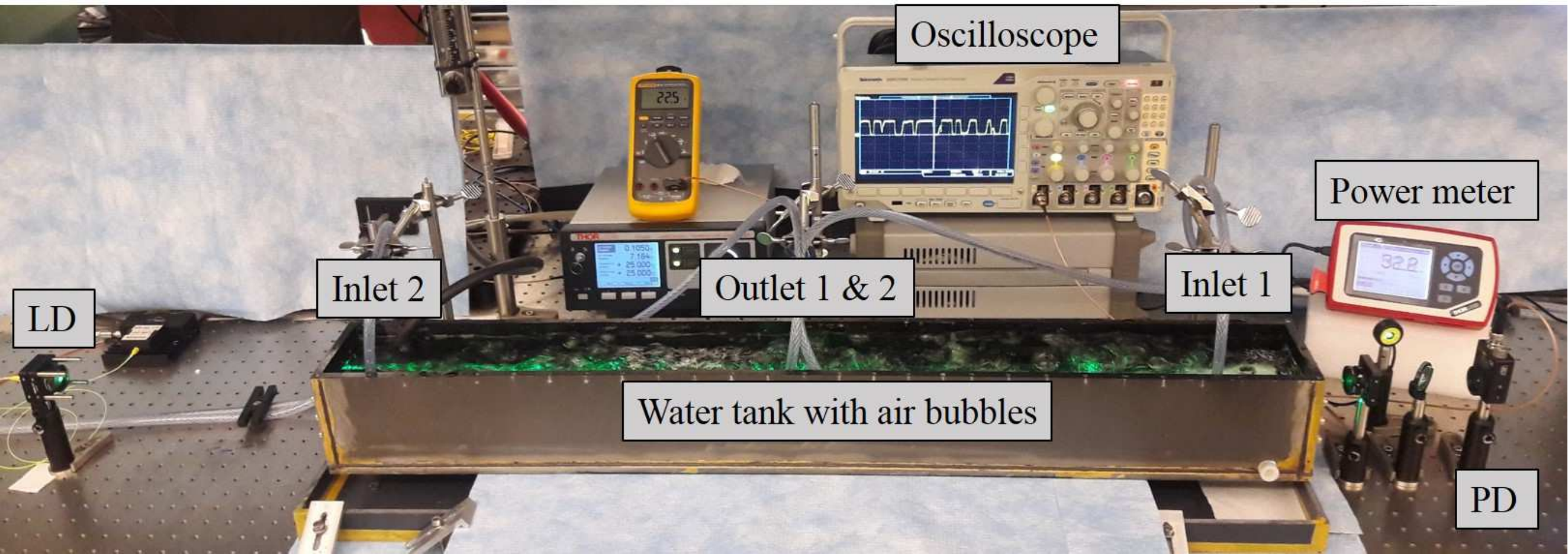}
\caption{Experimental setup used to study the statistics of temperature-induced turbulent underwater wireless optical channel in the presence of air bubbles: laser diode (LD), and photodetector (PD).}
\label{expsetup1}
\end{figure}

The experimental setup was kept identical to the one described in reference \cite{Oubei:17} including the different temperature values used to create temperature gradient in the water channel. Additionally, we have considered a water temperature gradient of $0.22\,^\circ$C.$cm^{-1}$ corresponding to two different temperature values, $17.3\,^{\circ}\mathrm{C}$ and $39.2\,^{\circ}\mathrm{C}$ for strong effect of the air bubbles.
The optical transmitter was a single-mode TO-can and fiber pigtailed green LD (Thorlabs LP520) operating at a wavelength of 515 nm with 25.4 mm diameter and 25.4 mm focal length. In room temperature operation, the threshold current of the LD was 58 mA. The transmission power was set at 5.7 mW.
Air bubbles were introduced in
the water tank via a 3/4 diameter and 0.9 m long PVC pipe with 2 mm holes placed in the bottom of the tank as shown in Fig.~\ref{expsetup1}.
\noindent
The air flow rate was measured in terms of liters per minute (L/min) that can be also expressed in terms of standard cubic feet per hour (SCFH).
Four levels of air bubbles were generated namely BL=2.4 L/min (5 SCFH), 4.7 L/min (10 SCFH), 16.5 L/min (35 SCFH), and 23.6 L/min (50 SCFH) throughout the experiment.
The size of the bubble was assumed to be uniform as no external force was introduced influence the size and speed of the bubble generation \cite{7879801}. The tank was filled with fresh municipal water with an estimated attenuation coefficient of 0.071 m-1 at 520 nm. After propagating through the turbulent and bubbly water, we collected 100000 samples of intensity fluctuations data for statistical analysis using a silicon photodiode receiver unit (Thorlabs DET36A) with 25.4 mm diameter and 75 mm focal length.

\subsection{Turbulent UWOC Channels with Uniform Temperature}
\begin{figure}[h]
\centering
  \includegraphics[scale=0.5]{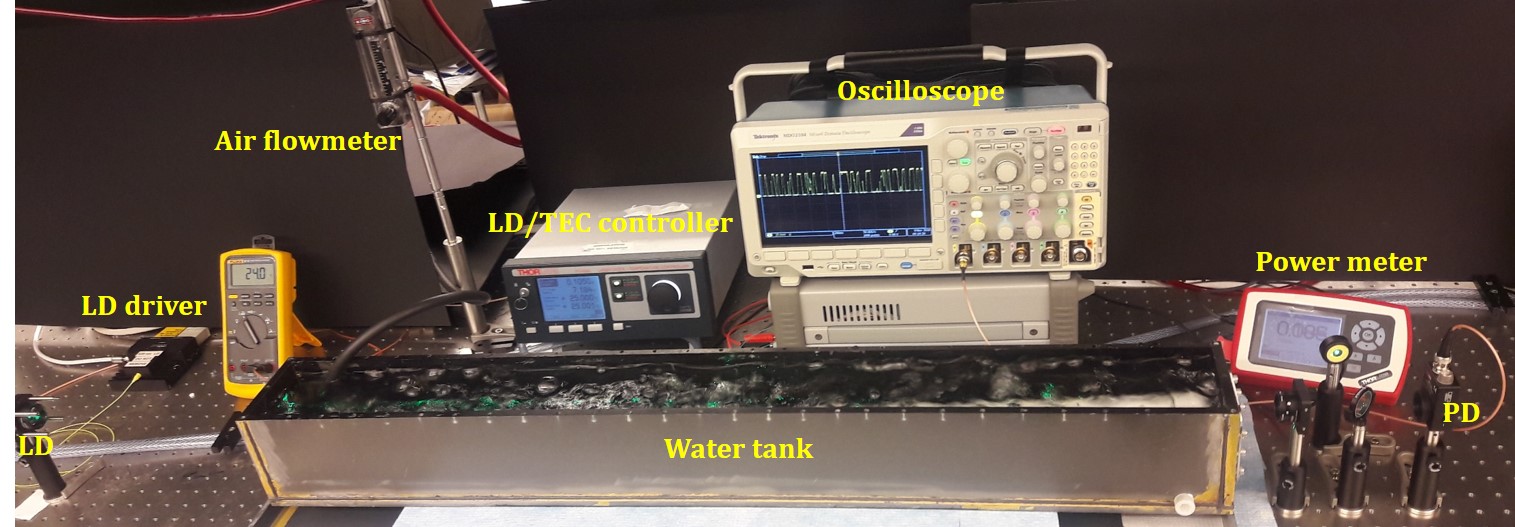}
  \caption{Actual Photograph of the experimental setup for intensity fluctuations measurements in 1 m underwater channel in the presence of air bubbles for uniform temperature.}
  \label{fig:expsetup2}
\end{figure}

Fig.~\ref{fig:expsetup2} illustrates the experimental setup used to measure and collect the intensity fluctuations data for underwater wireless optical channels under the combined effect of salinity as well as air bubbles induced turbulences.
\noindent
There was no attempt to control the temperature of the water. Using a thermometer, the temperature in the tank was kept constant at $25\,^{\circ}\mathrm{C}$.
The transmitter is a green LD with a peak emission wavelength of around 515 nm with 0.45nm full-width at half-maximum (FWHM) under 70mA current injection.
A plano-convex lens (Thorlabs LA1951-A) of 25.4 mm focal length is used to collimate and produce a parallel beam. The transmitted power is 7.5 mW (8.8 dBm). The underwater environment was simulated using $1$ m$\times 0.6$ m$\times 0.6$ water tank made of polyvinyl chloride (PVC) with 6 cm $\times$ 6 cm acrylic glass windows. The inside of the tank was painted black in order to minimize light reflecting off the sidewalls. Both fresh and salty waters are considered in our measurements.
For salinity, we added 118 g of table salt into the fresh water tank. Air bubbles were generated by a 3/4'' PVC pipe with 2 mm holes placed along the tank. The hole spacing is 5 cm. Five levels of air bubbles (BL) were generated, namely BL=0 L/min, 2.4 L/min, 4.7 L/min, 7.1 L/min (15 SCFH), and 16.5 L/min.

After propagating through the 1 m underwater channel, the 520 nm beam was focused into a biased silicon PIN photodiode (PD) receiver utilizing a 75 mm focal length lens (Thorlabs LA1608-A). The technical specifications of the PD (Thorlabs DET36A) include an active diameter of 13 mm$^2$, a responsivity of around 0.19 A/W at 520 nm and a noise equivalent power (NEP) of 0.016 pW/Hz$^{\frac{1}{2}}$. The output of the PD was captured by a 1 GHz bandwidth mixed domain oscilloscope (Tektronix, MDO 3104) with a maximum sampling rate of up to 5 GSa/s for power fluctuations monitoring and measurements.
For channel coherence time measurements, we collected 100000 samples with the sampling rate of 100 kS/s.
In the case of intensity fluctuations' distribution data, we also collected 100000 samples with sampling rate of 100 S/s. For all tests, measurements were taken under normal room illumination conditions.

\section{Modeling Underwater Turbulence With the Mixture EGG Model}
\subsection{Statistics of the New Model}
Throughout this paper, the irradiance fluctuations of the received optical wave due to air bubbles and temperature-induced fading in both fresh and salty (by adding salt into the fresh water tank) waters, $I$, is modeled by the mixture EGG distribution, which is a weighted sum of the Exponential and Generalized Gamma distributions and can be expressed as
\begin{equation}
f_I(I)=\omega f(I;\lambda)+(1-\omega)g(I;[a,b,c]),
\label{PDFmix}
\end{equation}
with
\begin{align}
    f(I;\lambda)&=\frac{1}{\lambda}\exp(-\frac{I}{\lambda})\\
    g(I;[a,b,c])&=c\frac{I^{ac-1}}{b^{ac}}\frac{\exp(-\left(\frac{I}{b}\right)^c)}{\Gamma(a)}
\end{align}
$f$ and $g$ being respectively the Exponential and Generalized Gamma distributions
where $\omega$ is the mixture weight or mixture coefficient of the distributions, satisfying $0<\omega<1$, $\lambda$ is the parameter associated with the Exponential distribution, and $a,b$ and $c$ are the parameters of the Generalized Gamma distribution and $\Gamma(.)$ denotes the Gamma function.

The $n$th moment of $I$ defined as $\mathbb{E}[I^n]\triangleq\int_{0}^{\infty}I^n f_I(I)dI$, can be obtained by substituting (\ref{PDFmix}) into the definition then utilizing \cite[Eqs.~(3.351/3)~and~(3.478/1)]{Tableofintegrals} yielding
\begin{align}\label{momentsIrradiance}
\mathbb{E}[I^n]=\omega\, \lambda^n \,n! +(1-\omega) \frac{b^n\,\Gamma(a+\frac{n}{c})}{\Gamma(a)},
\end{align}
where $\mathbb{E}$ represents the expected value.

The scintillation index $\sigma_I^2$, defined as the normalized variance of the intensity fluctuations can be expressed as
\begin{align}\label{SIdef}
\sigma_I^2  \triangleq   \frac{\mathbb{E}[I^2]-\mathbb{E}[I]^2}{\mathbb{E}[I]^2}.
\end{align}
Using (\ref{momentsIrradiance}), the scintillation index can be derived as
\begin{align}\label{SInewmodel}
\sigma_I^2=\frac{2\omega\lambda^2+(1-\omega)b^2\frac{\Gamma(a+\frac{2}{c})}{\Gamma(a)}}{\left[\omega \lambda +(1-\omega) \frac{b\Gamma(a+\frac{1}{c})}{\Gamma(a)})\right]^2}-1.
\end{align}

\textbf{Special Case} (Uniform Temperature). Thermally uniform UWOC channels are perfectly characterized by the simple EG mixture model which is a special case of EGG for $c=1$. The EG model is a weighted sum of the Exponential and Gamma distributions whose PDF is obtained by setting $c=1$ in (\ref{PDFmix}), that is,
\begin{align}\label{PDFmixSim}
f_I(I)&=\frac{\omega}{\lambda}\, \exp \left (-\frac{I}{\lambda}  \right ) +(1-\omega)\, I^{\alpha-1} \frac{\exp\left({-\frac{I}{\beta}}\right)}{\beta^\alpha \,\Gamma(\alpha)},
\end{align}
where $\alpha$ and $\beta$ represent the shape and scale parameters of the Gamma distribution, respectively.
We should emphasize that the distribution in (\ref{PDFmixSim}) has a simpler mathematical form than the Lognormal-based PDF given in \cite[Eq.(8)]{ExpLN} and thus lead to straightforward performance evaluation of UWOC systems, with closed-form and mathematically tractable results.

\subsection{ML Parameter Estimation of the New Model}
In this paper, we use the expectation maximization (EM) algorithm to find maximum
likelihood estimates of the model (\ref{PDFmix}) parameters, i.e. $\omega$, $\lambda$, $a$, $b$, and $c$. The EM algorithm is an effective iterative
method that starts from some arbitrarily initial values for
the model parameters and then proceeds iteratively to update
them until convergence. In other words, the EM algorithm provides us the parameters that realize the best fit with the measured data. These values vary depending on the water temperature, the water salinity, and the level of the air bubbles as shown by Table I and Table II.

Let $I_1,\ldots,I_n$ be the set of independent and identically distributed (iid) irradiance observations with $n$ being the number of measured samples,
using the experimental setups previously described.
As clearly shown in the experimental setup section, we have collected $n=100000$ samples of intensity fluctuations data with a sampling rate of 100 S/s. Therefore, it is important to mention here that for every channel condition specified by the rows of Table I and Table II, we have collected 100000 irradiance fluctuations.
In other words, for a specific channel condition, we use 100000 realizations and we run the EM algorithm to obtain the maximum likelihood estimates of $\omega$, $\lambda$, $a$, $b$ and $c$ which are obtained in Table I and Table II.

The EM algorithm, generally used for maximum likelihood estimation of models involving missing data, has also been applied  to estimate the parameters of  mixture models.
This is because data generated from the mixture model as in (\ref{PDFmix}) can be regarded as an incomplete data set. Indeed,  it is possible to associate each observed irradiance realization $I_i$  with a hidden unobserved binary random variable $z_i$ taking $1$ with probability $\omega$ when the data point is drawn from the Exponential distribution and $0$ with probability $1-\omega$ if drawn from the Generalized Gamma distribution.

The EM-algorithm seeks to determine the maximum likelihood estimates of the parameters of the mixture model in (\ref{PDFmix}) by alternating the following two steps
\begin{itemize}
    \item E-step: The E-step consists in computing the expected values of the hidden variables $\left\{z_i\right\}$ given the incomplete data set $\left\{I_i\right\}_{i=1}^n$. Using the Bayes' rule, these quantities are given by
        \begin{align}
            \gamma_i&\triangleq \mathbb{P}\left[z_i=1|\left\{I_i\right\}_{i=1}^n\right] \nonumber\\
                    &=\frac{\omega f(I_i;\lambda)}{\omega f(I_i;\lambda) +(1-\omega)g(I_i;\left[a,b,c\right])}. \label{eq:gamma}
    \end{align}
\item M-step: The M-step consists in selecting the parameters of the mixture model that maximize the following functional which coincides with the expected value of the log likelihood function of the complete data set $\left\{(I_i,z_i)\right\}_{i=1}^n$ with respect to the conditional distribution $(z_1,\ldots,z_n)$ given $I_1,\ldots,I_n$
    \begin{align}
    \nonumber & \ell\left(\left\{I_i\right\};\lambda,[a,b,c]\right)=\sum_{i=1}^n \gamma_i \log(f(I_i;\lambda)) + \gamma_i \log(\omega)\\
    &+(1-\gamma_i)\log(1-\omega) + (1-\gamma_i) \log(g(I_i;\left[a,b,c\right])).
    \end{align}
   As already shown in \cite{parameter}, when it comes to compute the maximum values of the above function, it is more handy to work with $\theta=b^c$ than $b$. We will thus maximize over the variables $a, \theta$ and $c$.
   Taking the derivatives of functional $\ell$ with respect to $\theta,c$ and $a$ results in the following set of equations
   \begin{align}
       \theta&=\frac{\sum_{i=1}^n(1-\gamma_i)I_i^c}{\sum_{i=1}^n(1-\gamma_i)a}\label{eq:theta}\\
       a&=\frac{\sum_{i=1}^n \frac{\gamma_i}{c}}{\frac{\sum_{i=1}^n\gamma_i\log(I_i)I_i^c\sum_{j=1}^n\log(\gamma_j)}{\sum_{j=1}^n \gamma_j I_j^c}-\sum_{i=1}^n\gamma_i\log(I_i)} \label{eq:a}\\
        &\sum_{i=1}^n (1-\gamma_i)\psi(a)+\sum_{i=1}^n(1-\gamma_i)\log(\theta)-\sum_{i=1}^n (1-\gamma_i)c\log(I_i)=0,  \label{eq:c}
   \end{align}\noindent

where $\psi$ is the digamma function \cite[Eq.~(8.360)]{Tableofintegrals}.
To find $a,\theta$ and $c$, it suffices to replace into \eqref{eq:c} $\theta$ and $a$ with their expressions in \eqref{eq:theta} and \eqref{eq:a}. In doing so, \eqref{eq:c} becomes a single variable function of $c$, the zero of which can be solved numerically using the MATLAB function fzero. Once $c$ is obtained, $a$ and $\theta$ are retrieved  using again \eqref{eq:a} and \eqref{eq:theta}. As for the Exponential distribution, the maximization over the parameter $\lambda$ leads to
\begin{equation}
\lambda=\frac{\sum_{i=1}^n\gamma_i I_i}{\sum_{i=1}^n I_i}.
\label{eq:lambda}
\end{equation}
Finally, the weight $\omega$ satisfies
\begin{equation}
\omega=\frac{1}{n}\sum_{i=1}^n \gamma_i.
\label{eq:omega}
\end{equation}
\end{itemize}
\begin{algorithm}[!h]
\begin{algorithmic}[1]
    \State Initialize, $t=0$ and  $p^{t}=\left[a,b,c,\lambda,\omega\right]$ and $\epsilon>0$.
    \Repeat
    \State $t:=t+1$
    \State {\bf E-Step}: Compute $\gamma_i^{t}$ as \eqref{eq:gamma}
    \State{\bf M-step}: Set $c^{t}$ to the positive zero of the following function
    \begin{align*}
    h(c)&=-\sum_{i=1}^n (1-\gamma_i^{t})\psi(a)-\sum_{i=1}^n(1-\gamma_{i}^{t})\log(\theta)\\
    &+\sum_{i=1}^n (1-\gamma_i^{t})c\log(I_i)
    \end{align*}
    where $a$, $\theta$ depend on $c$ through \eqref{eq:a} and \eqref{eq:theta}.
    \State Compute $a^{t}$, $\theta^{t}$ using \eqref{eq:a} and \eqref{eq:theta} with $c$ replaced by $c^{t}$, Set $b^{t}=\left(\theta^{t}\right)^{\frac{1}{c_t}}$
    \State Compute $\lambda^{t}$ and $\omega^{t}$ using \eqref{eq:lambda} and \eqref{eq:omega}
    \State Update $p^{t}=\left[a^{t},b^{t},c^{t},\lambda^{t},\omega^{t}\right]$
\Until{$\max(\left|p^{t}-p^{t-1}\right|)>\epsilon$}
\end{algorithmic}
\caption{EM algorithm to tune the EGG mixture model}
\end{algorithm}
For the sake of simplicity, we summarize in Algorithm 1 the EM algorithm for the EGG mixture model.

It is worth accentuating that, the EM algorithm is also used to estimate the parameters of the EG model and the ML estimates of $\alpha$ and $\beta$ parameters of the Gamma distribution may be determined utilizing \cite[Eqs.~(11) and (12)]{ConfUWOC}. Moreover, to compare the new proposed model with the Exponential-Lognormal model presented in \cite{ExpLN}, we have also applied the EM algorithm to obtain ML estimates of the Lognormal distribution parameters, $\mu$ and $\sigma^2$, that may be calculated using \cite[Eqs.~(14) and (15)]{ConfUWOC}.

\section{Goodness of Fit Tests}
The validity of the new proposed model may be verified statistically by conducting goodness of fit tests that
describe how well the new model fits to the measured data.
Specifically, we use the mean square error (MSE) test and the R-square (R$^2$) test that have been widely employed in evaluating the goodness of fit of a variety of fading distributions to channel measurements.
Additionally, by conducting these tests, we compare the proposed EGG distribution with the EG as well as the Exponential-Lognormal distributions and we demonstrate that our proposed model can efficiently describe the irradiance fluctuations under all channel conditions for both fresh and salty waters, providing analytical tractability as well.

The results of the MSE as well as the R$^2$ tests along with the estimated parameters of the proposed EGG, the EG, and the Exponential-Lognormal distributions for different levels of air bubbles for thermally uniform and
gradient-based UWOC channels are listed in Table~\ref{goodnessoffit1} and Table~\ref{goodnessoffit2}, respectively.

\subsection{MSE Test}
The MSE is a simple and efficient measure of how accurately the proposed EGG model predicts the measured irradiance fluctuations. It is defined as
\begin{align}
\text{MSE}=\frac{\sum_{i=1}^{N}\left (F_e(I_i)-F(I_i)\right)^{2}}{N},
\end{align}
where $F_e(I)$ denotes the empirical distribution function of $I$ and $F(I)$ stands for the theoretical CDF computed with parameters estimated from the measured data defined as $F(x)=\int_{-\infty}^{x} f_I(I,\theta)\,dI$.
It is important to mention here that lower values of MSE (i.e. MSE $\to 0$) indicate a better fit to the acquired experimental data and subsequently a better model.
\subsection{R$^2$ Test}
The coefficient of determination, $R^{2}$, is used to quantify the goodness of fit. $R^{2}$ is computed from the sum of squared errors, $SS_{\text{err}}$, and the sum of the squares of the distances of the measured points from their mean, $SS_{\text{tot}}$, and can be expressed as \cite{Rsquare}
\begin{align}
R^2=1-\frac{SS_{\text{err}}}{SS_{\text{tot}}},
\end{align}
where $SS_{\text{err}}=\sum_{i=1}^{M}\left (f_{m,i}-f_{p,i}  \right )^2$ and $SS_{\text{tot}}=\sum_{i=1}^{M}\left (f_{m,i}-\bar{f}  \right )^2$, with $f_{m,i}$ and $f_{p,i}$ are the measured and predicted probability values for a given received irradiance level, $M$ represents the number of bins of the data histogram, and $\bar{f}=\sum_{i=1}^{M}\frac{f_{m,i}}{M}$.

It is worth mentioning that the $R^2$ measure ranges from 0 to 1 and the higher the value of $R^2$ (i.e. $R^2 \to 1$), the better the proposed model fits the measured intensity through the experiment. Note that the value of $R^2$ depends on the number of bins of the acquired
data histogram.

\section{Experimental Validation}
\subsection{Turbulent UWOC Channels with Gradient Temperature}
\begin{figure}[htp]
\vspace{-0.5cm}
\centering
\subfigure[BL=2.4 L/min, 0.05 $^\circ$C.$cm^{-1}$.]{%
\includegraphics[width=0.455\textwidth]{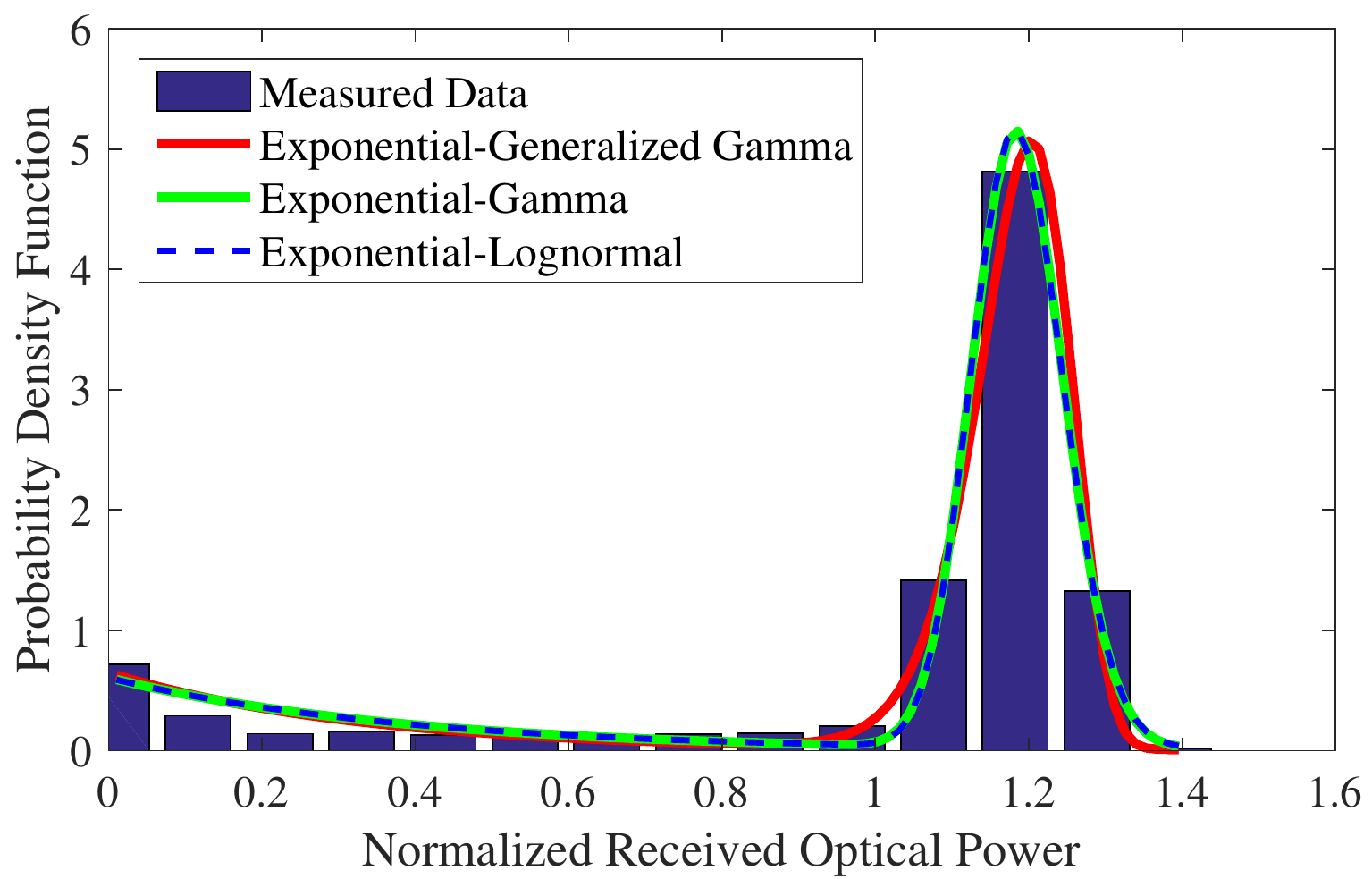}
\label{fig:subfigure1}}
\quad
\subfigure[BL=2.4 L/min, 0.10 $^\circ$C.$cm^{-1}$.]{%
\includegraphics[width=0.455\textwidth]{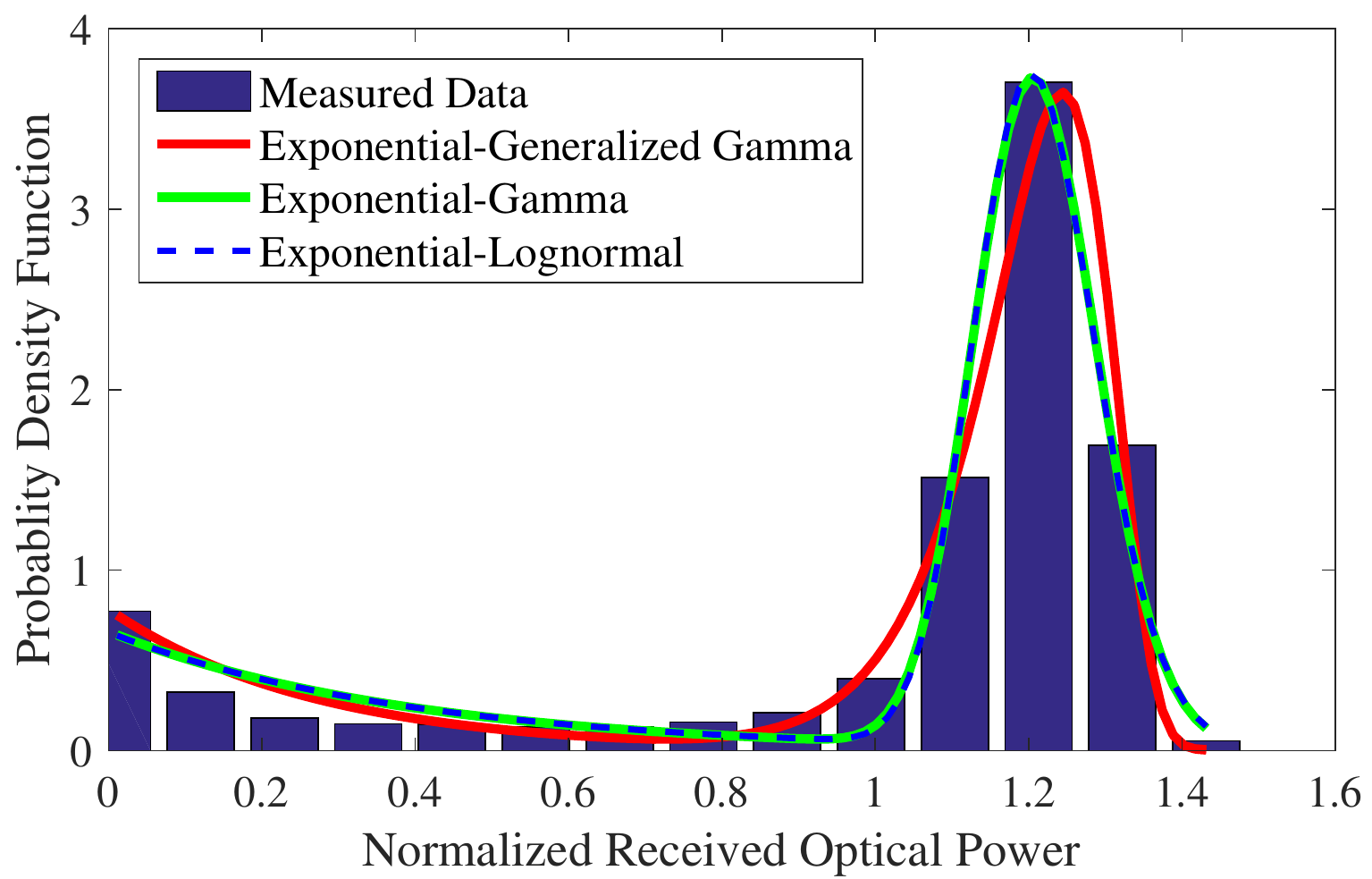}
\label{fig:subfigure2}}
\subfigure[BL=2.4 L/min, 0.15 $^\circ$C.$cm^{-1}$.]{%
\includegraphics[width=0.455\textwidth]{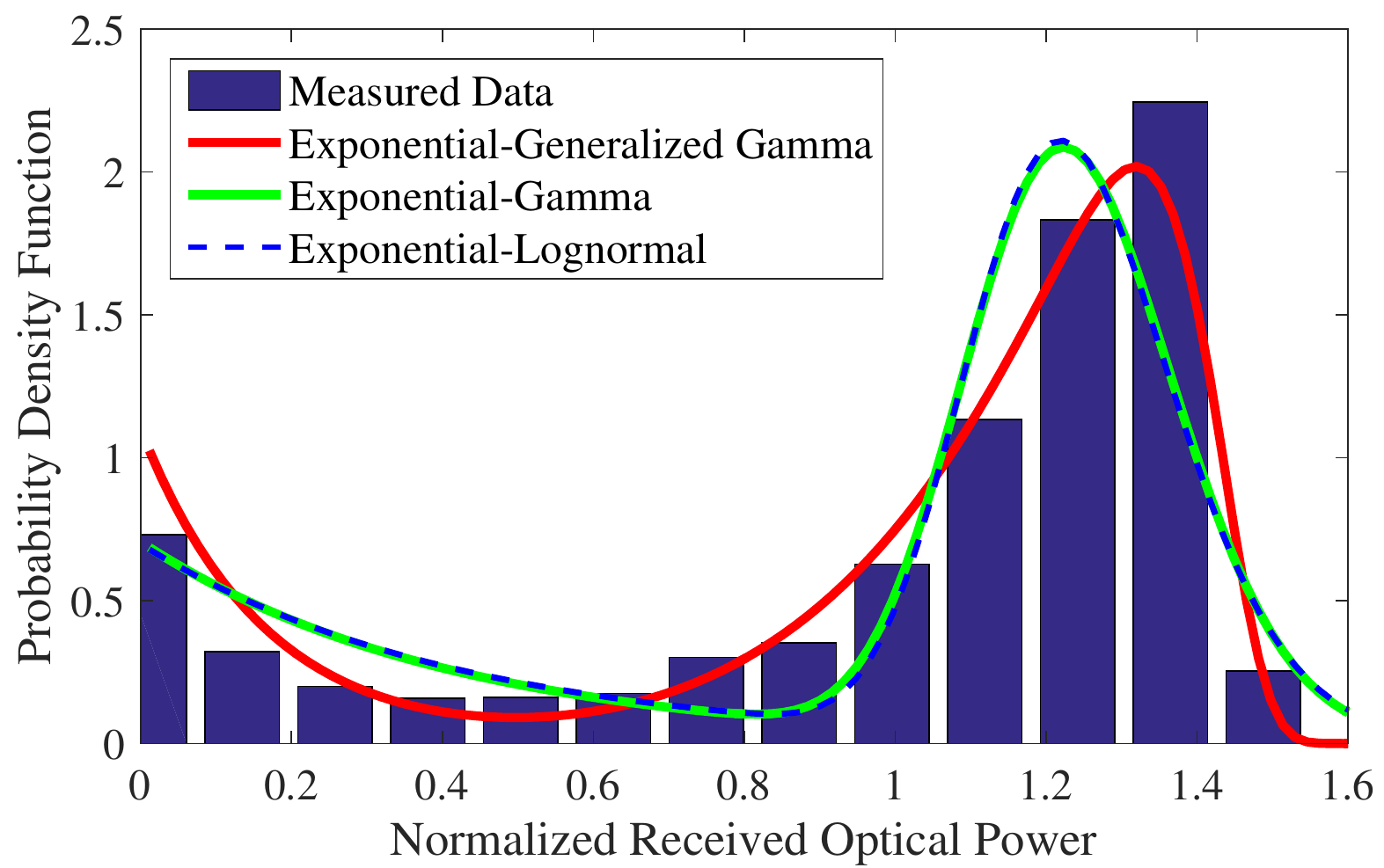}
\label{fig:subfigure3}}
\quad
\subfigure[BL=2.4 L/min, 0.20 $^\circ$C.$cm^{-1}$.]{%
\includegraphics[width=0.455\textwidth]{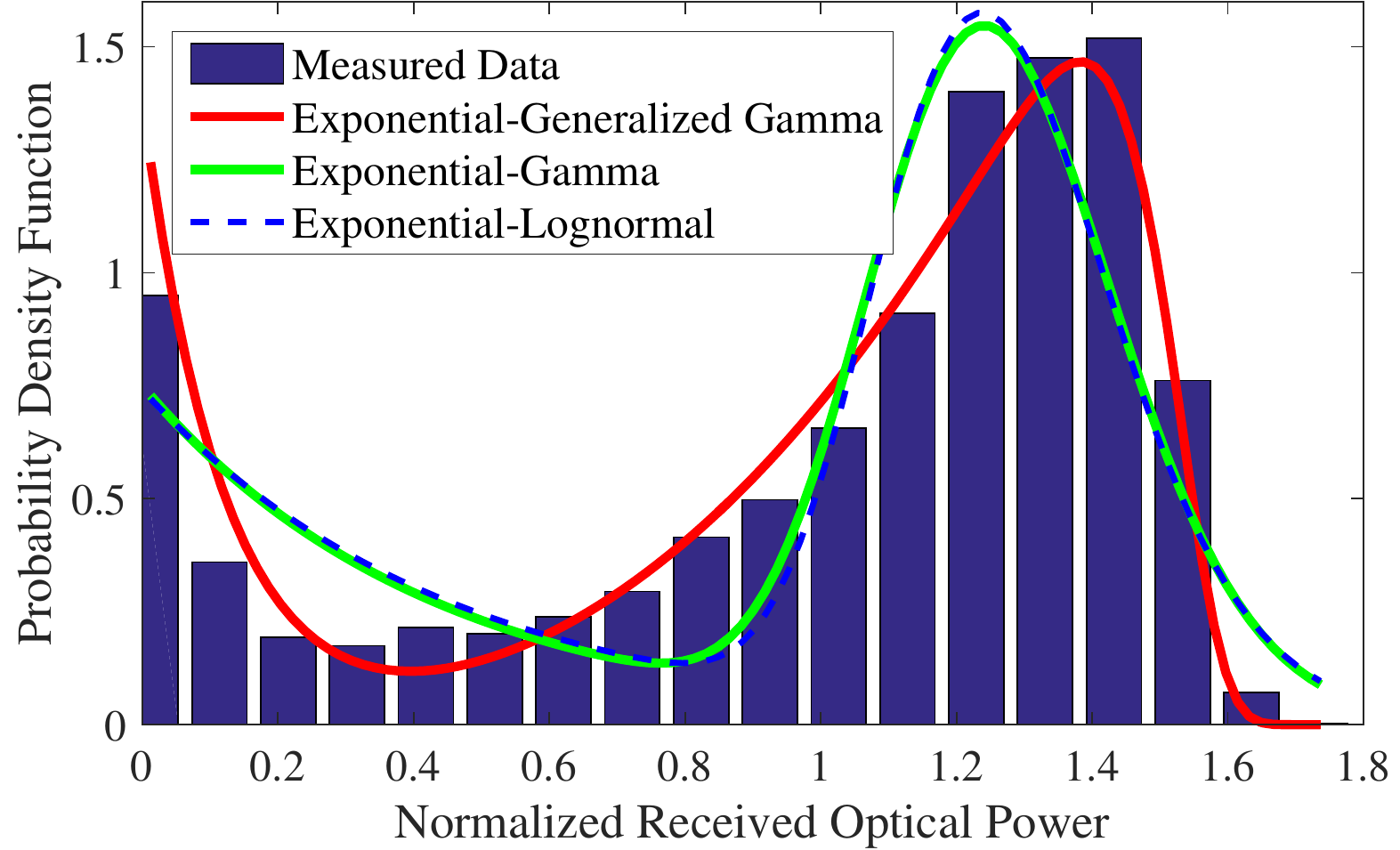}
\label{fig:subfigure4}}
\subfigure[BL=4.7 L/min, 0.05 $^\circ$C.$cm^{-1}$.]{%
\includegraphics[width=0.455\textwidth]{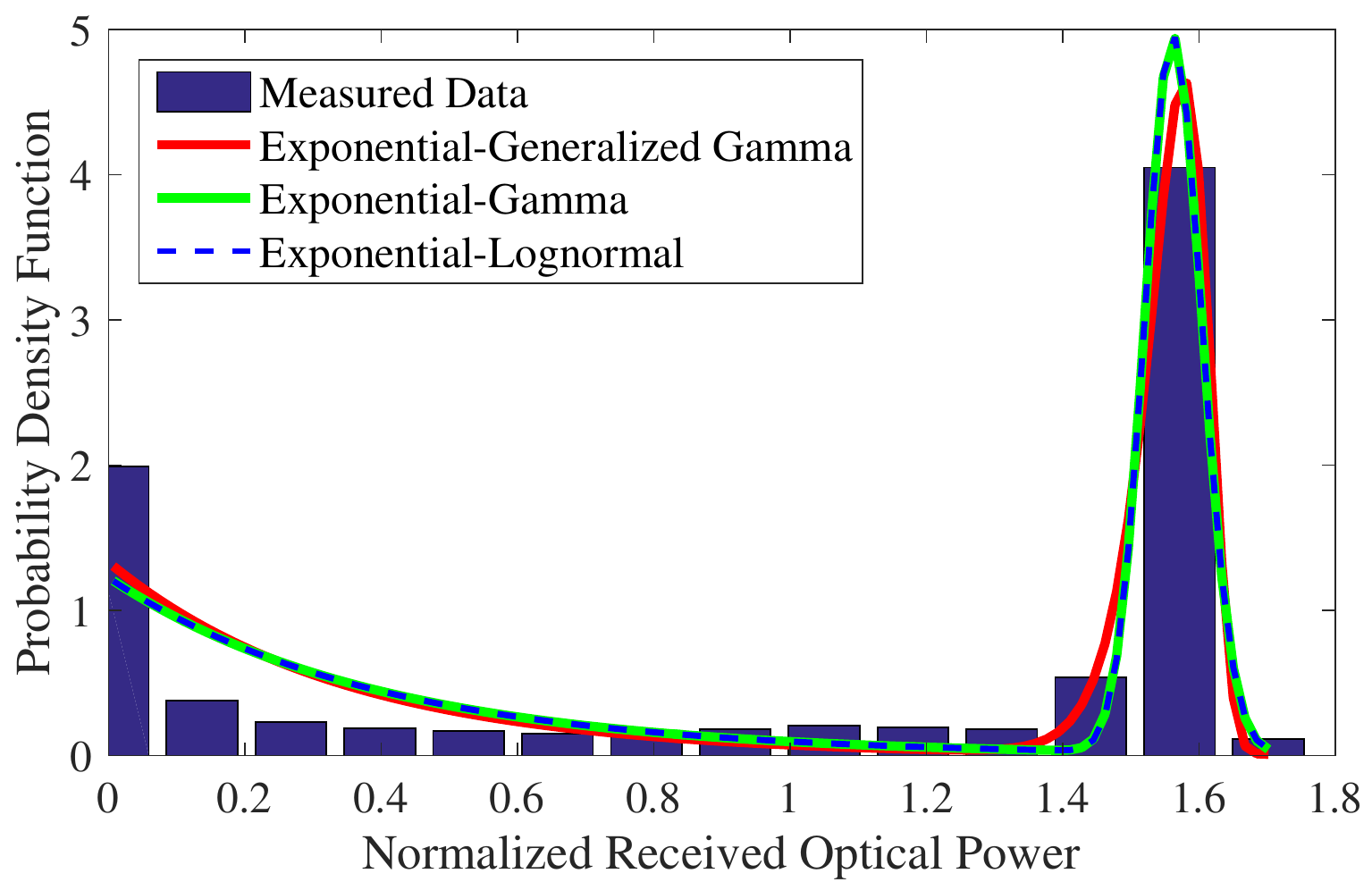}
\label{fig:subfigure11}}
\quad
\subfigure[BL=4.7 L/min, 0.10 $^\circ$C.$cm^{-1}$.]{%
\includegraphics[width=0.455\textwidth]{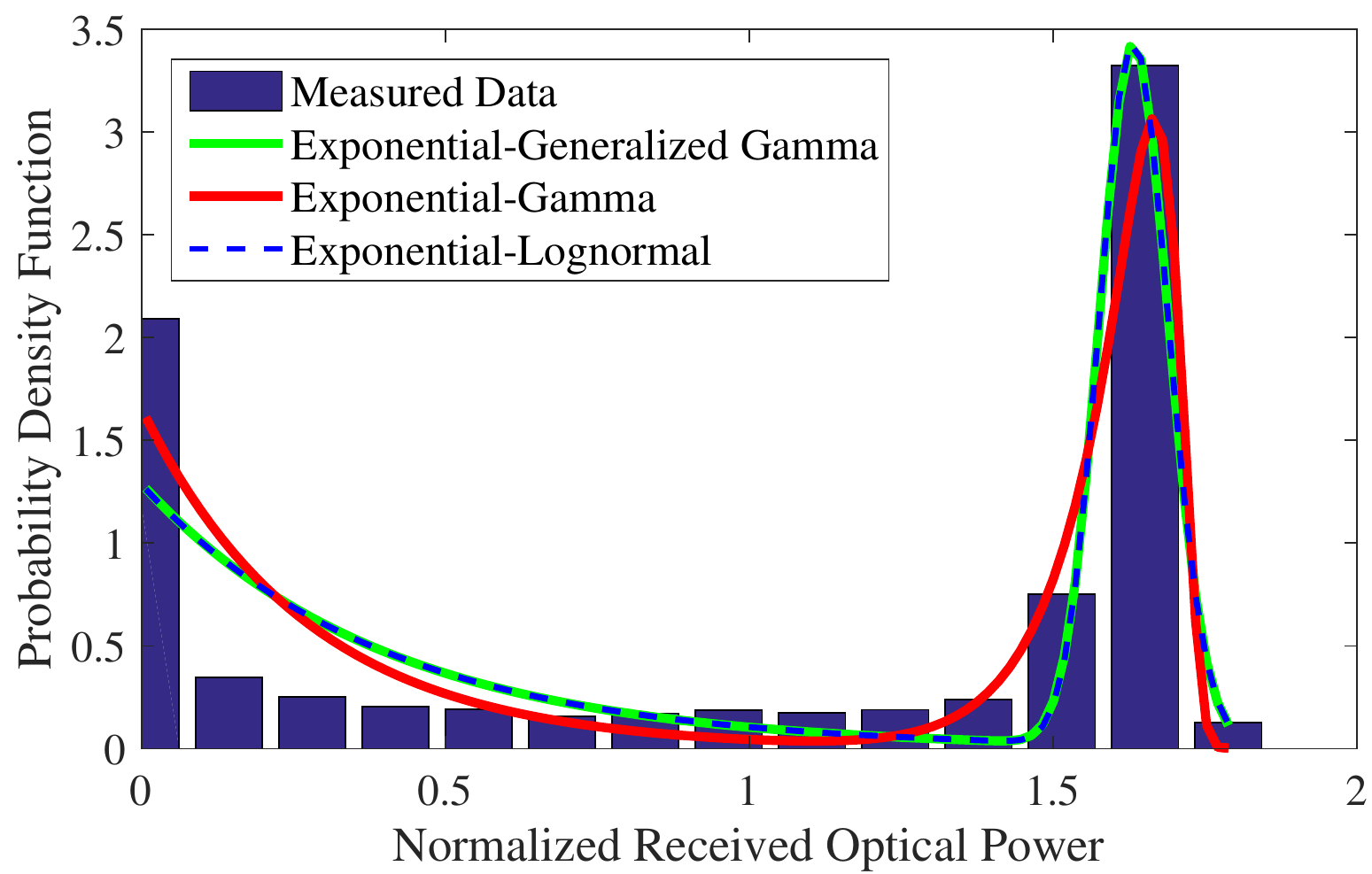}
\label{fig:subfigure22}}
\subfigure[BL=16.5 L/min, 0.22 $^\circ$C.$cm^{-1}$.]{%
\includegraphics[width=0.455\textwidth]{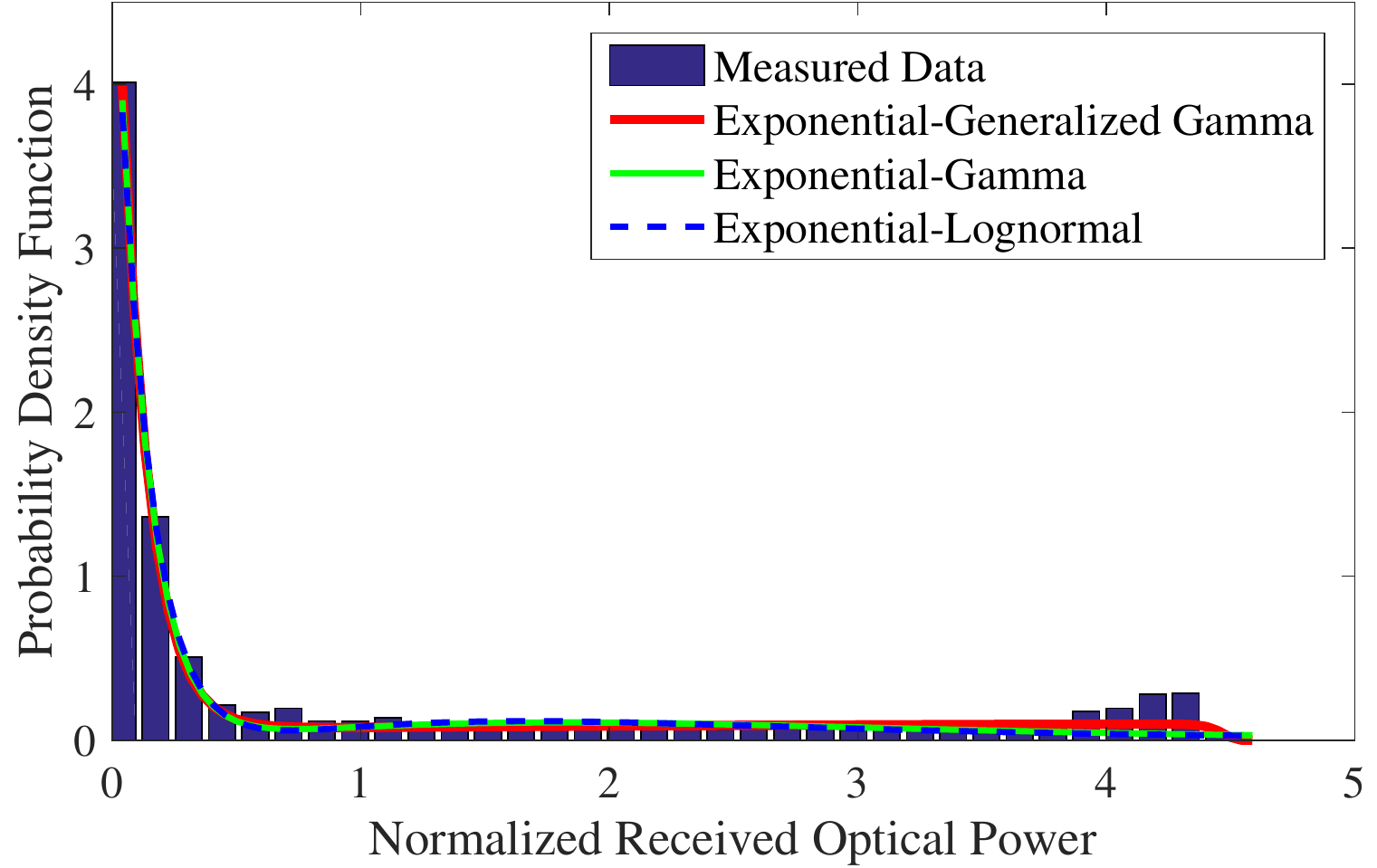}
\label{fig:subfigure33}}
\quad
\subfigure[BL=23.6 L/min, 0.22 $^\circ$C.$cm^{-1}$.]{%
\includegraphics[width=0.455\textwidth]{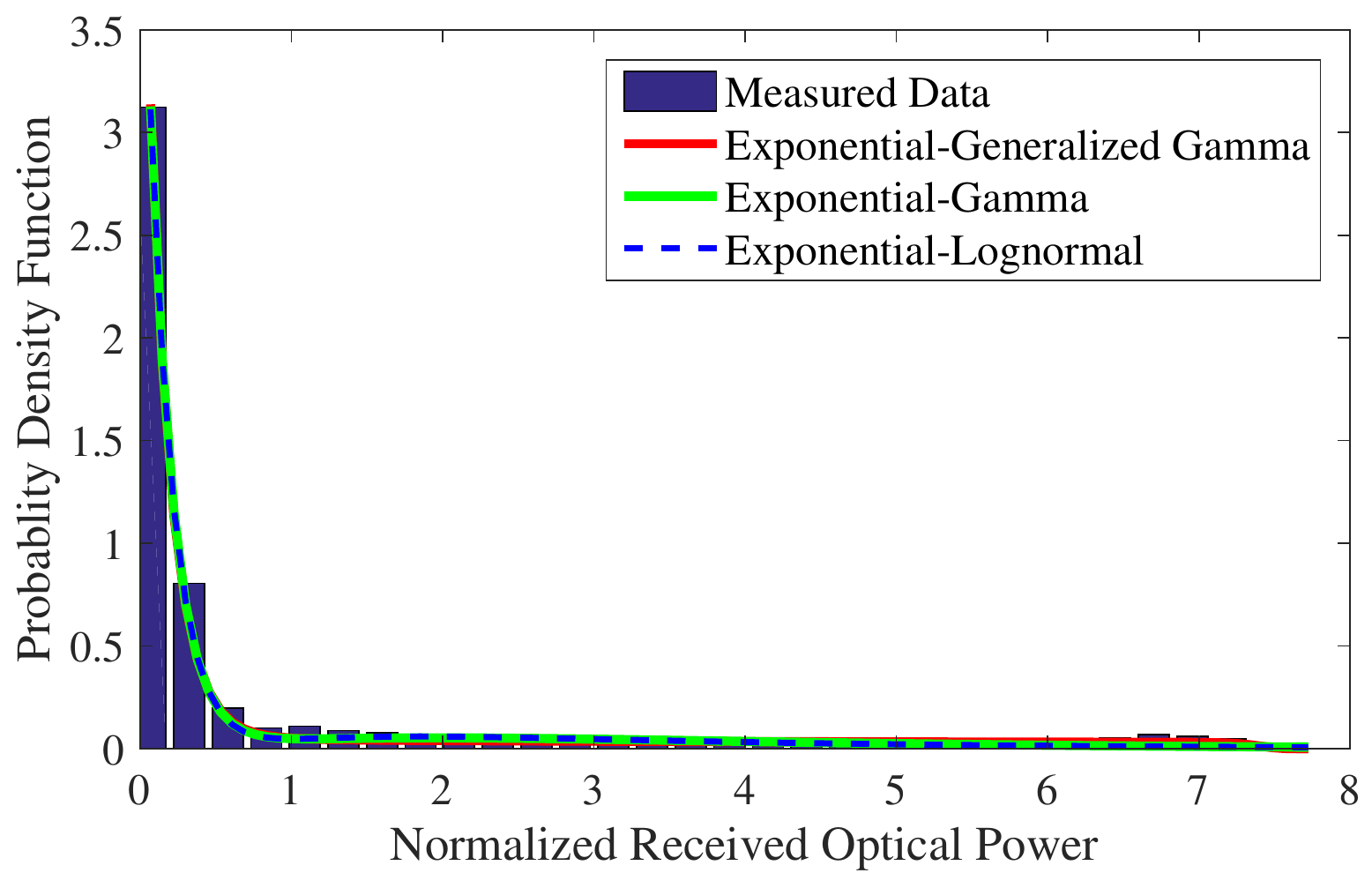}
\label{fig:subfigure44}}
\caption{Histograms of the measured data along with the new EGG, the EG, and the Exponential-Lognormal PDFs under various levels of air bubbles and different temperature gradients.}
\label{fig:gradientTemp}
\end{figure}

In this section, we show how the new proposed EGG model provides an excellent agreement with the measured data under all channel conditions.
 Fig.~\ref{fig:gradientTemp} shows the histograms of the experimental data along with the fitness of the new EGG probability distribution function under various levels of air bubbles and different levels of temperature gradient, based on the parameters of Table~\ref{goodnessoffit1}. For comparison purposes, we also show the PDFs of EG and Exponential-Lognormal. We can clearly observe that as the temperature-induced turbulence increases, the histogram is more skewed to the left (Figs.~\ref{fig:gradientTemp}(c) \& \ref{fig:gradientTemp}(d)), indicating a stretch and the shape of the peak becomes wider. Note that both EG and Exponential-Lognormal distributions fit the measured data quite well when the temperature gradient in the underwater channel is low (Figs.~\ref{fig:gradientTemp}(a) \& \ref{fig:gradientTemp}(b)). As the water temperature gradient increases, both distributions can not follow the stretching shape of the graph and start to loose accuracy. However, as clearly seen in Fig. \ref{fig:gradientTemp}, the proposed EGG model perfectly matches the measured data under all channel conditions from weak to strong turbulence. This excellent agreement clearly demonstrates that the EGG distribution is the most suitable probability distribution to characterize underwater optical signal irradiance fluctuations due to both air bubbles and temperature-induced turbulence. Interestingly, this new distribution not only provides excellent agreement with the measured data under all conditions of turbulence but also serves as a more tractable model that introduces a lot of analytical facilities in deriving easy-to-use expressions for several performance metrics of UWOC systems such as the outage probability and the average BER.

Table~\ref{goodnessoffit1} also compares the scintillation index of the experimental data to the scintillation index of the new EGG model as well as the EG and the Exponential-Lognormal models. The scintillation index of the measured data is computed according to (\ref{SIdef}), and the scintillation indices of the proposed EGG model as well as the EG and the Exponential-Lognormal model are calculated theoretically using (\ref{SInewmodel}),  \cite[Eq.(4)]{ConfUWOC}, and \cite[Eq.(8)]{ExpLN}, respectively. As shown in Table~\ref{goodnessoffit1}, the scintillation index calculated from the new PDF is the closest to the scintillation index obtained from the measured data. In addition, we can deduce from Table~\ref{goodnessoffit1} that as the level of air bubbles or temperature gradient increases, the strength of the turbulence increases, and therefore the scintillation index becomes larger, as expected.
In addition, the scale parameter of the EGG increases indicating the left-skewness or the stretching nature of the histogram as the channel turbulence increases. Overall, the EGG distribution gave the best performance in terms of quality of fit to the measured data.

The results of MSE and R$^{2}$ goodness of fit tests for the EGG, the EG, and the Exponential-Lognormal PDFs are also listed in Table~\ref{goodnessoffit1}. It is clearly illustrated that the MSE values corresponding to the new EGG model are the smallest, under all turbulence conditions. Furthermore, it can be observed that R$^{2}$ measures
associated with the EGG model have the highest values. These results indicate that the new PDF provides a better fit to the experimental data and therefore strongly support the application of the EGG model for turbulence induced-fading in UWOC channels, as a more accurate and simple alternative to the Exponential-Lognormal model.
\begin{table}[h]
\centering
\caption{Measured and estimated parameters of the EGG, the EG, and the Exponential-Lognormal distributions along with the goodness of fits tests for gradient temperature UWOC system.}
\begin{adjustbox}{max width=\linewidth}
\begin{tabular}{@{}|c|c|c|c|l|c|c|c|l|l|l|c|c|c|l|c|c|@{}}
\toprule
\multirow{2}{*}{\textbf{\begin{tabular}[c]{@{}c@{}}Bubbles Level\\ BL (L/min)\end{tabular}}} & \multirow{2}{*}{\textbf{\begin{tabular}[c]{@{}c@{}}Temperature\\ Gradient\\ ($\bm{^\circ}$C.$\bm{cm^{-1}}$)\end{tabular}}} & \multirow{2}{*}{$\bm{\sigma_{I,meas}^2}$} & \multicolumn{4}{c|}{\textbf{Exponential-Generalized Gamma Distribution}} & \multicolumn{6}{c|}{\textbf{Exponential-Gamma Distribution}} & \multicolumn{4}{c|}{\textbf{Exponential-Lognormal Distribution}} \\ \cmidrule(l){4-17}
 &  &  & $\bm{\sigma_{I}^2}$ & \multicolumn{1}{c|}{$\bm{(\omega, \lambda, a, b, c)}$} & \textbf{MSE} & $\bm{R^2}$ & $\bm{\sigma_{I}^2}$ & \multicolumn{3}{c|}{$\bm{(\omega, \lambda, \alpha, \beta)}$} & \textbf{MSE} & $\bm{R^2}$ & $\bm{\sigma_{I}^2}$ & \multicolumn{1}{c|}{$\bm{(\omega, \lambda, \mu, \sigma^2)}$} & \textbf{MSE} & \textbf{$\bm{R^2}$} \\ \midrule
$2.4$ & $0.05$ & $0.1494$ & $0.1484$ & $\begin{aligned}&\left(0.2130, 0.3291,\right. \\&\left. 1.4299, 1.1817, 17.1984\right)\end{aligned}$ & \begin{tabular}[c]{@{}c@{}}$3.5218$\\ $\times 10^{-6}$\end{tabular} & $0.9918$ & $0.1521$ & \multicolumn{3}{l|}{$\begin{aligned}&\left(0.2324, 0.3831,\right. \\&\left. 393.5944, 0.0030\right)\end{aligned}$} & \begin{tabular}[c]{@{}c@{}}$4.0131$\\ $\times 10^{-6}$\end{tabular} & $0.9772$ & $0.1521$ & $\begin{aligned}&\left(0.2338, 0.3869,\right. \\&\left. 0.1702, 0.0025\right)\end{aligned}$ & \begin{tabular}[c]{@{}c@{}}$4.0777$\\ $\times 10^{-6}$\end{tabular} & $0.9754$ \\ \midrule
$2.4$ & $0.10$ & $0.1693$ & $0.1659$ & $\begin{aligned}&\left(0.2108, 0.2694,\right. \\&\left. 0.6020, 1.2795, 21.1611\right)\end{aligned}$ & \begin{tabular}[c]{@{}c@{}}$2.3385$\\ $\times 10^{-6}$\end{tabular} & $0.9870$ & $0.1726$ & \multicolumn{3}{l|}{$\begin{aligned}&\left(0.2570, 0.3897,\right. \\&\left. 227.8358, 0.0053\right)\end{aligned}$} & \begin{tabular}[c]{@{}c@{}}$4.1168$\\ $\times 10^{-6}$\end{tabular} & $0.9770$ & $0.1734$ & $\begin{aligned}&\left(0.2596, 0.3962,\right. \\&\left. 0.1899, 0.0043\right)\end{aligned}$ & \begin{tabular}[c]{@{}c@{}}$4.2987$\\ $\times 10^{-6}$\end{tabular} & $0.9754$ \\ \midrule
$2.4$ & $0.15$ & $0.1953$ & $0.1915$ & $\begin{aligned}&\left(0.1807, 0.1641,\right. \\&\left. 0.2334, 1.4201, 22.5924\right)\end{aligned}$ & \begin{tabular}[c]{@{}c@{}}$1.2139$\\ $\times 10^{-6}$\end{tabular} & $0.9522$ & $0.2033$ & \multicolumn{3}{l|}{$\begin{aligned}&\left(0.2877, 0.4077,\right. \\&\left. 79.2682, 0.0156\right)\end{aligned}$} & \begin{tabular}[c]{@{}c@{}}$4.7317$\\ $\times 10^{-6}$\end{tabular} & $0.7690$ & $0.2066$ & $\begin{aligned}&\left(0.2963, 0.4244,\right. \\&\left. 0.2111, 0.0122\right)\end{aligned}$ & \begin{tabular}[c]{@{}c@{}}$5.2412$\\ $\times 10^{-6}$\end{tabular} & $0.7539$ \\ \midrule
$2.4$ & $0.20$ & $0.2221$ & $0.2178$ & $\begin{aligned}&\left(0.1665, 0.1207,\right. \\&\left. 0.1559, 1.5216, 22.8754\right)\end{aligned}$ & \begin{tabular}[c]{@{}c@{}}$9.6766$\\ $\times 10^{-7}$\end{tabular} & $0.9435$ & $0.2346$ & \multicolumn{3}{l|}{$\begin{aligned}&\left(0.3183, 0.4246,\right. \\&\left. 48.5897, 0.0261\right)\end{aligned}$} & \begin{tabular}[c]{@{}c@{}}$5.2964$\\ $\times 10^{-6}$\end{tabular} & $0.7739$ & $0.2413$ & $\begin{aligned}&\left(0.3344, 0.4511,\right. \\&\left. 0.2343, 0.0193\right)\end{aligned}$ & \begin{tabular}[c]{@{}c@{}}$6.0558$\\ $\times 10^{-6}$\end{tabular} & $0.7563$ \\ \midrule
$4.7$ & $0.05$ & $0.4523$ & $0.4201$ & $\begin{aligned}&\left(0.4589, 0.3449,\right. \\&\left. 1.0421, 1.5768, 35.9424\right)\end{aligned}$ & \begin{tabular}[c]{@{}c@{}}$3.6264$\\ $\times 10^{-5}$\end{tabular} & $0.9135$ & $0.4171$ & \multicolumn{3}{l|}{$\begin{aligned}&\left(0.4811, 0.3926,\right. \\&\left. 1.3828\times 10^{3}, 0.0011\right)\end{aligned}$} & \begin{tabular}[c]{@{}c@{}}$4.2544$\\ $\times 10^{-5}$\end{tabular} & $0.8747$ & $0.4171$ & $\begin{aligned}&\left(0.4817, 0.3939,\right. \\&\left.0.4464, 7.1846\times 10^{-4}\right)\end{aligned}$ & \begin{tabular}[c]{@{}c@{}}$4.2718$\\ $\times 10^{-5}$\end{tabular} & $0.8740$ \\ \midrule
$4.7$ & $0.10$ & $0.5059$ & $0.4769$ & $\begin{aligned}&\left(0.4539, 0.2744,\right. \\&\left. 0.3008, 1.7053, 54.1422\right)\end{aligned}$ & \begin{tabular}[c]{@{}c@{}}$3.1442$\\ $\times 10^{-5}$\end{tabular} & $0.9123$ & $0.4646$ & \multicolumn{3}{l|}{$\begin{aligned}&\left(0.5129, 0.3978,\right. \\&\left. 822.3038, 0.0020\right)\end{aligned}$} & \begin{tabular}[c]{@{}c@{}}$4.9171$\\ $\times 10^{-5}$\end{tabular} & $0.8758$ & $0.4645$ & $\begin{aligned}&\left(0.5140, 0.4001,\right. \\&\left.0.4907, 0.0012\right)\end{aligned}$ & \begin{tabular}[c]{@{}c@{}}$4.9545$\\ $\times 10^{-5}$\end{tabular} & $0.8739$ \\ \midrule
$16.5$ & $0.22$ & $2.0493$ & $1.9328$ & $\begin{aligned}&\left(0.6238, 0.1094,\right. \\&\left. 0.0111, 4.4750, 105.3550\right)\end{aligned}$ & \begin{tabular}[c]{@{}c@{}}$1.3212$\\ $\times 10^{-6}$\end{tabular} & $0.9909$ & $2.2447$ & \multicolumn{3}{l|}{$\begin{aligned}&\left(0.6527, 0.1194,\right. \\&\left. 3.1458, 0.8439\right)\end{aligned}$} & \begin{tabular}[c]{@{}c@{}}$1.7627$\\ $\times 10^{-6}$\end{tabular} & $0.9861$ & $2.7411$ & $\begin{aligned}&\left(0.6628, 0.1257,\right. \\&\left.0.8547, 0.3458\right)\end{aligned}$ & \begin{tabular}[c]{@{}c@{}}$1.9621$\\ $\times 10^{-6}$\end{tabular} & $0.9844$ \\ \midrule
$23.6$ & $0.22$ & $3.3238$ & $3.1952$ & $\begin{aligned}&\left(0.7210, 0.1479,\right. \\&\left. 0.0121, 7.4189, 65.6983\right)\end{aligned}$ & \begin{tabular}[c]{@{}c@{}}$1.8010$\\ $\times 10^{-6}$\end{tabular} & $0.9940$ & $3.5978$ & \multicolumn{3}{l|}{$\begin{aligned}&\left(0.7518, 0.1536,\right. \\&\left. 2.2364, 1.5937\right)\end{aligned}$} & \begin{tabular}[c]{@{}c@{}}$1.9224$\\ $\times 10^{-6}$\end{tabular} & $0.9938$ & $4.5424$ & $\begin{aligned}&\left(0.7602, 0.1577,\right. \\&\left.1.0955, 0.4713\right)\end{aligned}$ & \begin{tabular}[c]{@{}c@{}}$2.0011$\\ $\times 10^{-6}$\end{tabular} & $0.9940$ \\ \bottomrule
\end{tabular}
\end{adjustbox}
\label{goodnessoffit1}
\end{table}

\subsection{Turbulent UWOC Channels with Uniform Temperature}
\begin{figure}[htp]
\centering
\subfigure[BL=0 L/min, Salty Water.]{%
\includegraphics[width=0.47\textwidth]{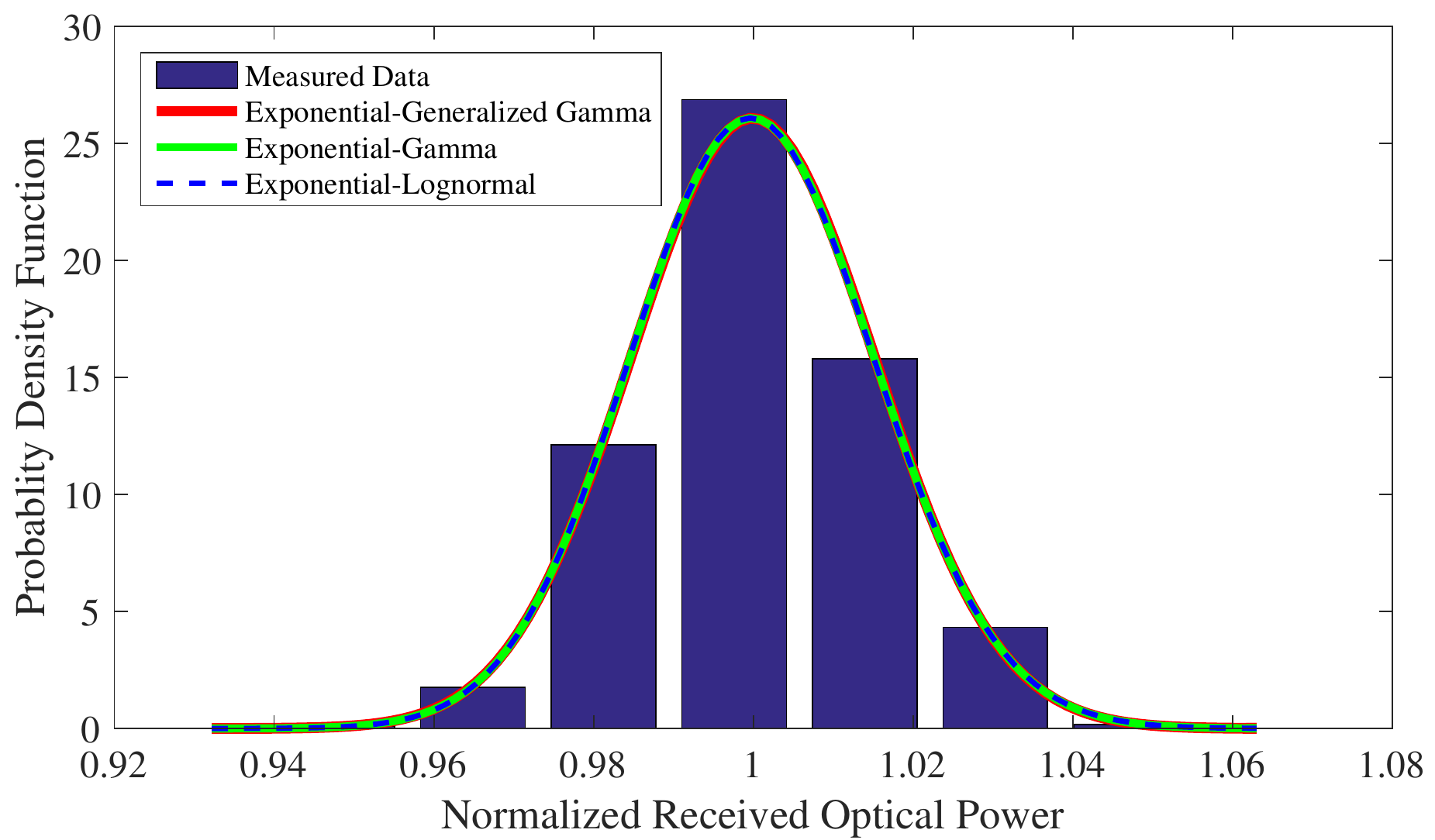}
\label{fig:subfigure1}}
\quad
\subfigure[BL=2.4 L/min, Salty Water.]{%
\includegraphics[width=0.465\textwidth]{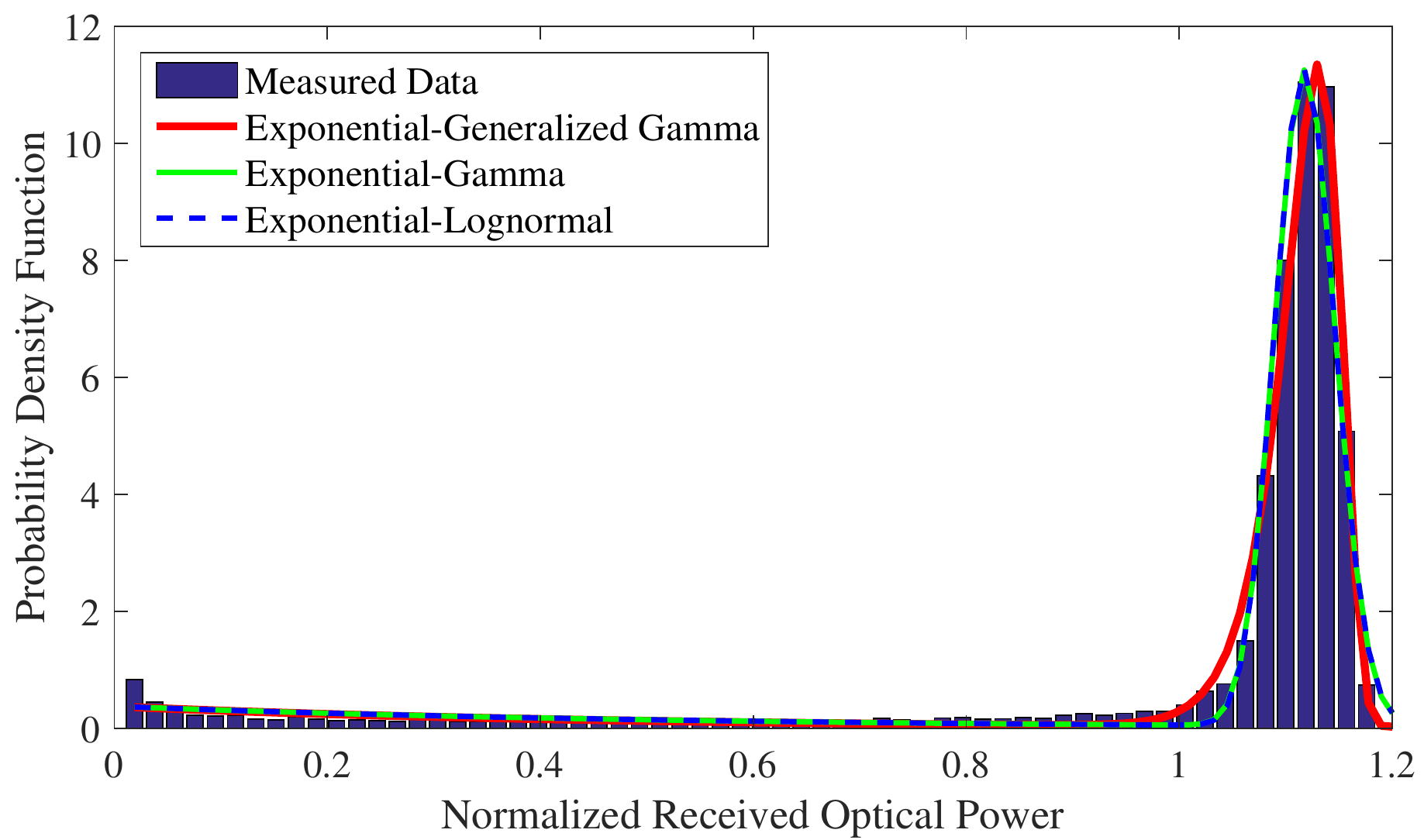}
\label{fig:subfigure2}}
\subfigure[BL=7.1 L/min, Salty Water.]{%
\includegraphics[width=0.465\textwidth]{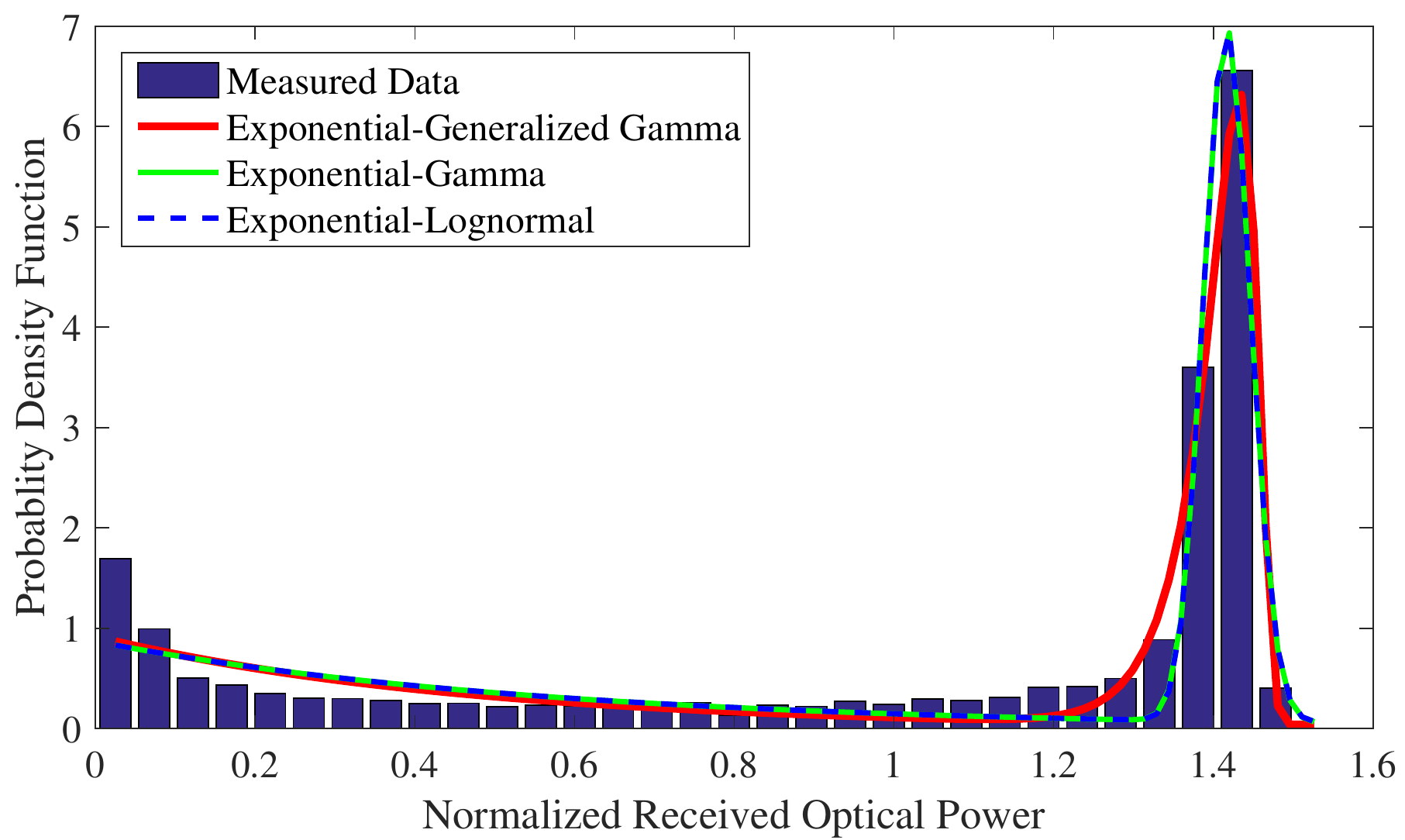}
\label{fig:subfigure3}}
\quad
\subfigure[BL=16.5 L/min, Salty Water.]{%
\includegraphics[width=0.47\textwidth]{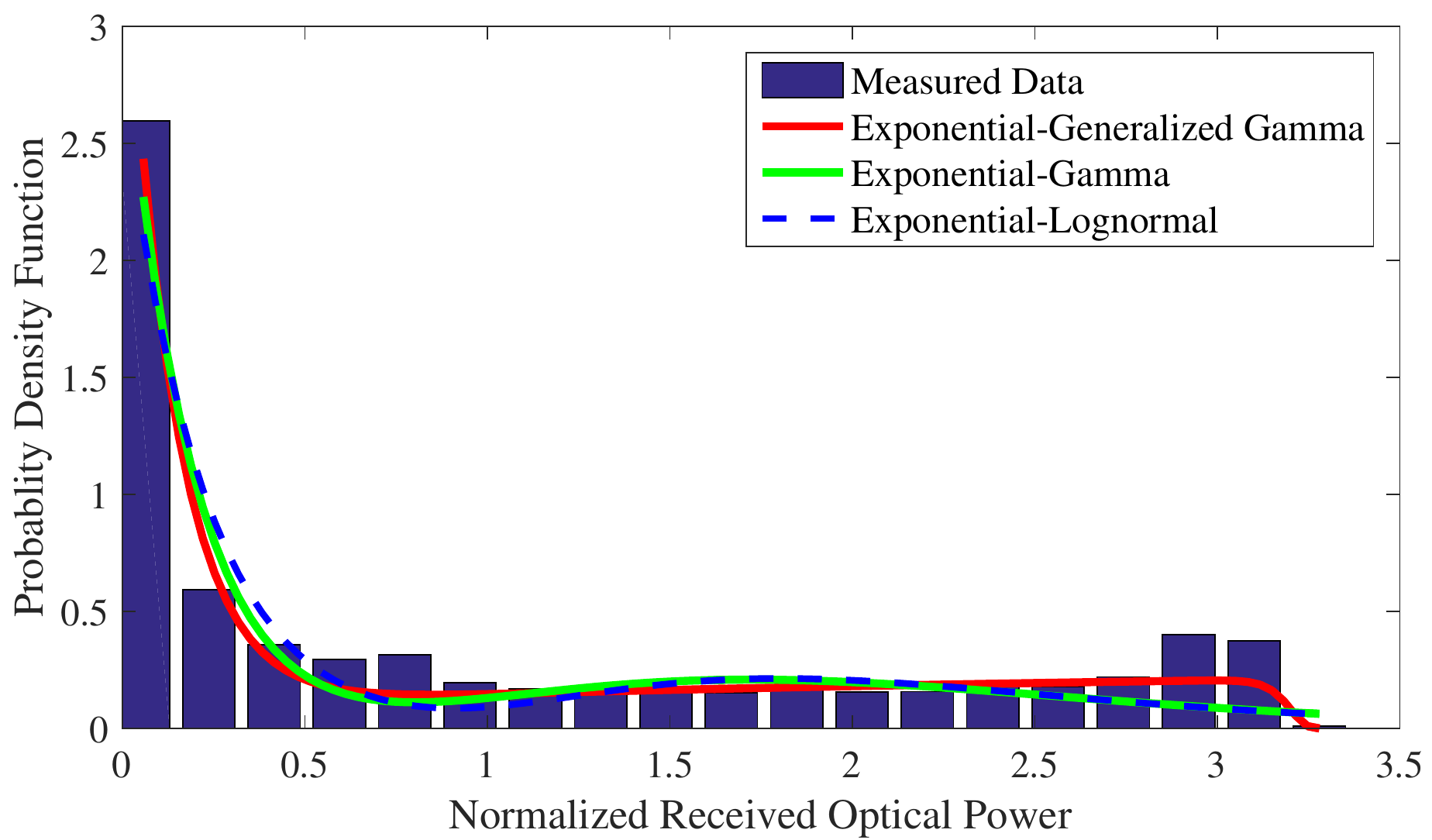}
\label{fig:subfigure4}}
\subfigure[BL=0 L/min, Fresh Water.]{%
\includegraphics[width=0.47\textwidth]{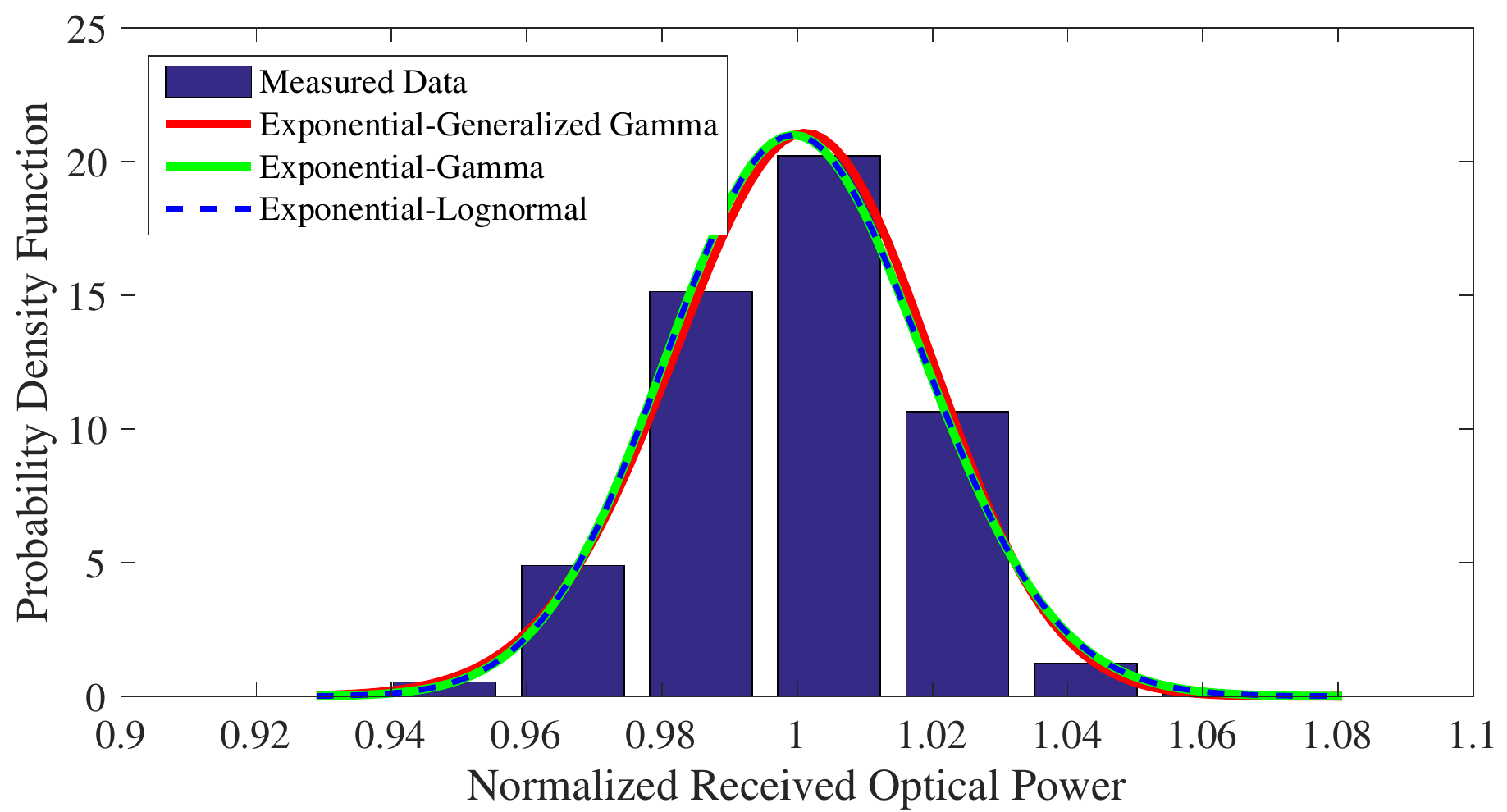}
\label{fig:subfigure11}}
\quad
\subfigure[BL=2.4 L/min, Fresh Water.]{%
\includegraphics[width=0.465\textwidth]{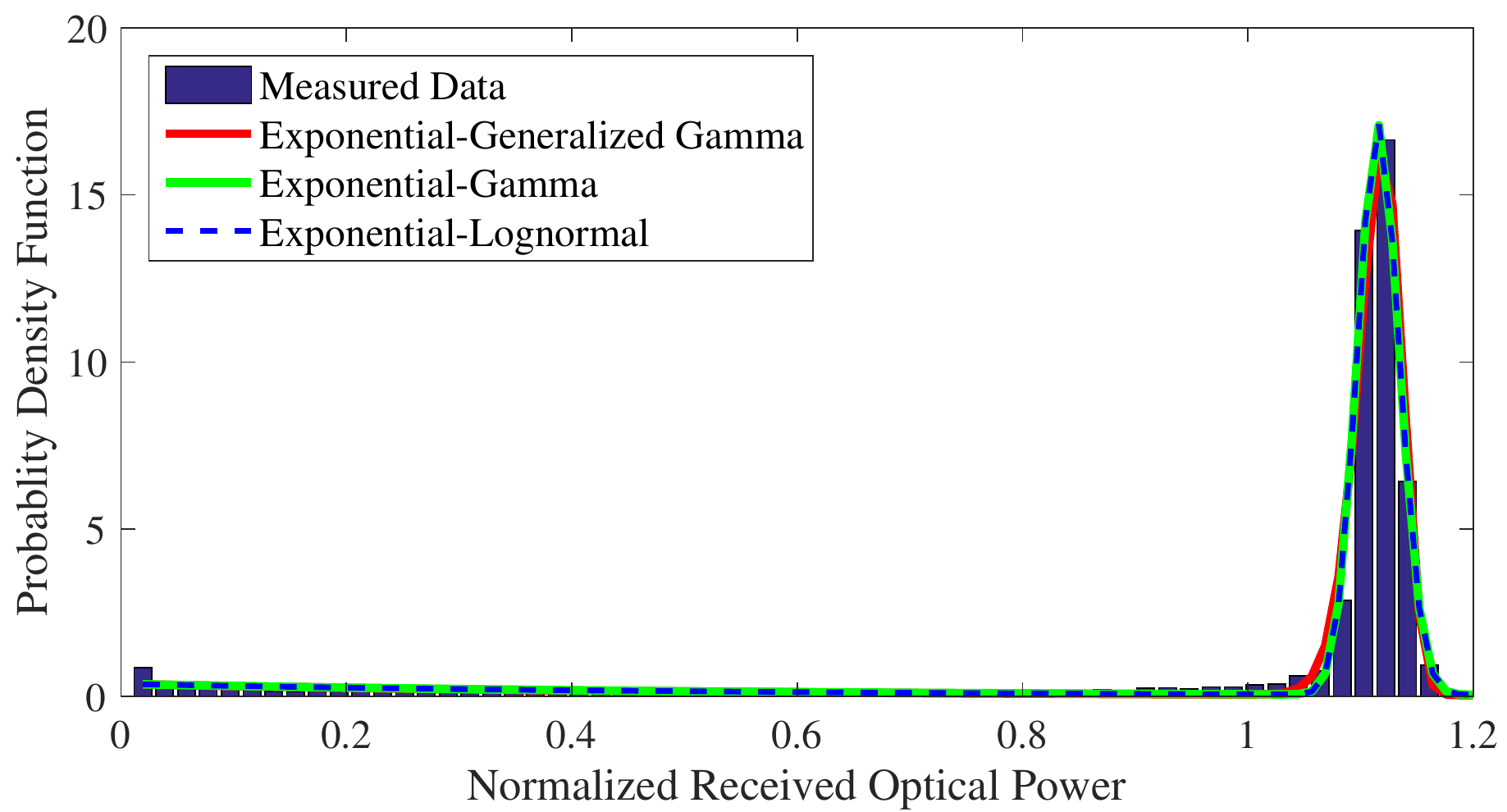}
\label{fig:subfigure22}}
\subfigure[BL=7.1 L/min, Fresh Water.]{%
\includegraphics[width=0.465\textwidth]{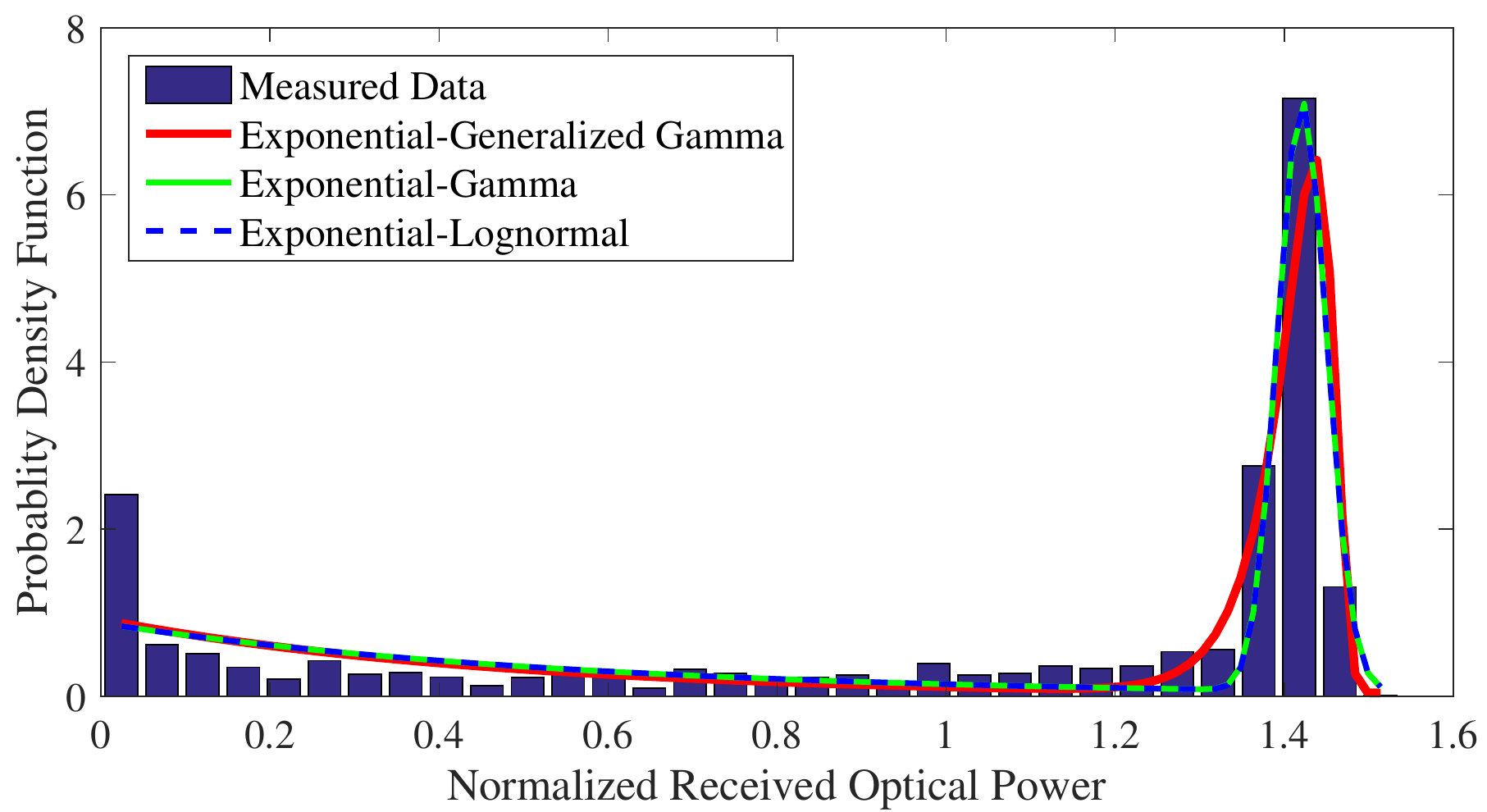}
\label{fig:subfigure33}}
\quad
\subfigure[BL=16.5 L/min, Fresh Water.]{%
\includegraphics[width=0.47\textwidth]{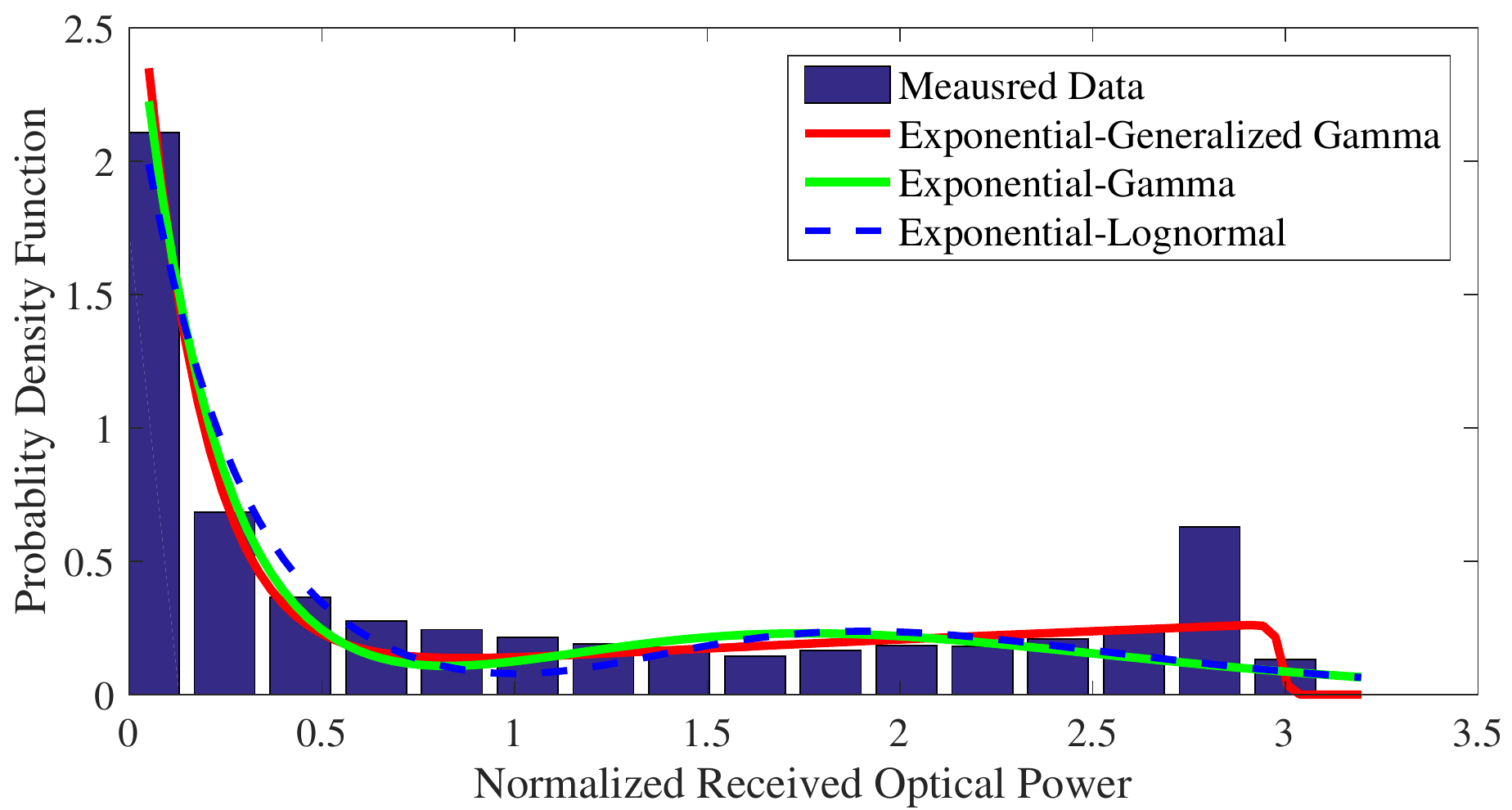}
\label{fig:subfigure44}}
\caption{Histograms of the measured data along with the new EGG, the EG,and the Exponential-Lognormal PDFs under different levels of air bubbles for fresh as well as salty waters.}
\label{fig:UniformTemp}
\end{figure}

In this section, we present experimental results for the proposed EGG distribution model under uniform temperature, for both salty and fresh waters. Fig.~\ref{fig:UniformTemp} illustrates histograms of the experimental data together with the EGG distribution as well as the EG and the Exponential-Lognormal distributions using different levels of air bubbles, based on the parameters of Table~\ref{goodnessoffit2}. Results corresponding to the third and the eighth rows of
Table~\ref{goodnessoffit2} are not included in Fig.~\ref{fig:UniformTemp} due to space limitation.

\normalsize

As shown in Fig.~\ref{fig:UniformTemp}, the proposed EGG model matches perfectly the measured data for
all bubbles levels which correspond to different turbulence conditions varying from weak to strong turbulence, for both fresh and salty waters. This excellent match indicates the effectiveness of our model to describe irradiance fluctuations in UWOC channels.
Moreover, an excellent agreement between the Exponential-Lognormal model and the EG model is depicted in Fig.~\ref{fig:UniformTemp}. Indeed, the plots of the two distributions are almost indistinguishable, and both fit very well to the measured data collected at different levels of air bubbles, for both types of water.
These facts make the EG distribution an attractive model to describe turbulence-induced fading in thermally uniform UWOC channels
operating under weak, moderate, and strong turbulence conditions.
\begin{table}[h]
\centering
\caption{Measured and estimated parameters of the EGG, the EG, and the Exponential-Lognormal distributions along with the goodness of fits tests for thermally uniform UWOC system.}
\begin{adjustbox}{max width=\linewidth}
\begin{tabular}{@{}c|c|c|c|l|c|c|c|l|l|l|c|c|c|l|c|c|@{}}
\cmidrule(l){2-17}
 & \multirow{2}{*}{\textbf{\begin{tabular}[c]{@{}c@{}}Bubbles Level\\ BL (L/min)\end{tabular}}} & \multirow{2}{*}{$\bm{\sigma_{I,meas}^2}$} & \multicolumn{4}{c|}{\textbf{Exponential-Generalized Gamma Distribution}} & \multicolumn{6}{c|}{\textbf{Exponential-Gamma Distribution}} & \multicolumn{4}{c|}{\textbf{Exponential-Lognormal Distribution}} \\ \cmidrule(l){4-17}
 &  &  & $\bm{\sigma_{I}^2}$ & \multicolumn{1}{c|}{$\bm{(\omega, \lambda, a, b, c)}$} & \textbf{MSE} & $\bm{R^2}$ & $\bm{\sigma_{I}^2}$ & \multicolumn{3}{c|}{$\bm{(\omega, \lambda, \alpha, \beta)}$} & \textbf{MSE} & $\bm{R^2}$ & $\bm{\sigma_{I}^2}$ & \multicolumn{1}{c|}{$\bm{(\omega, \lambda, \mu, \sigma^2)}$} & \textbf{MSE} & \bm{$R^2$} \\ \midrule
\multicolumn{1}{|c|}{\multirow{5}{*}{\rotatebox[origin=c]{90}{\textbf{Salty Water~~~~~~~~~~~~~~}}}} & $0$ & \begin{tabular}[c]{@{}c@{}}$2.3407$\\ $\times10^{-4}$\end{tabular} & \begin{tabular}[c]{@{}c@{}}$2.3408$\\ $\times10^{-4}$\end{tabular} & $\begin{aligned}&\left(1.4684 \times 10^{-23}, 0.9853,\right. \\&\left. 1.0126\times 10^{3}, 0.0344, 2.0541\right)\end{aligned}$ & \begin{tabular}[c]{@{}c@{}}$7.9274$\\ $\times 10^{-7}$\end{tabular} & $0.9953$ & \begin{tabular}[c]{@{}c@{}}$2.3404$\\ $\times10^{-4}$\end{tabular} & \multicolumn{3}{l|}{$\begin{aligned}&\left(1.5540 \times 10^{-18}, 0.9820,\right. \\&\left. 4.2727\times 10^{-3}, 2.3404 \times 10^{-4}\right)\end{aligned}$} & \begin{tabular}[c]{@{}c@{}}$7.8872$\\ $\times 10^{-7}$\end{tabular} & $0.9957$ & \begin{tabular}[c]{@{}c@{}}$2.3409$\\ $\times10^{-4}$\end{tabular} & $\begin{aligned}&\left(7.0109 \times 10^{-12}, 0.9786,\right. \\&\left. -1.1703\times 10^{-4}, 2.3406\times 10^{-4}\right)\end{aligned}$ & \begin{tabular}[c]{@{}c@{}}$7.8489$\\ $\times 10^{-7}$\end{tabular} & $0.9960$ \\ \cmidrule(l){2-17}
\multicolumn{1}{|c|}{} & $2.4$ & $0.0821$ & $0.1006$ & $\begin{aligned}&\left(0.1770, 0.4687,\right. \\&\left. 0.7736, 1.1372, 49.1773\right)\end{aligned}$ & \begin{tabular}[c]{@{}c@{}}$5.4258$\\ $\times 10^{-7}$\end{tabular} & $0.9913$ & $0.1142$ & \multicolumn{3}{l|}{$\begin{aligned}&\left(0.2037, 0.5369,\right. \\&\left. 1.5559\times 10^{3}, 7.1885 \times 10^{-4}\right)\end{aligned}$} & \begin{tabular}[c]{@{}c@{}}$6.9256$\\ $\times 10^{-7}$\end{tabular} & $0.9705$ & $0.1147$ & $\begin{aligned}&\left(0.2045, 0.5389,\right. \\&\left. 0.1117, 6.3979 \times 10^{-4}\right)\end{aligned}$ & \begin{tabular}[c]{@{}c@{}}$6.9907$\\ $\times 10^{-7}$\end{tabular} & $0.9692$ \\ \cmidrule(l){2-17}
\multicolumn{1}{|c|}{} & $4.7$ & $0.1216$ & $0.1308$ & $\begin{aligned}&\left(0.2064, 0.3953,\right. \\&\left. 0.5307, 1.2154, 35.7368\right)\end{aligned}$ & \begin{tabular}[c]{@{}c@{}}$3.3475$\\ $\times 10^{-7}$\end{tabular} & $0.9772$ & $0.1450$ & \multicolumn{3}{l|}{$\begin{aligned}&\left(0.2436, 0.4818,\right. \\&\left. 501.9905, 0.0023\right)\end{aligned}$} & \begin{tabular}[c]{@{}c@{}}$4.9395$\\ $\times 10^{-7}$\end{tabular} & $0.9317$ & $0.1458$ & $\begin{aligned}&\left(0.2451, 0.4854,\right. \\&\left. 0.1536, 0.0020\right)\end{aligned}$ & \begin{tabular}[c]{@{}c@{}}$5.0477$\\ $\times 10^{-7}$\end{tabular} & $0.9283$ \\ \cmidrule(l){2-17}
\multicolumn{1}{|c|}{} & $7.1$ & $0.2917$ & $0.3111$ & $\begin{aligned}&\left(0.4344, 0.4747,\right. \\&\left. 0.3935, 1.4506, 77.0245\right)\end{aligned}$ & \begin{tabular}[c]{@{}c@{}}$9.0251$\\ $\times 10^{-7}$\end{tabular} & $0.9657$ & $0.3372$ & \multicolumn{3}{l|}{$\begin{aligned}&\left(0.4876, 0.5612,\right. \\&\left. 2.2911\times 10^{3}, 6.1870\times 10^{-4}\right)\end{aligned}$} & \begin{tabular}[c]{@{}c@{}}$1.4489$\\ $\times 10^{-6}$\end{tabular} & $0.9484$ & $0.3376$ & $\begin{aligned}&\left(0.4882, 0.5622,\right. \\&\left. 0.3488, 4.3403\times 10^{-4}\right)\end{aligned}$ & \begin{tabular}[c]{@{}c@{}}$1.4562$\\ $\times 10^{-6}$\end{tabular} & $0.9480$ \\ \cmidrule(l){2-17}
\multicolumn{1}{|c|}{} & $16.5$ & $1.1847$ & $1.1273$ & $\begin{aligned}&\left(0.4951, 0.1368,\right. \\&\left. 0.0161, 3.2033, 82.1030\right)\end{aligned}$ & \begin{tabular}[c]{@{}c@{}}$1.2536$\\ $\times 10^{-6}$\end{tabular} & $0.9690$ & $1.2456$ & \multicolumn{3}{l|}{$\begin{aligned}&\left(0.5740, 0.1853,\right. \\&\left. 5.6545, 0.3710\right)\end{aligned}$} & \begin{tabular}[c]{@{}c@{}}$1.7021$\\ $\times 10^{-6}$\end{tabular} & $0.9191$ & $1.2995$ & $\begin{aligned}&\left(0.6113, 0.2240,\right. \\&\left. 0.7345, 0.1407\right)\end{aligned}$ & \begin{tabular}[c]{@{}c@{}}$1.8384$\\ $\times 10^{-6}$\end{tabular} & $0.8843$ \\ \midrule
\multicolumn{1}{|c|}{\multirow{5}{*}{\rotatebox[origin=c]{90}{\textbf{Fresh~Water~~~~~~~~~~~~~}}}} & $0$ & \begin{tabular}[c]{@{}c@{}}$3.6039$\\ $\times 10^{-4}$\end{tabular} & \begin{tabular}[c]{@{}c@{}}$3.6044$\\ $\times 10^{-4}$\end{tabular} & $\begin{aligned}&\left(4.0628 \times 10^{-21}, 1.0225,\right. \\&\left. 30.8432, 0.6993, 9.5461\right)\end{aligned}$ & \begin{tabular}[c]{@{}c@{}}$7.0429$\\ $\times 10^{-7}$\end{tabular} & $0.9982$ & \begin{tabular}[c]{@{}c@{}}$3.6108$\\ $\times10^{-4}$\end{tabular} & \multicolumn{3}{l|}{$\begin{aligned}&\left(8.2201\times 10^{-17}, 0.9912,\right. \\&\left. 2.7695\times 10^{3}, 3.6108\times 10^{-4}\right)\end{aligned}$} & \begin{tabular}[c]{@{}c@{}}$6.6882$\\ $\times 10^{-7}$\end{tabular} & $0.9948$ & \begin{tabular}[c]{@{}c@{}}$3.6195$\\ $\times10^{-4}$\end{tabular} & $\begin{aligned}&\left(1.3445\times 10^{-10}, 0.9884,\right. \\&\left. -1.8055\times 10^{-4}, 3.6149\times 10^{-4}\right)\end{aligned}$ & \begin{tabular}[c]{@{}c@{}}$6.6479$\\ $\times 10^{-7}$\end{tabular} & $0.9941$ \\ \cmidrule(l){2-17}
\multicolumn{1}{|c|}{} & $2.4$ & $0.0798$ & $0.1088$ & $\begin{aligned}&\left(0.1953, 0.5273,\right. \\&\left. 3.7291, 1.0721, 30.3214\right)\end{aligned}$ & \begin{tabular}[c]{@{}c@{}}$8.9304$\\ $\times 10^{-7}$\end{tabular} & $0.9822$ & $0.1157$ & \multicolumn{3}{l|}{$\begin{aligned}&\left(0.2069, 0.5560,\right. \\&\left. 3.6140\times 10^{3}, 3.0876\times 10^{-4}\right)\end{aligned}$} & \begin{tabular}[c]{@{}c@{}}$9.6972$\\ $\times 10^{-7}$\end{tabular} & $0.9944$ & $0.1159$ & $\begin{aligned}&\left(0.2073, 0.5567,\right. \\&\left.0.1095, 2.7575\times 10^{-4}\right)\end{aligned}$ & \begin{tabular}[c]{@{}c@{}}$9.7184$\\ $\times 10^{-7}$\end{tabular} & $0.9945$ \\ \cmidrule(l){2-17}
\multicolumn{1}{|c|}{} & $4.7$ & $0.1058$ & $0.1233$ & $\begin{aligned}&\left(0.2109, 0.4603,\right. \\&\left. 1.2526, 1.1501, 41.3258\right)\end{aligned}$ & \begin{tabular}[c]{@{}c@{}}$6.6032$\\ $\times 10^{-7}$\end{tabular} & $0.9827$ & $0.1320$ & \multicolumn{3}{l|}{$\begin{aligned}&\left(0.2298, 0.5075,\right. \\&\left. 2.0129\times 10^{3}, 5.6979\times 10^{-4}\right)\end{aligned}$} & \begin{tabular}[c]{@{}c@{}}$7.4577$\\ $\times 10^{-7}$\end{tabular} & $0.9822$ & $0.1323$ & $\begin{aligned}&\left(0.2302, 0.5085,\right. \\&\left.0.1369, 4.9552\times 10^{-4}\right)\end{aligned}$ & \begin{tabular}[c]{@{}c@{}}$7.4795$\\ $\times 10^{-7}$\end{tabular} & $0.9815$ \\ \cmidrule(l){2-17}
\multicolumn{1}{|c|}{} & $7.1$ & $0.2963$ & $0.3150$ & $\begin{aligned}&\left(0.3489, 0.4771,\right. \\&\left. 0.4319, 1.4531, 74.3650\right)\end{aligned}$ & \begin{tabular}[c]{@{}c@{}}$9.4207$\\ $\times 10^{-7}$\end{tabular} & $0.9612$ & $0.3380$ & \multicolumn{3}{l|}{$\begin{aligned}&\left(0.4866, 0.5549,\right. \\&\left. 2.3951\times 10^{3}, 5.9365\times 10^{-4}\right)\end{aligned}$} & \begin{tabular}[c]{@{}c@{}}$1.4547$\\ $\times 10^{-6}$\end{tabular} & $0.9533$ & $0.3383$ & $\begin{aligned}&\left(0.4870, 0.5556,\right. \\&\left.0.3518, 4.1578\times 10^{-4}\right)\end{aligned}$ & \begin{tabular}[c]{@{}c@{}}$1.4597$\\ $\times 10^{-6}$\end{tabular} & $0.9533$ \\ \cmidrule(l){2-17}
\multicolumn{1}{|c|}{} & $16.5$ & $1.1030$ & $1.0409$ & $\begin{aligned}&\left(0.5117, 0.1602,\right. \\&\left. 0.0075, 2.9963, 216.8356\right)\end{aligned}$ & \begin{tabular}[c]{@{}c@{}}$1.2822$\\ $\times 10^{-6}$\end{tabular} & $0.9625$ & $1.1495$ & \multicolumn{3}{l|}{$\begin{aligned}&\left(0.5717, 0.1992,\right. \\&\left. 6.7615, 0.3059\right)\end{aligned}$} & \begin{tabular}[c]{@{}c@{}}$1.8421$\\ $\times 10^{-6}$\end{tabular} & $0.9254$ & $1.1646$ & $\begin{aligned}&\left(0.6207, 0.2561,\right. \\&\left.0.7502, 0.1014\right)\end{aligned}$ & \begin{tabular}[c]{@{}c@{}}$2.0354$\\ $\times 10^{-6}$\end{tabular} & $0.8933$ \\ \bottomrule
\end{tabular}
\end{adjustbox}
\label{goodnessoffit2}
\end{table}

In addition, Table~\ref{goodnessoffit2} shows that increasing the salinity of the water by adding 118g of table salt into the fresh water tank does not have a significant impact on the scintillation index.

Table~\ref{goodnessoffit2} presents also the results of R$^2$ goodness of fit test
and estimated parameters of EGG, EG, and Exponential-Lognormal distributions. It is evident that R$^2$ values corresponding to the EGG model are the highest, under all turbulence conditions.
Overall, the EGG distribution gave the best performance in terms of quality of fit to the measured data.
Moreover, the EGG distribution is mathematically simple and attractive from the system performance analysis standpoint because it leads to closed-form and analytically tractable expressions for the outage probability and the average BER.


\section{Performance Analysis Based on the New Model}
In this section, we demonstrate the utility of the EGG PDF in modeling turbulence-induced fading for UWOC channels. The easy-to-use
expression of the PDF derived in (\ref{PDFmix}) can greatly simplify the analytical calculations of various performance metrics of interest over UWOC channels.
By using the EGG model, we can easily obtain tractable and closed-form expressions for the outage probability, the average BER, and the ergodic capacity over UWOC channels, and their analytical accuracy are verified by means of Monte Carlo simulations. The competing  Exponential-Lognormal proposed in \cite{ExpLN} is not handy when it comes to performance analysis, as, being based on the Lognormal distribution, it would lead to integral expressions that are intractable and hence need to be solved numerically.

In what follows, we study the performance of an UWOC system which employs either IM/DD or heterodyne techniques using a variety of modulation schemes. Note that, in this work we consider only the case in which
the scattering/absorption effects and the inter-symbol interference are not significant, and the underwater optical turbulence dominates the fading characteristics of the channel.

Assuming that the laser beam propagates through a mixture EGG turbulence channel with additive white Gaussian noise (AWGN), the received signal can be given as \cite{Simodetection}
\begin{align}
y=\eta I\,x+n,
\end{align}
where $\eta$ represents the optical-to-electrical conversion coefficient, $I$ is the normalized irradiance, $x \in {0, 1}$ denotes the transmitted information bit, and $n$ is the AWGN with zero mean and variance $N_0/2$.

\subsection{Probability Density Function}
Considering both types of detection techniques (IM/DD as well as heterodyne detection), the instantaneous signal-to-noise ratio (SNR) can be given as $\gamma=\left ( \eta \,I \right )^{r}/N_0$, with $r$ being the parameter specifying the type of detection technique (i.e. $r = 1$ for heterodyne detection and $r = 2$ for IM/DD). The average electrical SNR can be expressed as $\mu_r=\left (\eta\,\mathbb{E}[I]  \right )^{r}/N_0$ and is related to the average SNR ,$\overline{\gamma}$, such that $\overline{\gamma}=\left (\mu_r \,\mathbb{E}[I^r] \right )/\mathbb{E}[I]^r$.
\subsubsection{Heterodyne Detection}
In the case of heterodyne detection, the average electrical SNR, $\mu_1$, is defined as $\mu_1=\overline{\gamma}$. By using the transformation of the random variable $I$ along with \cite[Eq.(2.9.4)]{HTranforms} then \cite[Eq.(2.1.4)]{HTranforms} and \cite[Eq.(2.1.9)]{HTranforms}, the PDF of the SNR when the UWOC system is operating under the heterodyne detection can be given as
\begin{align}\label{SNRPDFHeterodyne}
f_{\gamma}(\gamma)=\frac{\omega}{\lambda \mu_1}e^{-\frac{\gamma}{\lambda\mu_1}}
+\frac{c(1-\omega)}{\Gamma(a)\gamma}{\rm{G}}_{0,1}^{1,0}\left[ \left ( \frac{\gamma}{b\mu_1} \right )^c\left| \begin{matrix} {-} \\ {a} \\ \end{matrix} \right. \right].
\end{align}
\subsubsection{Intensity Modulation/Direct Detection}
Under this type of detection, the average electrical SNR $\mu_2$ is given as
\begin{align}
\mu_2=\frac{\bar{\gamma}}{2 w \lambda^2+b^2(1-w)\Gamma\left ( a+2/c \right )/\Gamma(a)}.
\end{align}
Now, applying \cite[Eqs.(2.9.4), (2.1.4), and (2.1.9)]{HTranforms}, (\ref{PDFmix}) is easily transformed into
\begin{align}\label{SNRPDFIMDD}
f_{\gamma}(\gamma)=\frac{\omega}{2\lambda \sqrt{\mu_2\gamma}}e^{-\sqrt{\frac{\gamma}{\lambda^2\mu_2}}}
+\frac{c(1-\omega)}{2\Gamma(a)\gamma}{\rm{G}}_{0,1}^{1,0}\left[ \left ( \frac{\gamma}{b^2\mu_2} \right )^{\frac{c}{2}}\left| \begin{matrix} {-} \\ {a} \\ \end{matrix} \right. \right].
\end{align}
\subsubsection{Unified PDF Expression}
From (\ref{SNRPDFHeterodyne}) and (\ref{SNRPDFIMDD}) along with utilizing \cite[Eqs.~(8.4.3/1) and (8.2.2/15)]{PrudinkovVol3} we get the following unified PDF

\begin{align}\label{SNRPDFunified}
\nonumber f_{\gamma}(\gamma)&=\frac{\omega}{r\,\gamma} \,{\rm{G}}_{0,1}^{1,0}\left[ \frac{1}{\lambda}\left ( \frac{\gamma}{\mu_r} \right )^{\frac{1}{r}}\left| \begin{matrix} {-} \\ {1} \\ \end{matrix} \right. \right]\\
&+\frac{c(1-\omega)}{r\,\gamma\Gamma(a)}{\rm{G}}_{0,1}^{1,0}\left[\frac{1}{b^c} \left ( \frac{\gamma}{\mu_r} \right )^{\frac{c}{r}}\left| \begin{matrix} {-} \\ {a} \\ \end{matrix} \right. \right].
\end{align}

It is worthy to mention that this resulting PDF reduces to the EG fading model with uniform temperature case by setting $c=1$ in (\ref{SNRPDFunified}).

\subsection{Cumulative Distribution Function}
The CDF of $\gamma$ defined as $F_{\gamma}(\gamma)=\int_{0}^{\gamma}f_{\gamma}(\gamma)\,d\gamma$ can be obtained by using the definition of the Meijer's G function in \cite[Eq.(2.9.1)]{HTranforms} as

\begin{align}\label{SNCDFunified}
\nonumber F_{\gamma}(\gamma)&=\omega \,{\rm{G}}_{1,2}^{1,1}\left[ \frac{1}{\lambda}\left ( \frac{\gamma}{\mu_r} \right )^{\frac{1}{r}}\left| \begin{matrix} {1} \\ {1,0} \\ \end{matrix} \right. \right]\\
&+\frac{(1-\omega)}{\Gamma(a)}{\rm{G}}_{1,2}^{1,1}\left[ \frac{1}{b^c}\left ( \frac{\gamma}{\mu_r} \right )^{\frac{c}{r}}\left| \begin{matrix} {1} \\ {a,0} \\ \end{matrix} \right. \right].
\end{align}

At high SNR, a very tight asymptotic expression for the CDF in (\ref{SNCDFunified}) can be obtained in a simpler form by means of using \cite[Eq.(2.9.1)]{HTranforms} then \cite[Eq.(1.8.4)]{HTranforms} yielding
\begin{align}\label{CDFasymp}
F_{\gamma}(\gamma)\underset{\mu_{r}\gg 1}{\mathop{\approx }}\,\frac{\omega}{\lambda}\left ( \frac{\gamma}{\mu_r} \right )^{\frac{1}{r}}+\frac{(1-\omega)}{\Gamma(a+1)}\left ( \frac{\gamma}{b^r\mu_r} \right )^{\frac{ac}{r}}.
\end{align}

\subsection{Moments}
The moments $\mathbb{E}[\gamma^n]$, defined as $\mathbb{E}[\gamma^{n}]=\int_{0}^{\infty }\gamma^n f_{\gamma}(\gamma)\,d\gamma$, can be obtained in closed-form by substituting (\ref{SNRPDFunified}) into the definition, utilizing \cite[Eqs.(2.9.1) and (2.1.4)]{HTranforms}, and applying
\cite[Eq.~(2.25.2/1)]{PrudinkovVol3} as

\begin{align}\label{moments}
\mathbb{E}[\gamma^n]=\omega \left(\lambda^r \mu_r\right)^n\,\Gamma(rn+1)+\frac{(1-\omega)\left(b^r \mu_r\right)^n}{\Gamma(a)}\Gamma\left (\frac{rn}{c}+a  \right ).
\end{align}
It is worth accentuating that the expression in (\ref{moments}) is useful to derive very tight asymptotic approximations of the ergodic capacity at high SNR regime, as will be shown in the next section.

\subsection{Applications to Performance Analysis}
\subsubsection{Outage Probability}
The outage probability, $P_{\text{out}}$, is defined as the probability that the instantaneous SNR, $\gamma$, falls below a certain specified threshold, $\gamma_{\text{th}}$, which is considered as a protection value of the SNR above which the channel quality is satisfactory.
Mathematically speaking, $P_{\text{out}}$ is the CDF of $\gamma$ given in (\ref{SNCDFunified}) evaluated at $\gamma_{\text{th}}$, that is,

\begin{align}\label{OP}
P_{\text{out}}=\text{Pr}\left [ \gamma <\gamma_{\text{th}} \right ]=F_{\gamma}(\gamma_{\text{th}}).
\end{align}

\subsubsection{Average BER}
A unified expression for the average BER for a variety of modulation schemes can be given as \cite{dualhopFSO}
\begin{align}\label{BERdef}
P_e=\frac{\delta}{2 \Gamma(p)}\sum_{k=1}^{n}\int_{0}^{\infty}\Gamma(p,q_k\,\gamma) f_\gamma(\gamma)\,d\gamma,
\end{align}
where $n$, $\delta$, $p$, and $q_k$ vary depending on the modulation technique being used and the type of detection (i.e IM/DD or heterodyne detection) and are summarized in Table~\ref{modulation}. It is worthy to mention that this expression is general enough to be used for both heterodyne and IM/DD techniques and can be applicable to different modulation schemes.

By substituting (\ref{SNRPDFunified}) into (\ref{BERdef}), utilizing \cite[Eq.(2.9.1)]{HTranforms}, applying the integral identity \cite[Eq.~(6.455/1)]{Tableofintegrals} then \cite[Eq.(1.1.2)]{HTranforms} followed by \cite[Eq.(2.1.4)]{HTranforms}, a general expression of the average BER for OOK, BPSK, M-QAM, and M-PSK modulations can be derived in closed-form in terms of the Fox's H function as

\begin{align}\label{BER}
\nonumber P_e&=\frac{\delta}{2 \Gamma(p)}\sum_{k=1}^{n}\left(\omega\,{\rm{H}}_{2,2}^{1,2}\left[\frac{1}{\lambda}\left ( \frac{1}{q_k \mu_r} \right )^{\frac{1}{r}}\left| \begin{matrix} {(1,1)(1-p,\frac{1}{r})} \\ {(1,1)(0,1)} \\ \end{matrix} \right. \right]\right.\\
&\left.+\frac{(1-\omega)}{\Gamma(a)}{\rm{H}}_{2,2}^{1,2}\left[\frac{1}{b^c}\left ( \frac{1}{q_k \mu_r} \right )^{\frac{c}{r}}\left| \begin{matrix} {(1,1)(1-p,\frac{c}{r})} \\ {(a,1)(0,1)} \\ \end{matrix} \right. \right]\right).
\end{align}
Note that an efficient MATHEMATICA implementation for evaluating the Fox's H function $\rm{H}_{\cdot,\cdot}^{\cdot,\cdot}\left(\cdot\right)$ is presented in \cite{FerkanHFox}.
\begin{table}[h]
\begin{minipage}{1\textwidth}
\normalsize
\begin{center}
\caption{Parameters for Different Modulations}\label{modulation}
\begin{tabular}{@{}llllll@{}}
\toprule
\multicolumn{1}{c}{\textbf{Modulation}} & \multicolumn{1}{c}{$\bm{\delta}$} & \multicolumn{1}{c}{$\bm{p}$} & \multicolumn{1}{c}{$\bm{q_k}$} & \multicolumn{1}{c}{$\bm{n}$} & \multicolumn{1}{c}{\textbf{Detection Type}} \\ \midrule
\textbf{OOK}        & $1$                         & $1/2$    & $1/4$      & $1$      & IM/DD                   \\
\textbf{BPSK}       & $1$                         & $1/2$    & $1$        & $1$      & Heterodyne              \\
\textbf{M-PSK}      & $\frac{2}{\max(\log_2M,2)}$ & $1/2$    &     $\sin^2\left ( \frac{(2k-1)\pi} {M} \right )$       & $\max\left(\frac{M}{4},1\right)$         &          Heterodyne               \\
\textbf{M-QAM}      & $\frac{4}{\log_2M}\left ( 1-\frac{1}{\sqrt{M}} \right )$  &   $1/2$       & $\frac{3(2k-1)^2}{2(M-1)}$  & $\frac{\sqrt{M}}{2}$         & Heterodyne                        \\ \bottomrule
\end{tabular}
\end{center}
\end{minipage}
\end{table}

In the special case when the UWOC system is operating under uniform temperature, the average BER in (\ref{BER}) can be obtained in a simpler form in terms of the Meijer's G function as

\begin{align}\label{BERExpG}
\nonumber P_e&=\frac{\delta}{2 \Gamma(p)}\sum_{k=1}^{n} \left( \frac{\omega\, r^{\frac{1}{2}}}{(2 \pi )^{\frac{r-1}{2}}}\,{\rm{G}}_{2,r+1}^{r,2}\left[\frac{1}{q_k (r \lambda)^r\mu_r} \left| \begin{matrix} {1,1-p} \\ {\Delta(r,1),0} \\ \end{matrix} \right. \right]\right.\\
&\left.+\frac{(1-\omega)r^{a-\frac{1}{2}}}{\Gamma(a)(2 \pi )^{\frac{r-1}{2}}}{\rm{G}}_{2,r+1}^{r,2}\left[\frac{1}{q_k (rb)^r\mu_r} \left| \begin{matrix} {1,1-p} \\ {\Delta(r,a),0} \\ \end{matrix} \right. \right]\right).
\end{align}

Furthermore and similar to the CDF, the average BER can be expressed asymptotically at high SNR by means of using \cite[Eq.(1.8.4)]{HTranforms} as
\begin{align}\label{BERasym}
\nonumber  P_e\underset{\mu_{r}\gg 1}{\mathop{\approx }}\,&\frac{\delta}{2 \Gamma(p)}\sum_{k=1}^{n}  \left[  \omega\,\Gamma \left( p+\frac{1}{r}  \right)\left ( \frac{1}{\lambda^r q_k  \mu_r} \right )^{\frac{1}{r}}\right.\\
& \left.+\frac{(1-\omega)}{\Gamma(a+1)}\Gamma \left( p+\frac{ac}{r}  \right)
\left (\frac{1}{b^rq_k  \mu_r}  \right )^{\frac{a c}{r}} \right ].
\end{align}

\subsubsection{Ergodic Capacity}
The ergodic capacity is defined as
 \begin{align}\label{capacitydef}
 \overline{C}\triangleq \mathbb{E}[\ln(1+\tau\,\gamma)],
  \end{align}
 where $\tau$ is a constant equal to $\tau=e/(2\,\pi)$ \cite[Eq.~(26)]{capacityFSOlinks}, \cite{AnasTCOM}.
 Substituting (\ref{SNRPDFunified}) into (\ref{capacitydef}), using \cite[Eq.(2.9.1)]{HTranforms}, utilizing the Meijers's G function representation of $\ln(1+\tau\,\gamma)$ as $\scriptsize{{\rm{G}}_{2,2}^{1,2}\left[\tau\, \gamma\left| \begin{matrix} 1,1 \\ 1,0  \end{matrix} \right. \right]}$\normalsize \cite[Eq.~(8.4.6/5)]{PrudinkovVol3} and applying \cite[Eq.~(2.24.2/1)]{PrudinkovVol3}, then utilizing \cite[Eq.(1.1.3)]{HTranforms}, the ergodic capacity of the UWOC system can be expressed in closed-form as

\begin{align}\label{capacity}
\nonumber \overline{C}&=\omega\,{\rm{H}}_{1,2}^{2,1}\left[\frac{1}{\lambda}\left ( \frac{1}{\tau \mu_r} \right )^{\frac{1}{r}}\left| \begin{matrix} {(0,\frac{1}{r})} \\ {(0,1)(0,\frac{1}{r})} \\ \end{matrix} \right. \right]\\
&+\frac{(1-\omega)}{\Gamma(a)}
{\rm{H}}_{2,3}^{3,1}\left[\frac{1}{b^c}\left ( \frac{1}{\tau \mu_r} \right )^{\frac{c}{r}}\left| \begin{matrix} {(0,\frac{c}{r})(1,1)} \\ {(a,1)(0,1)(0,\frac{c}{r})} \\ \end{matrix} \right. \right].
\end{align}

When $c=1$, (\ref{capacity}) becomes the capacity of UWOC systems under uniform temperature and can be simplified in terms of the Meijer's G function as

\begin{align}\label{CapacityExpG}
\nonumber \overline{C}&=\frac{\omega\, r^{\frac{1}{2}}}{(2 \pi )^{\frac{r-1}{2}}}\,{\rm{G}}_{2,r+2}^{r+2,1}\left[\frac{1}{\tau (r \lambda)^r\mu_r} \left| \begin{matrix} {0,1} \\ {\Delta(r,1),0,0} \\ \end{matrix} \right. \right]\\
&+\frac{(1-\omega)r^{a-\frac{1}{2}}}{\Gamma(a)(2 \pi )^{\frac{r-1}{2}}}{\rm{G}}_{2,r+2}^{r+2,1}\left[\frac{1}{\tau (r b)^r\mu_r} \left| \begin{matrix} {0,1} \\ {\Delta(r,a),0,0} \\ \end{matrix} \right. \right].
\end{align}

Furthermore, the ergodic capacity in (\ref{capacity}) can be asymptotically approximated at high SNR by utilizing the first derivative of the $n$th order moment of $\gamma$ \cite[Eqs.~(8) and (9)]{AFN} as
\begin{align}\label{capacityasymmom}
\overline{C}\underset{\mu_r\gg 1}{\mathop{\approx }}\log(\tau)+\frac{\partial }{\partial {n}}\mathbb{E}[\gamma^n]\Big|_{n=0}.
\end{align}
By substituting (\ref{moments}) into (\ref{capacityasymmom}) and after some algebraic manipulations, we get an accurate simple closed-form approximation of the ergodic capacity at high SNR as

\begin{align}\label{Asymcapacity}
\nonumber \overline{C}\underset{\mu_r\gg 1}{\mathop{\approx }}&\log(\tau)+
\omega\left[\log(\lambda^r \mu_r)+r\,\psi(1)\right]\\
&+(1-\omega)\left[\log(b^r \mu_r)+\frac{r}{c}\,\psi(a)\right],
\end{align}
where $\psi(\cdot)$ is the psi function \cite[Eq.~(8.360/1)]{Tableofintegrals}.

\section{Numerical Results}
In this section, we provide some numerical results to illustrate the
outage probability, the average BER, and the ergodic capacity of the UWOC link modeled as EGG turbulent channel in the presence of air bubbles and temperature gradients for both fresh and salty waters, based on the values of $\omega$, $\lambda$, $a$, $b$, and $c$ listed in Table I and Table II. Monte Carlo simulations are also included to validate the obtained results.

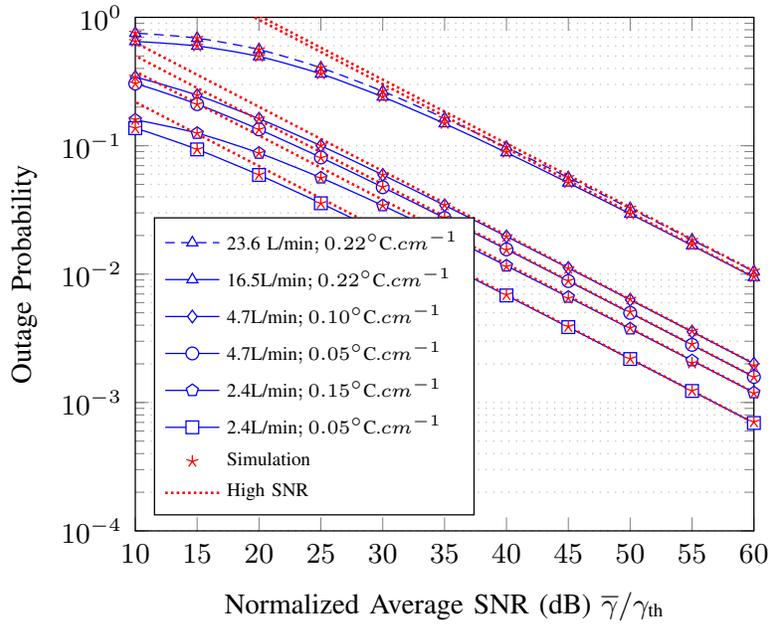
\begin{figure}[h]
   \begin{center}
\begin{tikzpicture}[scale=1.2]
    \begin{axis}[xtick=data,xmin=10,xmax=60,font=\footnotesize,
      ymin=1e-04,ymax=1, ymode=log, xlabel=Normalized Average SNR (dB) $\overline{\gamma}/\gamma_{\text{th}}$, ylabel= Outage Probability,
       legend style={nodes=right},legend style={font=\tiny},
    legend pos= south west,
    xtick=data,
    xminorgrids,
    grid style={dotted},
    yminorgrids,]


    \addplot[smooth,blue,mark=triangle*,mark options={solid},every mark/.append style={solid, fill=white},densely dashed] plot coordinates {
(0.000000,8.232470e-01)(5.000000,7.854370e-01)(10.000000,7.535100e-01)(15.000000,6.878200e-01)(20.000000,5.607960e-01)(25.000000,4.045050e-01)(30.000000,2.654640e-01)(35.000000,1.637290e-01)(40.000000,9.724490e-02)(45.000000,5.650250e-02)(50.000000,3.242380e-02)(55.000000,1.847900e-02)(60.000000,1.049320e-02)
   };
    \addplot[smooth,blue,mark=triangle*,mark options={solid},every mark/.append style={solid, fill=white}] plot coordinates {
(0.000000,7.427770e-01)(5.000000,6.843670e-01)(10.000000,6.496600e-01)(15.000000,5.978110e-01)(20.000000,4.954560e-01)(25.000000,3.625910e-01)(30.000000,2.401870e-01)(35.000000,1.488360e-01)(40.000000,8.852200e-02)(45.000000,5.139090e-02)(50.000000,2.942140e-02)(55.000000,1.671050e-02)(60.000000,9.448530e-03)
   };
    \addplot[smooth,blue,mark=diamond*,mark options={solid},every mark/.append style={solid, fill=white}] plot coordinates {
(0.000000,4.508910e-01)(5.000000,4.162290e-01)(10.000000,3.419310e-01)(15.000000,2.472980e-01)(20.000000,1.623330e-01)(25.000000,1.000120e-01)(30.000000,5.928450e-02)(35.000000,3.435580e-02)(40.000000,1.965190e-02)(45.000000,1.115800e-02)(50.000000,6.308760e-03)(55.000000,3.558540e-03)(60.000000,2.004560e-03)
   };
    \addplot[smooth,blue,mark=*,mark options={solid},every mark/.append style={solid, fill=white}] plot coordinates {
(0.000000,4.444180e-01)(5.000000,3.931420e-01)(10.000000,3.050030e-01)(15.000000,2.106470e-01)(20.000000,1.340580e-01)(25.000000,8.103130e-02)(30.000000,4.749530e-02)(35.000000,2.734550e-02)(40.000000,1.558400e-02)(45.000000,8.829710e-03)(50.000000,4.986410e-03)(55.000000,2.810770e-03)(60.000000,1.582740e-03)
   };

    \addplot[smooth,blue,mark=pentagon*,mark options={solid},every mark/.append style={solid, fill=white}] plot coordinates {
(0.000000,4.030120e-01)(5.000000,1.870800e-01)(10.000000,1.590780e-01)(15.000000,1.252380e-01)(20.000000,8.767490e-02)(25.000000,5.630420e-02)(30.000000,3.422170e-02)(35.000000,2.012380e-02)(40.000000,1.160810e-02)(45.000000,6.622390e-03)(50.000000,3.754450e-03)(55.000000,2.120980e-03)(60.000000,1.195790e-03)
   };
    \addplot[smooth,blue,mark=square*,mark options={solid},every mark/.append style={solid, fill=white}] plot coordinates {
(0.000000,2.550850e-01)(5.000000,1.788670e-01)(10.000000,1.369320e-01)(15.000000,9.362540e-02)(20.000000,5.919550e-02)(25.000000,3.563880e-02)(30.000000,2.084060e-02)(35.000000,1.198300e-02)(40.000000,6.823790e-03)(45.000000,3.864620e-03)(50.000000,2.181940e-03)(55.000000,1.229760e-03)(60.000000,6.924210e-04)
   };

      \addplot[smooth, mark=star,red,only marks] plot coordinates {
(0.000000,8.233860e-01)(5.000000,7.853090e-01)(10.000000,7.532280e-01)(15.000000,6.876950e-01)(20.000000,5.606510e-01)(25.000000,4.044590e-01)(30.000000,2.656680e-01)(35.000000,1.639710e-01)(40.000000,9.740300e-02)(45.000000,5.666800e-02)(50.000000,3.273000e-02)(55.000000,1.834700e-02)(60.000000,1.059700e-02)
   };
      \addplot[smooth, mark=star,red,only marks] plot coordinates {
(0.000000,7.424080e-01)(5.000000,6.851490e-01)(10.000000,6.495370e-01)(15.000000,5.978290e-01)(20.000000,4.956990e-01)(25.000000,3.623040e-01)(30.000000,2.401270e-01)(35.000000,1.489160e-01)(40.000000,8.875500e-02)(45.000000,5.150100e-02)(50.000000,2.938100e-02)(55.000000,1.690500e-02)(60.000000,9.550000e-03)
   };
      \addplot[smooth, mark=star,red,only marks] plot coordinates {
(0.000000,4.503860e-01)(5.000000,4.162720e-01)(10.000000,3.425230e-01)(15.000000,2.474030e-01)(20.000000,1.625440e-01)(25.000000,1.001860e-01)(30.000000,5.925800e-02)(35.000000,3.449500e-02)(40.000000,1.962000e-02)(45.000000,1.106600e-02)(50.000000,6.390000e-03)(55.000000,3.607000e-03)(60.000000,1.938000e-03)
   };
      \addplot[smooth, mark=star,red,only marks] plot coordinates {
(0.000000,4.448300e-01)(5.000000,3.930020e-01)(10.000000,3.048870e-01)(15.000000,2.100750e-01)(20.000000,1.339830e-01)(25.000000,8.053300e-02)(30.000000,4.771000e-02)(35.000000,2.755100e-02)(40.000000,1.543300e-02)(45.000000,8.871000e-03)(50.000000,5.086000e-03)(55.000000,2.847000e-03)(60.000000,1.580000e-03)
   };
      \addplot[smooth, mark=star,red,only marks] plot coordinates {
(0.000000,4.031330e-01)(5.000000,1.866000e-01)(10.000000,1.597370e-01)(15.000000,1.251600e-01)(20.000000,8.761500e-02)(25.000000,5.598200e-02)(30.000000,3.423500e-02)(35.000000,2.025300e-02)(40.000000,1.145300e-02)(45.000000,6.489000e-03)(50.000000,3.813000e-03)(55.000000,2.027000e-03)(60.000000,1.180000e-03)
   };
      \addplot[smooth, mark=star,red,only marks] plot coordinates {
(0.000000,2.549320e-01)(5.000000,1.785430e-01)(10.000000,1.370260e-01)(15.000000,9.367000e-02)(20.000000,5.946000e-02)(25.000000,3.555600e-02)(30.000000,2.091600e-02)(35.000000,1.191200e-02)(40.000000,6.904000e-03)(45.000000,3.906000e-03)(50.000000,2.192000e-03)(55.000000,1.236000e-03)(60.000000,7.060000e-04)
   };


      \addplot[smooth,red,densely dotted, thick] plot coordinates {
(0.000000,1.024500e+01)(5.000000,5.768370e+00)(10.000000,3.248350e+00)(15.000000,1.829570e+00)(20.000000,1.030670e+00)(25.000000,5.807420e-01)(30.000000,3.273060e-01)(35.000000,1.845200e-01)(40.000000,1.040560e-01)(45.000000,5.870020e-02)(50.000000,3.312670e-02)(55.000000,1.870270e-02)(60.000000,1.056420e-02)
   };
      \addplot[smooth,red,densely dotted, thick] plot coordinates {
(0.000000,9.602320e+00)(5.000000,5.393570e+00)(10.000000,3.029850e+00)(15.000000,1.702190e+00)(20.000000,9.563890e-01)(25.000000,5.373960e-01)(30.000000,3.019850e-01)(35.000000,1.697090e-01)(40.000000,9.537870e-02)(45.000000,5.360690e-02)(50.000000,3.013080e-02)(55.000000,1.693640e-02)(60.000000,9.520260e-03)
   };
      \addplot[smooth,red,densely dotted, thick] plot coordinates {
(0.000000,2.011420e+00)(5.000000,1.129740e+00)(10.000000,6.353010e-01)(15.000000,3.572560e-01)(20.000000,2.009000e-01)(25.000000,1.129740e-01)(30.000000,6.353010e-02)(35.000000,3.572560e-02)(40.000000,2.009000e-02)(45.000000,1.129740e-02)(50.000000,6.353010e-03)(55.000000,3.572560e-03)(60.000000,2.009000e-03)
   };
      \addplot[smooth,red,densely dotted, thick] plot coordinates {
(0.000000,1.585490e+00)(5.000000,8.915770e-01)(10.000000,5.013700e-01)(15.000000,2.819410e-01)(20.000000,1.585470e-01)(25.000000,8.915770e-02)(30.000000,5.013700e-02)(35.000000,2.819410e-02)(40.000000,1.585470e-02)(45.000000,8.915770e-03)(50.000000,5.013700e-03)(55.000000,2.819410e-03)(60.000000,1.585470e-03)
   };
      \addplot[smooth,red,densely dotted, thick] plot coordinates {
(0.000000,1.422420e+00)(5.000000,6.853790e-01)(10.000000,3.799140e-01)(15.000000,2.133770e-01)(20.000000,1.199780e-01)(25.000000,6.746800e-02)(30.000000,3.794000e-02)(35.000000,2.133520e-02)(40.000000,1.199770e-02)(45.000000,6.746790e-03)(50.000000,3.794000e-03)(55.000000,2.133520e-03)(60.000000,1.199770e-03)
   };
      \addplot[smooth,red,densely dotted, thick] plot coordinates {
(0.000000,7.495900e-01)(5.000000,3.900110e-01)(10.000000,2.193190e-01)(15.000000,1.233320e-01)(20.000000,6.935490e-02)(25.000000,3.900110e-02)(30.000000,2.193190e-02)(35.000000,1.233320e-02)(40.000000,6.935490e-03)(45.000000,3.900110e-03)(50.000000,2.193190e-03)(55.000000,1.233320e-03)(60.000000,6.935490e-04)
   };
   \legend{23.6 L/min$;0.22^\circ$C.$cm^{-1}$,16.5L/min$;0.22^\circ$C.$cm^{-1}$,4.7L/min$;0.10^\circ$C.$cm^{-1}$,4.7L/min$;0.05^\circ$C.$cm^{-1}$,
   2.4L/min$;0.15^\circ$C.$cm^{-1}$,2.4L/min$;0.05^\circ$C.$cm^{-1}$, Simulation,,,,,,High SNR};

%
  \end{axis}
    \end{tikzpicture}
   \caption{Outage probability for different levels of air bubbles and gradient temperatures as given by Table~\ref{goodnessoffit1} in the case of IM/DD technique along with the asymptotic results at high SNR.}
      \label{fig:OPfig}
          \end{center}
\end{figure}

The outage probability is presented in Fig.~\ref{fig:OPfig} as a function of the normalized average SNR under different turbulence conditions in the case of IM/DD technique.
\noindent
Clearly, we can observe from Fig.~\ref{fig:OPfig} that the simulation results provide
a perfect match to the analytical results, confirming the accuracy of our derivation. In addition, it can be seen from Fig.~\ref{fig:OPfig} that the higher is the level of the air bubbles and/or the temperature gradient, the higher is the value of the scintillation index and therefore, the stronger is the turbulence leading to a performance deterioration.
For example, at SNR=30 dB, $P_{\text{out}}=3.075850 \times 10^{-2}$ for a temperature gradient equal to $0.05^\circ$C.$cm^{-1}$ and $\sigma_I^2=0.1484$ and it increases to $P_{\text{out}}=3.422170 \times 10^{-2}$ for a temperature gradient of $0.15^\circ$C.$cm^{-1}$ $\sigma_I^2=0.1915$, for a fixed level of air bubbles BL=2.4 L/min. This observation shows the role of the temperature gradient in introducing severe irradiance fluctuations and hence severe turbulence conditions.
The asymptotic results of the outage probability at high SNR values obtained by using (\ref{CDFasymp}) are also shown in Fig.~\ref{fig:OPfig}. As clearly seen from this figure, the asymptotic results of the outage probability are in a perfect match with the analytical results in the high SNR regime. This justifies the accuracy and the tightness of the derived asymptotic expression in (\ref{CDFasymp}).
\begin{figure}[!h]
   \begin{center}
\begin{tikzpicture}[scale=1.2]
    \begin{axis}[xtick=data,xmin=5,xmax=60,font=\footnotesize,
      ymin=1e-05,ymax=1, ymode=log, xlabel=Normalized Average SNR (dB) $\overline{\gamma}/\gamma_{\text{th}}$, ylabel= Outage Probability,
       legend style={nodes=right},legend style={font=\tiny},
    legend pos= south west,
    xtick=data,
    xminorgrids,
    grid style={dotted},
    yminorgrids,]
\addplot[smooth,blue,mark=triangle*, mark options={solid},every mark/.append style={solid, fill=white}] plot coordinates {
(0.000000,6.710290e-01)(5.000000,5.759380e-01)(10.000000,5.154780e-01)(15.000000,4.361900e-01)(20.000000,3.298050e-01)(25.000000,2.243170e-01)(30.000000,1.414400e-01)(35.000000,8.500290e-02)(40.000000,4.963940e-02)(45.000000,2.850950e-02)(50.000000,1.621940e-02)(55.000000,9.178240e-03)(60.000000,5.178360e-03)
   };
\addplot[smooth,blue,mark=triangle*,mark options={solid},every mark/.append style={solid, fill=white}, densely dashed] plot coordinates {
(0.000000,6.552090e-01)(5.000000,5.643520e-01)(10.000000,5.020070e-01)(15.000000,4.131350e-01)(20.000000,3.026050e-01)(25.000000,2.008680e-01)(30.000000,1.246800e-01)(35.000000,7.423900e-02)(40.000000,4.313130e-02)(45.000000,2.470550e-02)(50.000000,1.403750e-02)(55.000000,7.939680e-03)(60.000000,4.479240e-03)
   };
\addplot[smooth,green!70!black,mark=diamond*,mark options={solid},every mark/.append style={solid, fill=white}] plot coordinates {
(0.000000,0.262834)(5.000000,1.609160e-01)(10.000000,1.182260e-01)(15.000000,7.846220e-02)(20.000000,4.867330e-02)(25.000000,2.897050e-02)(30.000000,1.682860e-02)(35.000000,9.639180e-03)(40.000000,5.477180e-03)(45.000000,3.098160e-03)(50.000000,1.747990e-03)(55.000000,9.847960e-04)(60.000000,5.543710e-04)
   };
\addplot[smooth,green!70!black,mark=diamond*,mark options={solid},every mark/.append style={solid, fill=white},densely dashed] plot coordinates {
(0.000000,1.997360e-01)(5.000000,1.531240e-01)(10.000000,1.090740e-01)(15.000000,7.085950e-02)(20.000000,4.337510e-02)(25.000000,2.561340e-02)(30.000000,1.481030e-02)(35.000000,8.460960e-03)(40.000000,4.800540e-03)(45.000000,2.713140e-03)(50.000000,1.530030e-03)(55.000000,8.617700e-04)(60.000000,4.850430e-04)
   };

\addplot[smooth,blue,mark=triangle, mark options={solid},every mark/.append style={solid, fill=white}] plot coordinates {
(0.000000,6.041190e-01)(5.000000,4.699070e-01)(10.000000,2.619550e-01)(15.000000,1.033200e-01)(20.000000,3.514880e-02)(25.000000,1.136770e-02)(30.000000,3.617790e-03)(35.000000,1.145740e-03)(40.000000,3.623470e-04)(45.000000,1.145580e-04)(50.000000,3.621710e-05)(55.000000,1.145050e-05)(60.000000,3.620420e-06)
   };
\addplot[smooth,blue,mark=triangle,mark options={solid},every mark/.append style={solid, fill=white}, densely dashed] plot coordinates {
(0.000000,5.930220e-01)(5.000000,4.532790e-01)(10.000000,2.395390e-01)(15.000000,9.196220e-02)(20.000000,3.101080e-02)(25.000000,1.000880e-02)(30.000000,3.185270e-03)(35.000000,1.009240e-03)(40.000000,3.193390e-04)(45.000000,1.010010e-04)(50.000000,3.194090e-05)(55.000000,1.010070e-05)(60.000000,3.194140e-06)
   };
\addplot[smooth,green!70!black,mark=diamond,mark options={solid},every mark/.append style={solid, fill=white}] plot coordinates {
(0.000000,2.120590e-01)(5.000000,1.136560e-01)(10.000000,4.613260e-02)(15.000000,1.586820e-02)(20.000000,5.155860e-03)(25.000000,1.644550e-03)(30.000000,5.214750e-04)(35.000000,1.650480e-04)(40.000000,5.220690e-05)(45.000000,1.651070e-05)(50.000000,5.221280e-06)(55.000000,1.651130e-06)(60.000000,5.221340e-07)
   };
\addplot[smooth,green!70!black,mark=diamond,mark options={solid},every mark/.append style={solid, fill=white},densely dashed] plot coordinates {
(0.000000,1.873780e-01)(5.000000,1.048000e-01)(10.000000,4.118260e-02)(15.000000,1.400240e-02)(20.000000,4.532380e-03)(25.000000,1.443930e-03)(30.000000,4.576820e-04)(35.000000,1.448390e-04)(40.000000,4.581300e-05)(45.000000,1.448840e-05)(50.000000,4.581740e-06)(55.000000,1.448890e-06)(60.000000,4.581790e-07)
   };

\addplot[smooth, mark=star,red,only marks] plot coordinates {
(0.000000,6.713910e-01)(5.000000,5.764140e-01)(10.000000,5.160320e-01)(15.000000,4.365450e-01)(20.000000,3.297110e-01)(25.000000,2.237810e-01)(30.000000,1.412690e-01)(35.000000,8.522700e-02)(40.000000,4.945300e-02)(45.000000,2.828900e-02)(50.000000,1.614900e-02)(55.000000,9.104000e-03)(60.000000,5.284000e-03)
   };
\addplot[smooth, mark=star,red,only marks] plot coordinates {
(0.000000,6.552090e-01)(5.000000,5.643520e-01)(10.000000,5.020070e-01)(15.000000,4.131350e-01)(20.000000,3.026050e-01)(25.000000,2.008680e-01)(30.000000,1.246800e-01)(35.000000,7.423900e-02)(40.000000,4.313130e-02)(45.000000,2.470550e-02)(50.000000,1.403750e-02)(55.000000,7.939680e-03)(60.000000,4.479240e-03)
   };
\addplot[smooth, mark=star,only marks,red] plot coordinates {
(0.000000,2.630660e-01)(5.000000,1.616780e-01)(10.000000,1.179210e-01)(15.000000,7.803500e-02)(20.000000,4.914200e-02)(25.000000,2.882600e-02)(30.000000,1.704100e-02)(35.000000,9.664000e-03)(40.000000,5.513000e-03)(45.000000,3.122000e-03)(50.000000,1.761000e-03)(55.000000,9.850000e-04)(60.000000,5.280000e-04)
   };
\addplot[smooth, mark=star,only marks,red] plot coordinates {
(0.000000,1.997470e-01)(5.000000,1.529110e-01)(10.000000,1.084430e-01)(15.000000,7.118400e-02)(20.000000,4.334000e-02)(25.000000,2.572300e-02)(30.000000,1.477300e-02)(35.000000,8.634000e-03)(40.000000,4.879000e-03)(45.000000,2.629000e-03)(50.000000,1.506000e-03)(55.000000,9.100000e-04)(60.000000,4.680000e-04)
   };

\addplot[smooth, mark=star,red,only marks] plot coordinates {
(0.000000,6.013876e-01)(5.000000,4.673056e-01)(10.000000,2.587721e-01)(15.000000,1.018029e-01)(20.000000,3.456850e-02)(25.000000,1.114760e-02)(30.000000,3.547700e-03)(35.000000,1.136000e-03)(40.000000,3.623470e-04)(45.000000,1.145580e-04)(50.000000,3.621710e-05)(55.000000,1.145050e-05)(60.000000,3.620420e-06)
   };
\addplot[smooth, mark=star,red,only marks] plot coordinates {
(0.000000,5.930220e-01)(5.000000,4.532790e-01)(10.000000,2.395390e-01)(15.000000,9.196220e-02)(20.000000,3.101080e-02)(25.000000,1.000880e-02)(30.000000,3.185270e-03)(35.000000,1.009240e-03)(40.000000,3.193390e-04)(45.000000,1.010010e-04)(50.000000,3.194090e-05)(55.000000,1.010070e-05)(60.000000,3.194140e-06)
   };
\addplot[smooth, mark=star,only marks,red] plot coordinates {
(0.000000,2.120590e-01)(5.000000,1.136560e-01)(10.000000,4.613260e-02)(15.000000,1.586820e-02)(20.000000,5.155860e-03)(25.000000,1.644550e-03)(30.000000,5.214750e-04)(35.000000,1.650480e-04)(40.000000,5.220690e-05)(45.000000,1.651070e-05)(50.000000,5.221280e-06)(55.000000,1.651130e-06)(60.000000,5.221340e-07)
   };
\addplot[smooth, mark=star,only marks,red] plot coordinates {
(0.000000,1.873780e-01)(5.000000,1.048000e-01)(10.000000,4.118260e-02)(15.000000,1.400240e-02)(20.000000,4.532380e-03)(25.000000,1.443930e-03)(30.000000,4.576820e-04)(35.000000,1.448390e-04)(40.000000,4.581300e-05)(45.000000,1.448840e-05)(50.000000,4.581740e-06)(55.000000,1.448890e-06)(60.000000,4.581790e-07)
   };

\addplot[smooth,red,densely dotted, thick] plot coordinates {
(0.000000,5.362330e+00)(5.000000,2.998730e+00)(10.000000,1.678490e+00)(15.000000,9.402310e-01)(20.000000,5.270240e-01)(25.000000,2.955700e-01)(30.000000,1.658390e-01)(35.000000,9.308380e-02)(40.000000,5.226350e-02)(45.000000,2.935190e-02)(50.000000,1.648800e-02)(55.000000,9.263610e-03)(60.000000,5.205430e-03)
};
\addplot[smooth,red,densely dotted, thick] plot coordinates {
(0.000000,4.640630e+00)(5.000000,2.585180e+00)(10.000000,1.444170e+00)(15.000000,8.083590e-01)(20.000000,4.531000e-01)(25.000000,2.542190e-01)(30.000000,1.427310e-01)(35.000000,8.017480e-02)(40.000000,4.505080e-02)(45.000000,2.532020e-02)(50.000000,1.423330e-02)(55.000000,8.001850e-03)(60.000000,4.498950e-03)
};
\addplot[smooth,red,densely dotted, thick] plot coordinates {
(0.000000,6.257720e-01)(5.000000,3.121660e-01)(10.000000,1.755430e-01)(15.000000,9.871530e-02)(20.000000,5.551170e-02)(25.000000,3.121650e-02)(30.000000,1.755430e-02)(35.000000,9.871530e-03)(40.000000,5.551170e-03)(45.000000,3.121650e-03)(50.000000,1.755430e-03)(55.000000,9.871530e-04)(60.000000,5.551170e-04)
};
\addplot[smooth,red,densely dotted, thick] plot coordinates {
(0.000000,4.957200e-01)(5.000000,2.730740e-01)(10.000000,1.535610e-01)(15.000000,8.635360e-02)(20.000000,4.856020e-02)(25.000000,2.730740e-02)(30.000000,1.535610e-02)(35.000000,8.635360e-03)(40.000000,4.856020e-03)(45.000000,2.730740e-03)(50.000000,1.535610e-03)(55.000000,8.635360e-04)(60.000000,4.856020e-04)
};
\addplot[smooth,red,densely dotted, thick] plot coordinates {
(0.000000,3.728500e+00)(5.000000,1.168350e+00)(10.000000,3.671270e-01)(15.000000,1.155850e-01)(20.000000,3.643990e-02)(25.000000,1.149900e-02)(30.000000,3.630990e-03)(35.000000,1.147060e-03)(40.000000,3.624790e-04)(45.000000,1.145710e-04)(50.000000,3.621840e-05)(55.000000,1.145060e-05)(60.000000,3.620430e-06)
};
\addplot[smooth,red,densely dotted, thick] plot coordinates {
(0.000000,3.276450e+00)(5.000000,1.022730e+00)(10.000000,3.213600e-01)(15.000000,1.013070e-01)(20.000000,3.198730e-02)(25.000000,1.010780e-02)(30.000000,3.195220e-03)(35.000000,1.010240e-03)(40.000000,3.194390e-04)(45.000000,1.010110e-04)(50.000000,3.194190e-05)(55.000000,1.010080e-05)(60.000000,3.194150e-06)
};
\addplot[smooth,red,densely dotted, thick] plot coordinates {
(0.000000,5.442480e-01)(5.000000,1.651140e-01)(10.000000,5.221350e-02)(15.000000,1.651140e-02)(20.000000,5.221350e-03)(25.000000,1.651140e-03)(30.000000,5.221350e-04)(35.000000,1.651140e-04)(40.000000,5.221350e-05)(45.000000,1.651140e-05)(50.000000,5.221350e-06)(55.000000,1.651140e-06)(60.000000,5.221350e-07)
};
\addplot[smooth,red,densely dotted, thick] plot coordinates {
(0.000000,4.586790e-01)(5.000000,1.448890e-01)(10.000000,4.581790e-02)(15.000000,1.448890e-02)(20.000000,4.581790e-03)(25.000000,1.448890e-03)(30.000000,4.581790e-04)(35.000000,1.448890e-04)(40.000000,4.581790e-05)(45.000000,1.448890e-05)(50.000000,4.581790e-06)(55.000000,1.448890e-06)(60.000000,4.581790e-07)
};
 \legend{Salty Water BL=16.5 L/min, Fresh Water BL=16.5 L/min, Salty Water BL=4.7 L/min, Fresh Water BL=4.7 L/min,,,,, Simulation,,,,,,,,High SNR};

  \draw \boundellipse{ axis cs:50,0.6e-002}{12}{1.8};
\node at (axis cs:50,0.5e-001){IM/DD};

\draw \boundellipse{ axis cs:39.5,1.5e-004}{64}{0.24};
\node at (axis cs:53,1.4e-004){Heterodyne};

 \end{axis}
    \end{tikzpicture}
   \caption{Outage probability for two different levels of air bubbles using both salty and fresh waters for thermally uniform UWOC channels for both IM/DD as well as heterodyne detection, along with the asymptotic results at high SNR.}
      \label{fig:OPfig2}
          \end{center}
\end{figure}
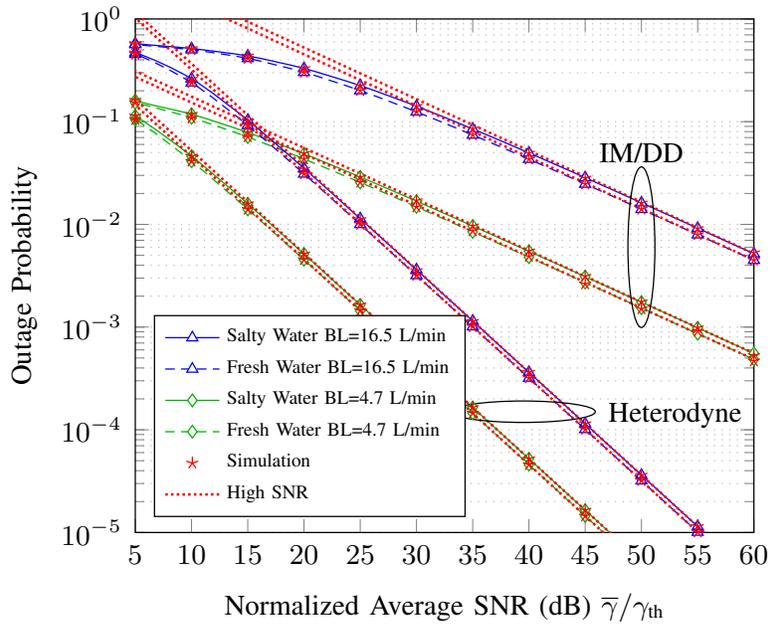\noindent

Fig.~\ref{fig:OPfig2} depicts the outage probability for the UWOC system under uniform temperature, various levels of air bubbles, and for both fresh as well as salty waters.
Expectedly, it can be observed that for a given type of water, $P_{\text{out}}$ increases as the severity of the turbulence increases
(i.e. the higher the level of air bubbles, the higher will be the outage probability for both types of water and under both IM/DD and heterodyne detection).
In addition, it can be inferred from Fig.~\ref{fig:OPfig2} that the water salinity affects the UOWC system performance but in a much lesser degree than air bubbles, which cause rapid intensity fluctuations. Furthermore, it can also be observed that implementing heterodyne detection results in a significant improvement in the UWOC system performance compared to IM/DD, as expected.
This performance
enhancement is due the fact that heterodyne technique can better
overcome the turbulence effects which comes at the
expense of complexity in implementing coherent receivers relative
to the IM/DD technique  \cite{heterodyne1}.

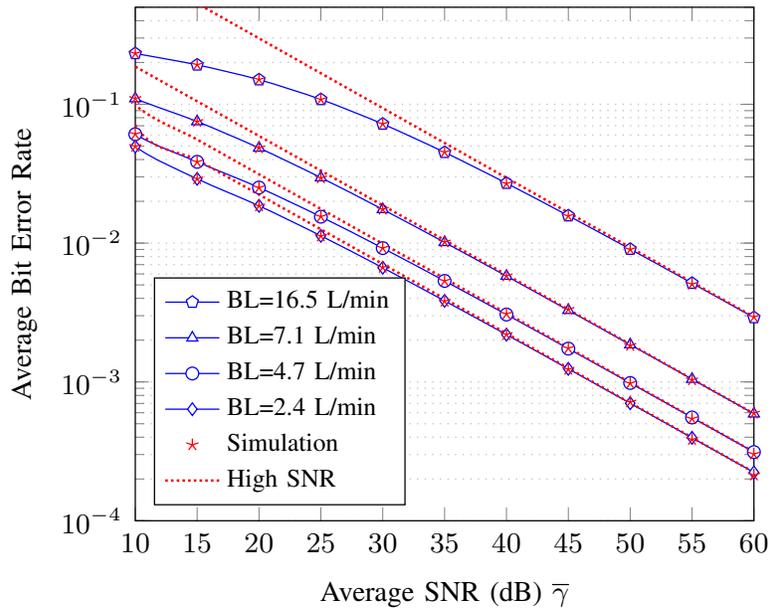
\begin{figure}[h]
   \begin{center}
\begin{tikzpicture}[scale=1.2]
    \begin{axis}[xtick=data,xmin=10,xmax=60,font=\footnotesize,
      ymin=1e-04,ymax=0.5, ymode=log, xlabel=Average SNR (dB) $\overline{\gamma}$, ylabel=Average Bit Error Rate,
       legend style={nodes=right},legend style={font=\scriptsize},
    legend pos= south west,
    xtick=data,
    xminorgrids,
    grid style={dotted},
    yminorgrids,]
 \addplot[smooth,blue,mark=pentagon*,mark options={solid},every mark/.append style={solid, fill=white}] plot coordinates {
(0.000000,3.382490e-01)(5.000000,2.778600e-01)(10.000000,2.322700e-01)(15.000000,1.922590e-01)(20.000000,1.503290e-01)(25.000000,1.083400e-01)(30.000000,7.209610e-02)(35.000000,4.507460e-02)(40.000000,2.700660e-02)(45.000000,1.575470e-02)(50.000000,9.045320e-03)(55.000000,5.145440e-03)(60.000000,2.911640e-03)
   };
 \addplot[smooth,blue,mark=triangle*,mark options={solid},every mark/.append style={solid, fill=white}] plot coordinates {
(0.000000,2.792030e-01)(5.000000,1.768950e-01)(10.000000,1.099040e-01)(15.000000,7.513910e-02)(20.000000,4.844580e-02)(25.000000,2.961720e-02)(30.000000,1.749710e-02)(35.000000,1.012400e-02)(40.000000,5.786680e-03)(45.000000,3.284320e-03)(50.000000,1.856580e-03)(55.000000,1.047120e-03)(60.000000,5.898140e-04)
   };
  \addplot[smooth,blue,mark=*,mark options={solid},every mark/.append style={solid, fill=white}] plot coordinates {
(0.000000,2.582000e-01)(5.000000,1.386620e-01)(10.000000,6.083830e-02)(15.000000,3.850750e-02)(20.000000,2.513240e-02)(25.000000,1.549520e-02)(30.000000,9.202670e-03)(35.000000,5.341580e-03)(40.000000,3.058740e-03)(45.000000,1.737860e-03)(50.000000,9.829800e-04)(55.000000,5.545890e-04)(60.000000,3.124460e-04)
   };
  \addplot[smooth,blue,mark=diamond*,mark options={solid},every mark/.append style={solid, fill=white}] plot coordinates {
(0.000000,2.542630e-01)(5.000000,1.310290e-01)(10.000000,4.957960e-02)(15.000000,2.904360e-02)(20.000000,1.856650e-02)(25.000000,1.128950e-02)(30.000000,6.647310e-03)(35.000000,3.838590e-03)(40.000000,2.191540e-03)(45.000000,1.243020e-03)(50.000000,7.024030e-04)(55.000000,3.960720e-04)(60.000000,2.230710e-04)
   };

      \addplot[smooth, mark=star,red,only marks] plot coordinates {
(0.000000,3.382913e-01)(5.000000,2.777775e-01)(10.000000,2.322517e-01)(15.000000,1.924030e-01)(20.000000,1.505812e-01)(25.000000,1.082036e-01)(30.000000,7.220981e-02)(35.000000,4.514611e-02)(40.000000,2.682906e-02)(45.000000,1.567875e-02)(50.000000,8.998576e-03)(55.000000,5.061797e-03)(60.000000,2.937190e-03)
   };
      \addplot[smooth, mark=star,only marks,red] plot coordinates {
(0.000000,2.791650e-01)(5.000000,1.768274e-01)(10.000000,1.099082e-01)(15.000000,7.507963e-02)(20.000000,4.851653e-02)(25.000000,2.946024e-02)(30.000000,1.750476e-02)(35.000000,1.018142e-02)(40.000000,5.797421e-03)(45.000000,3.327755e-03)(50.000000,1.830292e-03)(55.000000,1.034441e-03)(60.000000,5.954456e-04)
};
      \addplot[smooth, mark=star,only marks,red] plot coordinates {
(0.000000,2.582076e-01)(5.000000,1.386055e-01)(10.000000,6.094127e-02)(15.000000,3.844972e-02)(20.000000,2.507344e-02)(25.000000,1.552233e-02)(30.000000,9.211734e-03)(35.000000,5.339565e-03)(40.000000,3.087411e-03)(45.000000,1.749090e-03)(50.000000,9.773193e-04)(55.000000,5.431262e-04)(60.000000,3.020880e-04)
   };
      \addplot[smooth, mark=star,only marks,red] plot coordinates {
(0.000000,2.541712e-01)(5.000000,1.310086e-01)(10.000000,4.966568e-02)(15.000000,2.901641e-02)(20.000000,1.851518e-02)(25.000000,1.123227e-02)(30.000000,6.685196e-03)(35.000000,3.808696e-03)(40.000000,2.194965e-03)(45.000000,1.232027e-03)(50.000000,7.163414e-04)(55.000000,3.846881e-04)(60.000000,2.130819e-04)
   };
\addplot[smooth,red,densely dotted, thick] plot coordinates {
(0.000000,3.050180e+00)(5.000000,1.703440e+00)(10.000000,9.524030e-01)(15.000000,5.329990e-01)(20.000000,2.985240e-01)(25.000000,1.673100e-01)(30.000000,9.382250e-02)(35.000000,5.263750e-02)(40.000000,2.954290e-02)(45.000000,1.658640e-02)(50.000000,9.314680e-03)(55.000000,5.232180e-03)(60.000000,2.939540e-03)
};
  \addplot[smooth,red,densely dotted, thick] plot coordinates {
(0.000000,1.831630e+05)(5.000000,3.372340e-01)(10.000000,1.869140e-01)(15.000000,1.051100e-01)(20.000000,5.910750e-02)(25.000000,3.323860e-02)(30.000000,1.869140e-02)(35.000000,1.051100e-02)(40.000000,5.910750e-03)(45.000000,3.323860e-03)(50.000000,1.869140e-03)(55.000000,1.051100e-03)(60.000000,5.910750e-04)
};
\addplot[smooth,red,densely dotted, thick] plot coordinates {
(0.000000,1.020340e+04)(5.000000,3.611970e-01)(10.000000,9.904310e-02)(15.000000,5.569410e-02)(20.000000,3.131910e-02)(25.000000,1.761200e-02)(30.000000,9.903970e-03)(35.000000,5.569410e-03)(40.000000,3.131910e-03)(45.000000,1.761200e-03)(50.000000,9.903970e-04)(55.000000,5.569410e-04)(60.000000,3.131910e-04)
   };
     \addplot[smooth,red,densely dotted, thick] plot coordinates {
(0.000000,3.305180e+09)(5.000000,1.145000e+00)(10.000000,7.068130e-02)(15.000000,3.974700e-02)(20.000000,2.235140e-02)(25.000000,1.256910e-02)(30.000000,7.068130e-03)(35.000000,3.974700e-03)(40.000000,2.235140e-03)(45.000000,1.256910e-03)(50.000000,7.068130e-04)(55.000000,3.974700e-04)(60.000000,2.235140e-04)
   };

\legend{BL=16.5 L/min, BL=7.1 L/min, BL=4.7 L/min, BL=2.4 L/min, Simulation,,,,High SNR};
      \end{axis}
    \end{tikzpicture}
   \caption{Average BER for OOK under various levels of air bubbles using salty water for thermally uniform UWOC channels operating under IM/DD technique along with the asymptotic results at high SNR.}
      \label{fig:BER1fig}
          \end{center}
\end{figure}\noindent

In Fig.~\ref{fig:BER1fig}, the average BER of IM/DD with OOK is presented versus the average SNR under different channel conditions varying from weak to strong turbulence conditions in the case of uniform temperature and salty water.
We can see from this figure that the analytical results of the average BER are in an excellent match with the simulated results.
Interestingly, it can be seen from this figure that as the level of the air bubbles increases, the intensity of the received signal undergoes severe fluctuations and the scintillation index value becomes higher, resulting in an increase in the average BER. It is worth mentioning that the average BER shows similar behaviour when we use fresh water in our UWOC system.
Moreover, it can be observed from Fig.~\ref{fig:BER1fig} that the asymptotic expression of the average BER at high SNR given in (\ref{BERasym}) matches exactly the analytical expression derived in (\ref{BER}) proving the accuracy of the proposed asymptotic results at high SNR regime.

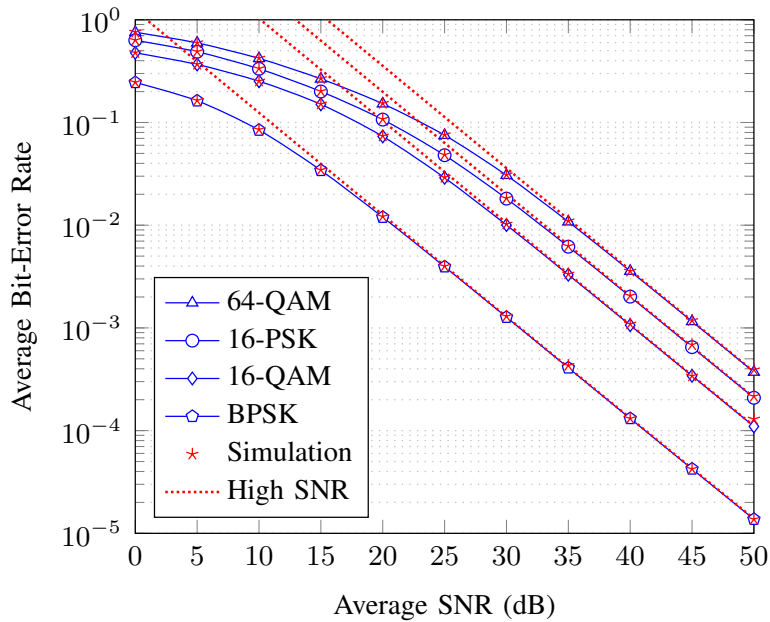
\begin{figure}[h]
  \begin{center}
\begin{tikzpicture}[scale=1.2]
    \begin{axis}[xtick=data,font=\footnotesize,
    ymode=log, xlabel= Average SNR (dB), ylabel= Average Bit-Error Rate,
  xmin=0,xmax=50,ymin=1e-05, ymax=1,
    legend style={nodes=right},legend pos= south west,legend style={font=\footnotesize},
    xtick=data,
      xminorgrids,
    grid style={dotted},
    yminorgrids,
   ]

     \addplot[smooth,blue,mark=triangle*,mark options={solid},every mark/.append style={solid, fill=white}] plot coordinates {
(0.000000,7.563360e-01)(5.000000,5.954630e-01)(10.000000,4.204510e-01)(15.000000,2.658320e-01)(20.000000,1.522680e-01)(25.000000,7.519530e-02)(30.000000,3.061940e-02)(35.000000,1.081360e-02)(40.000000,3.576240e-03)(45.000000,1.156330e-03)(50.000000,3.718070e-04)(55.000000,1.196220e-04)(60.000000,3.860530e-05)
};
    \addplot[smooth,blue,mark=*,mark options={solid},every mark/.append style={solid, fill=white}] plot coordinates {
(0.000000,6.316940e-01)(5.000000,4.885420e-01)(10.000000,3.331830e-01)(15.000000,2.003510e-01)(20.000000,1.067220e-01)(25.000000,4.801510e-02)(30.000000,1.814600e-02)(35.000000,6.177620e-03)(40.000000,2.016330e-03)(45.000000,6.496300e-04)(50.000000,2.088590e-04)(55.000000,6.727950e-05)(60.000000,2.175510e-05)
};
    \addplot[smooth,blue,mark=diamond*,mark options={solid},every mark/.append style={solid, fill=white}] plot coordinates {
(0.000000,4.771410e-01)(5.000000,3.673450e-01)(10.000000,2.521330e-01)(15.000000,1.501840e-01)(20.000000,7.319680e-02)(25.000000,2.893970e-02)(30.000000,1.003920e-02)(35.000000,3.296740e-03)(40.000000,1.063480e-03)(45.000000,3.417350e-04)(50.000000,1.099390e-04)(55.000000,3.548520e-05)(60.000000,1.150650e-05)
};
    \addplot[smooth,blue,mark=pentagon*,mark options={solid},every mark/.append style={solid, fill=white}] plot coordinates {
(0.000000,2.455700e-01)(5.000000,1.627920e-01)(10.000000,8.436790e-02)(15.000000,3.421690e-02)(20.000000,1.197420e-02)(25.000000,3.942220e-03)(30.000000,1.272240e-03)(35.000000,4.086670e-04)(40.000000,1.313770e-04)(45.000000,4.236370e-05)(50.000000,1.372040e-05)(55.000000,4.468060e-06)(60.000000,1.464650e-06)
};

    \addplot[smooth, mark=star,only marks,red] plot coordinates {
(0.000000,7.583851e-01)(5.000000,5.976892e-01)(10.000000,4.229249e-01)(15.000000,2.677108e-01)(20.000000,1.535849e-01)(25.000000,7.586701e-02)(30.000000,3.101213e-02)(35.000000,1.101408e-02)(40.000000,3.572282e-03)(45.000000,1.180261e-03)(50.000000,3.905968e-04)(55.000000,1.476935e-04)(60.000000,7.457049e-05)
};
   \addplot[smooth, mark=star,only marks,red] plot coordinates {
(0.000000,6.332048e-01)(5.000000,4.907370e-01)(10.000000,3.353116e-01)(15.000000,2.017602e-01)(20.000000,1.080476e-01)(25.000000,4.858922e-02)(30.000000,1.832784e-02)(35.000000,6.269151e-03)(40.000000,2.061427e-03)(45.000000,6.866316e-04)(50.000000,2.159851e-04)(55.000000,1.029977e-04)(60.000000,5.761884e-05)
};
   \addplot[smooth, mark=star,only marks,red] plot coordinates {
(0.000000,4.785877e-01)(5.000000,3.687258e-01)(10.000000,2.534596e-01)(15.000000,1.515135e-01)(20.000000,7.395313e-02)(25.000000,2.926254e-02)(30.000000,1.024349e-02)(35.000000,3.348700e-03)(40.000000,1.091291e-03)(45.000000,3.409269e-04)(50.000000,1.294739e-04)(55.000000,5.803900e-05)(60.000000,1.150650e-05)
};
   \addplot[smooth, mark=star,only marks,red] plot coordinates {
(0.000000,2.465538e-01)(5.000000,1.640477e-01)(10.000000,8.541767e-02)(15.000000,3.460751e-02)(20.000000,1.215269e-02)(25.000000,3.989836e-03)(30.000000,1.299842e-03)(35.000000,4.310133e-04)(40.000000,1.313770e-04)(45.000000,4.236370e-05)(50.000000,1.372040e-05)(55.000000,4.468060e-06)(60.000000,1.464650e-06)
};

     \addplot[smooth,red,densely dotted, thick] plot coordinates {
(0.000000,3.521810e+01)(5.000000,1.115700e+01)(10.000000,3.536170e+00)(15.000000,1.121450e+00)(20.000000,3.559190e-01)(25.000000,1.130660e-01)(30.000000,3.596090e-02)(35.000000,1.145440e-02)(40.000000,3.655270e-03)(45.000000,1.169140e-03)(50.000000,3.750160e-04)(55.000000,1.207140e-04)(60.000000,3.902310e-05)
};
    \addplot[smooth,red,densely dotted, thick] plot coordinates {
(0.000000,1.964760e+01)(5.000000,6.225550e+00)(10.000000,1.973680e+00)(15.000000,6.261290e-01)(20.000000,1.987990e-01)(25.000000,6.318590e-02)(30.000000,2.010930e-02)(35.000000,6.410470e-03)(40.000000,2.047720e-03)(45.000000,6.557780e-04)(50.000000,2.106710e-04)(55.000000,6.793990e-05)(60.000000,2.201300e-05)
};
   \addplot[smooth,red,densely dotted,  thick] plot coordinates {
(0.000000,1.024460e+01)(5.000000,3.247070e+00)(10.000000,1.029800e+00)(15.000000,3.268450e-01)(20.000000,1.038360e-01)(25.000000,3.302740e-02)(30.000000,1.052090e-02)(35.000000,3.357710e-03)(40.000000,1.074100e-03)(45.000000,3.445860e-04)(50.000000,1.109400e-04)(55.000000,3.587200e-05)(60.000000,1.166000e-05)
};
 \addplot[smooth,red,densely dotted, thick] plot coordinates {
(0.000000,1.234840e+00)(5.000000,3.918470e-01)(10.000000,1.244560e-01)(15.000000,3.957400e-02)(20.000000,1.260150e-02)(25.000000,4.019820e-03)(30.000000,1.285140e-03)(35.000000,4.119900e-04)(40.000000,1.325220e-04)(45.000000,4.280370e-05)(50.000000,1.389480e-05)(55.000000,4.537690e-06)(60.000000,1.492510e-06)
};
\legend{64-QAM, 16-PSK,16-QAM,BPSK,Simulation,,,,,High SNR};
%
%

 \end{axis}
  \end{tikzpicture}
     \caption{Average BER for different modulation schemes of UWOC systems operating under heterodyne detection along with the asymptotic results at high SNR for a bubbles level of 23.6 L/min and a temperature gradient of 0.22 $^\circ$C.$cm^{-1}$.}
          \label{fig:BER2}
     \end{center}
  \end{figure}
  The analytical accuracy of (\ref{BER}) is checked by simulations for various modulation
techniques including 64-QAM, 16-QAM, 16-PSK, and BPSK for UWOC system operating under the heterodyne detection in the case of strong turbulence conditions corresponding to a bubbles level of 23.6 L/min, a temperature gradient of 0.22 $^\circ$C.$cm^{-1}$, and a scintillation index $\sigma_I^2=3.1952$ in Fig.~\ref{fig:BER2}.
Obviously, it can be seen from this figure that BPSK modulation outperforms the other modulation techniques
Moreover, it can be observed from Fig.~\ref{fig:BER2} that 16-QAM outperforms 16-PSK, as expected when $M > 4$ \cite{proakis2008digital}.

\begin{figure}[!h]
   \begin{center}
\begin{tikzpicture}[scale=1.2]
    \begin{axis}[xtick=data,font=\footnotesize,xmin=10,xmax=55,
   ymin=0,ymax=12, xlabel=Average SNR (dB) $\overline{\gamma}$, ylabel= Ergodic Capacity (Nats/Sec/Hz),
       legend style={nodes=right},legend style={font=\scriptsize},
    legend pos= north west,
    xtick=data,
    xminorgrids,
    grid style={dotted},
    yminorgrids,]

    \addplot[smooth,blue,mark=triangle*,mark options={solid},every mark/.append style={solid, fill=white}] plot coordinates {
(0.000000,3.475708e-01)(5.000000,8.083910e-01)(10.000000,1.527470e+00)(15.000000,2.423910e+00)(20.000000,3.419650e+00)(25.000000,4.473130e+00)(30.000000,5.563110e+00)(35.000000,6.676720e+00)(40.000000,7.805360e+00)(45.000000,8.943290e+00)(50.000000,1.008680e+01)(55.000000,1.123360e+01)(60.000000,1.238230e+01)
   };
%
     \addplot[smooth,blue,mark=*,mark options={solid},every mark/.append style={solid, fill=white}] plot coordinates {
(0.000000,3.415470e-01)(5.000000,7.845820e-01)(10.000000,1.472960e+00)(15.000000,2.333680e+00)(20.000000,3.291230e+00)(25.000000,4.306280e+00)(30.000000,5.362020e+00)(35.000000,6.448720e+00)(40.000000,7.558250e+00)(45.000000,8.683590e+00)(50.000000,9.819240e+00)(55.000000,1.096130e+01)(60.000000,1.210730e+01)
   };

    \addplot[smooth,blue,mark=pentagon*,mark options={solid},every mark/.append style={solid, fill=white}] plot coordinates {
(0.000000,3.274200e-01)(5.000000,7.254700e-01)(10.000000,1.320060e+00)(15.000000,2.065550e+00)(20.000000,2.925490e+00)(25.000000,3.875730e+00)(30.000000,4.894790e+00)(35.000000,5.962820e+00)(40.000000,7.063450e+00)(45.000000,8.184700e+00)(50.000000,9.318490e+00)(55.000000,1.045970e+01)(60.000000,1.160530e+01)
   };

    \addplot[smooth,blue,mark=diamond*,mark options={solid},every mark/.append style={solid, fill=white}] plot coordinates {
(0.000000,3.246320e-01)(5.000000,7.138880e-01)(10.000000,1.289050e+00)(15.000000,2.005460e+00)(20.000000,2.832450e+00)(25.000000,3.752170e+00)(30.000000,4.746910e+00)(35.000000,5.797540e+00)(40.000000,6.886660e+00)(45.000000,8.000680e+00)(50.000000,9.130100e+00)(55.000000,1.026870e+01)(60.000000,1.141280e+01)
   };
    \addplot[smooth,blue,mark=square*,mark options={solid},every mark/.append style={solid, fill=white}] plot coordinates {
(0.000000,2.511910e-01)(5.000000,4.884420e-01)(10.000000,8.125720e-01)(15.000000,1.213360e+00)(20.000000,1.695600e+00)(25.000000,2.277960e+00)(30.000000,2.977590e+00)(35.000000,3.796220e+00)(40.000000,4.718840e+00)(45.000000,5.721650e+00)(50.000000,6.780670e+00)(55.000000,7.876430e+00)(60.000000,8.995050e+00)
   };
    \addplot[smooth,blue,mark=square*,mark options={solid},every mark/.append style={solid, fill=white},densely dashed] plot coordinates {
(0.000000,2.080880e-01)(5.000000,3.822350e-01)(10.000000,6.172490e-01)(15.000000,9.184330e-01)(20.000000,1.306110e+00)(25.000000,1.809710e+00)(30.000000,2.450480e+00)(35.000000,3.228720e+00)(40.000000,4.125250e+00)(45.000000,5.111860e+00)(50.000000,6.161040e+00)(55.000000,7.250910e+00)(60.000000,8.366010e+00)
   };
 \addplot[smooth,color=red,solid, mark=star,only marks] plot coordinates {
(0.000000,3.475708e-01)(5.000000,8.085062e-01)(10.000000,1.527527e+00)(15.000000,2.424912e+00)(20.000000,3.419513e+00)(25.000000,4.474579e+00)(30.000000,5.564598e+00)(35.000000,6.674069e+00)(40.000000,7.805378e+00)(45.000000,8.946361e+00)(50.000000,1.008849e+01)(55.000000,1.123390e+01)(60.000000,1.238304e+01)
   };
 \addplot[smooth,color=red,solid, mark=star,only marks] plot coordinates {
(0.000000,3.409222e-01)(5.000000,7.832046e-01)(10.000000,1.470734e+00)(15.000000,2.331252e+00)(20.000000,3.287939e+00)(25.000000,4.306071e+00)(30.000000,5.360042e+00)(35.000000,6.449980e+00)(40.000000,7.551866e+00)(45.000000,8.682647e+00)(50.000000,9.819884e+00)(55.000000,1.095997e+01)(60.000000,1.210454e+01)
   };
 \addplot[smooth,color=red,solid, mark=star,only marks] plot coordinates {
(0.000000,3.273697e-01)(5.000000,7.251754e-01)(10.000000,1.320257e+00)(15.000000,2.064036e+00)(20.000000,2.926274e+00)(25.000000,3.877545e+00)(30.000000,4.897867e+00)(35.000000,5.961163e+00)(40.000000,7.063983e+00)(45.000000,8.188317e+00)(50.000000,9.318088e+00)(55.000000,1.045731e+01)(60.000000,1.160516e+01)
   };
 \addplot[smooth,color=red,solid, mark=star,only marks] plot coordinates {
(0.000000,3.246320e-01)(5.000000,7.138380e-01)(10.000000,1.288895e+00)(15.000000,2.006863e+00)(20.000000,2.832614e+00)(25.000000,3.754256e+00)(30.000000,4.748106e+00)(35.000000,5.796470e+00)(40.000000,6.887058e+00)(45.000000,7.999024e+00)(50.000000,9.131854e+00)(55.000000,1.026545e+01)(60.000000,1.141518e+01)
   };
 \addplot[smooth,color=red,solid, mark=star,only marks] plot coordinates {
(0.000000,2.528051e-01)(5.000000,4.920288e-01)(10.000000,8.168823e-01)(15.000000,1.219872e+00)(20.000000,1.703044e+00)(25.000000,2.287044e+00)(30.000000,2.982946e+00)(35.000000,3.804916e+00)(40.000000,4.727591e+00)(45.000000,5.729727e+00)(50.000000,6.786020e+00)(55.000000,7.886614e+00)(60.000000,9.004081e+00)
   };
 \addplot[smooth,color=red,solid, mark=star,only marks] plot coordinates {
(0.000000,2.114659e-01)(5.000000,3.879468e-01)(10.000000,6.250097e-01)(15.000000,9.285668e-01)(20.000000,1.321470e+00)(25.000000,1.823128e+00)(30.000000,2.467395e+00)(35.000000,3.243676e+00)(40.000000,4.140558e+00)(45.000000,5.131314e+00)(50.000000,6.173849e+00)(55.000000,7.267449e+00)(60.000000,8.386165e+00)
   };
 \addplot[smooth,red,densely dotted, thick] plot coordinates {
(0.000000,-1.435560e+00)(5.000000,-2.842710e-01)(10.000000,8.670220e-01)(15.000000,2.018310e+00)(20.000000,3.169610e+00)(25.000000,4.320900e+00)(30.000000,5.472190e+00)(35.000000,6.623480e+00)(40.000000,7.774780e+00)(45.000000,8.926070e+00)(50.000000,1.007740e+01)(55.000000,1.122870e+01)(60.000000,1.237990e+01)
   };
 \addplot[smooth,red,densely dotted, thick] plot coordinates {
(0.000000,-1.718000e+00)(5.000000,-5.667040e-01)(10.000000,5.845890e-01)(15.000000,1.735880e+00)(20.000000,2.887170e+00)(25.000000,4.038470e+00)(30.000000,5.189760e+00)(35.000000,6.341050e+00)(40.000000,7.492340e+00)(45.000000,8.643640e+00)(50.000000,9.794930e+00)(55.000000,1.094620e+01)(60.000000,1.209750e+01)
   };
 \addplot[smooth,red,densely dotted, thick] plot coordinates {
(0.000000,-2.217780e+00)(5.000000,-1.066490e+00)(10.000000,8.480650e-02)(15.000000,1.236100e+00)(20.000000,2.387390e+00)(25.000000,3.538680e+00)(30.000000,4.689980e+00)(35.000000,5.841270e+00)(40.000000,6.992560e+00)(45.000000,8.143850e+00)(50.000000,9.295150e+00)(55.000000,1.044640e+01)(60.000000,1.159770e+01)
   };
\addplot[smooth,red,densely dotted, thick] plot coordinates {
(0.000000,-2.412010e+00)(5.000000,-1.260720e+00)(10.000000,-1.094270e-01)(15.000000,1.041870e+00)(20.000000,2.193160e+00)(25.000000,3.344450e+00)(30.000000,4.495740e+00)(35.000000,5.647040e+00)(40.000000,6.798330e+00)(45.000000,7.949620e+00)(50.000000,9.100910e+00)(55.000000,1.025220e+01)(60.000000,1.140350e+01)
   };
\addplot[smooth,red,densely dotted, thick] plot coordinates {
(0.000000,-4.855970e+00)(5.000000,-3.704680e+00)(10.000000,-2.553380e+00)(15.000000,-1.402090e+00)(20.000000,-2.507990e-01)(25.000000,9.004930e-01)(30.000000,2.051790e+00)(35.000000,3.203080e+00)(40.000000,4.354370e+00)(45.000000,5.505660e+00)(50.000000,6.656960e+00)(55.000000,7.808250e+00)(60.000000,8.959540e+00)
   };
\addplot[smooth,red,densely dotted,  thick] plot coordinates {
(0.000000,-5.479920e+00)(5.000000,-4.328630e+00)(10.000000,-3.177340e+00)(15.000000,-2.026050e+00)(20.000000,-8.747540e-01)(25.000000,2.765380e-01)(30.000000,1.427830e+00)(35.000000,2.579120e+00)(40.000000,3.730420e+00)(45.000000,4.881710e+00)(50.000000,6.033000e+00)(55.000000,7.184290e+00)(60.000000,8.335590e+00)
   };
\legend{$\sigma_I^2=0.1484$,$\sigma_I^2=0.2178$,$\sigma_I^2=0.4201$, $\sigma_I^2=0.4769$,$\sigma_I^2=1.9328$, $\sigma_I^2=3.1952$,Simulation,,,,,,High SNR};
      \end{axis}
    \end{tikzpicture}
    \caption{Ergodic capacity for different levels of air bubbles and temperature gradients for IM/DD technique along with the high SNR results based on the moments method.}
      \label{fig:capacityexact}
          \end{center}
    \end{figure}
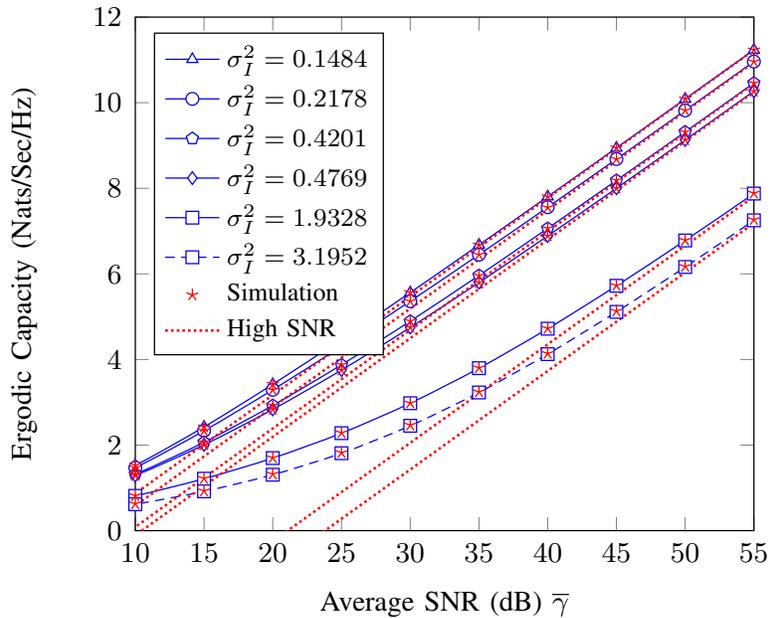

Fig.~\ref{fig:capacityexact} shows the ergodic capacity for different gradient temperatures and various levels of air bubbles under IM/DD technique. Equivalent results obtained via Monte-Carlo simulations are also included showing a perfect agreement with the obtained analytical results.
  Clearly, Fig.~\ref{fig:capacityexact} demonstrates the significant impact of the air bubbles and the gradient temperature on the system performance.
Moreover, as seen in this figure, when the level of air bubbles or temperature gradient decreases, the average BER decreases leading to a system performance improvement, as expected.
Additionally, one of the most important outcomes of Fig.~\ref{fig:capacityexact} are the accuracy and the tightness of the asymptotic results at high SNR regime, obtained via the moments-based approach in (\ref{Asymcapacity}).

\section{Conclusion}
In this paper, based on experimental data, we have proposed a
new model for UWOC channels, in which the irradiance fluctuations caused by air bubbles and gradient of temperature are characterized by the mixture EGG model.
We have demonstrated that this model perfectly matches the measured data, collected under different channel conditions ranging from weak to strong turbulence conditions, for both salty as well as fresh waters.
In addition, based on reference \cite{Oubei:17sg} where Weibull distribution which is a special case of Generalized Gamma distribution was used to fit irradiance fluctuations data due to underwater salinity gradient, this model is expected to accurately capture a combination of air bubbles, gradient of temperature, and gradient of salinity fluctuations, making it a unified model that can address the statistics of optical beam irradiance fluctuations in all types of turbulent underwater wireless optical channels. Moreover, when the water temperature is uniform, the received intensity of
the laser beam is best described by the EG distribution which is a special case of the EGG model.
In addition, our new model being simple and analytically tractable, is convenient for performance analysis and design of UWOC systems.
 Therefore, we
have studied the performance of the UWOC system operating under both IM/DD and heterodyne detection over EGG fading channels in the presence of both temperature gradients as well as air bubbles induced turbulences. We have derived simple and exact closed-form expressions for fundamental system performance metrics such as the outage probability, the average BER of different modulation scheme, and the ergodic capacity under different turbulence conditions for both fresh and salty waters.
Furthermore, by applying the moments-based method, we have derived very tight asymptotic results for the ergodic capacity at high SNR in terms of simple functions. We have also demonstrated that the performance of UWOC systems is degraded with an increase in the gradient of temperature or the level air bubbles.
In the quest to improve the reliability of UWOC links, we anticipate that our findings will catalyze the development of robust and reliable underwater communication systems and help push the frontiers of UWOC research towards the goal of seamless and high-speed underwater wireless networks.

\section*{Acknowledgment}
Authors at KAUST would like to acknowledge the funding support from King Abdulaziz City for Science and Technology (KACST) Grant KACST TIC R2-FP-008; King Abdullah University of Science and Technology (KAUST) BAS/1/1614-01-01, KCR/1/2081-01-01, GEN/1/6607-01-01, and REP/1/2878-01-01.

\bibliographystyle{IEEEtran}
\bibliography{IEEEabrv,IEEEexample}

\begin{thebibliography}{10}
\providecommand{\url}[1]{#1}
\csname url@samestyle\endcsname
\providecommand{\newblock}{\relax}
\providecommand{\bibinfo}[2]{#2}
\providecommand{\BIBentrySTDinterwordspacing}{\spaceskip=0pt\relax}
\providecommand{\BIBentryALTinterwordstretchfactor}{4}
\providecommand{\BIBentryALTinterwordspacing}{\spaceskip=\fontdimen2\font plus
\BIBentryALTinterwordstretchfactor\fontdimen3\font minus
  \fontdimen4\font\relax}
\providecommand{\BIBforeignlanguage}[2]{{%
\expandafter\ifx\csname l@#1\endcsname\relax
\typeout{** WARNING: IEEEtran.bst: No hyphenation pattern has been}%
\typeout{** loaded for the language `#1'. Using the pattern for}%
\typeout{** the default language instead.}%
\else
\language=\csname l@#1\endcsname
\fi
#2}}
\providecommand{\BIBdecl}{\relax}
\BIBdecl

\bibitem{SurveyZeng}
Z.~Zeng, S.~Fu, H.~Zhang, Y.~Dong, and J.~Cheng, ``A survey of underwater
  optical wireless communications,'' \emph{IEEE Communications Surveys
  Tutorials}, vol.~19, no.~1, pp. 204--238, Firstquarter 2017.

\bibitem{LEDadvances}
J.~Xu, M.~Kong, A.~Lin, Y.~Song, X.~Yu, F.~Qu, J.~Han, and N.~Deng,
  ``{OFDM}-based broadband underwater wireless optical communication system
  using a compact blue {LED},'' \emph{Optics Communications}, vol. 369, pp. 100
  -- 105, 2016.

\bibitem{HassanHighDR}
H.~M. Oubei, C.~Li, K.-H. Park, T.~K. Ng, M.-S. Alouini, and B.~S. Ooi, ``2.3
  {G}bit/s underwater wireless optical communications using directly modulated
  520 nm laser diode,'' \emph{Opt. Express}, vol.~23, no.~16, pp.
  20\,743--20\,748, Aug. 2015.

\bibitem{OubeiQAMOFDM}
H.~M. Oubei, J.~R. Duran, B.~Janjua, H.-Y. Wang, C.-T. Tsai, Y.-C. Chi, T.~K.
  Ng, H.-C. Kuo, J.-H. He, M.-S. Alouini, G.-R. Lin, and B.~S. Ooi, ``4.8
  {G}bit/s 16-{QAM}-{OFDM} transmission based on compact 450-nm laser for
  underwater wireless optical communication,'' \emph{Opt. Express}, vol.~23,
  no.~18, pp. 23\,302--23\,309, Sep. 2015.

\bibitem{Oubei20m}
C.~Shen, Y.~Guo, H.~M. Oubei, T.~K. Ng, G.~Liu, K.-H. Park, K.-T. Ho, M.-S.
  Alouini, and B.~S. Ooi, ``20-meter underwater wireless optical communication
  link with 1.5 {G}bps data rate,'' \emph{Opt. Express}, vol.~24, no.~22, pp.
  25\,502--25\,509, Oct. 2016.

\bibitem{Simodetection}
W.~Liu, Z.~Xu, and L.~Yang, ``{SIMO} detection schemes for underwater optical
  wireless communication under turbulence,'' \emph{Photon. Res.}, vol.~3,
  no.~3, pp. 48--53, Jun. 2015.

\bibitem{HighBdW}
F.~Hanson and S.~Radic, ``High bandwidth underwater optical communication,''
  \emph{Appl. Opt.}, vol.~47, no.~2, pp. 277--283, Jan. 2008.

\bibitem{Jaruwatanadilok}
S.~Jaruwatanadilok, ``Underwater wireless optical communication channel
  modeling and performance evaluation using vector radiative transfer theory,''
  \emph{IEEE Journal on Selected Areas in Communications}, vol.~26, no.~9, pp.
  1620--1627, Dec. 2008.

\bibitem{Gabriel2}
C.~Gabriel, M.~A. Khalighi, S.~Bourennane, P.~Leon, and V.~Rigaud,
  ``Monte-{C}arlo-{B}ased channel characterization for underwater optical
  communication systems,'' \emph{IEEE/OSA Journal of Optical Communications and
  Networking}, vol.~5, no.~1, pp. 1--12, Jan 2013.

\bibitem{KihongWCL}
C.~Li, K.~H. Park, and M.~S. Alouini, ``On the use of a direct radiative
  transfer equation solver for path loss calculation in underwater optical
  wireless channels,'' \emph{IEEE Wireless Communications Letters}, vol.~4,
  no.~5, pp. 561--564, Oct. 2015.

\bibitem{UOT}
R.~J. Hill, ``Optical propagation in turbulent water,'' \emph{J. Opt. Soc.
  Am.}, vol.~68, no.~8, pp. 1067--1072, Aug. 1978.

\bibitem{Nikishov2000}
V.~V. Nikishov and V.~I. Nikishov, ``Spectrum of turbulent fluctuations of the
  sea-water refraction index,'' \emph{International Journal of Fluid Mechanics
  Research}, vol.~27, no.~1, pp. 82--98, 2000.

\bibitem{Korotkova2012}
O.~Korotkova, N.~Farwell, and E.~Shchepakina, ``Light scintillation in oceanic
  turbulence,'' \emph{Waves in Random and Complex Media}, vol.~22, no.~2, pp.
  260--266, 2012.

\bibitem{blanchard}
D.~C. Blanchard and A.~H. Woodcock, ``Bubble formation and modification in the
  sea and its meteorological significance,'' \emph{Tellus}, vol.~9, no.~2, pp.
  145--158, May 1957.

\bibitem{Zhang}
X.~Zhang, M.~Lewis, and B.~Johnson, ``{Influence of Bubbles on Scattering of
  Light in the Ocean},'' \emph{Applied Optics}, vol.~37, no.~27, Sep. 1998.

\bibitem{MOP:MOP26664}
R.~M. Hagem, D.~V. Thiel, S.~G. O'Keefe, and T.~Fickenscher, ``The effect of
  air bubbles on an underwater optical communications system for wireless
  sensor network applications,'' \emph{Microwave and Optical Technology
  Letters}, vol.~54, no.~3, pp. 729--732, 2012.

\bibitem{Woolf:01}
D.~K. Woolf, \emph{Encyclopedia of Ocean Sciences}, S.~T. J.H.~Steele and
  K.~Turekian, Eds.\hskip 1em plus 0.5em minus 0.4em\relax Academic Press,
  2001.

\bibitem{Farmer:84}
D.~M. Farmer and D.~D. Lemon, ``The influence of bubbles on ambient noise in
  the ocean at high wind speeds,'' \emph{J. Phys. Oceanogr.}, vol.~14, no.~11,
  pp. 1762--1778, 1984.

\bibitem{Boyle:87}
E.~A. Boyle and L.~Keigwin, ``North atlantic thermohaline circulation during
  the past 20,000 years linked to high-latitude surface temperature,''
  \emph{Nature}, vol. 330, pp. 35 -- 40, 1987.

\bibitem{Wang2017}
W.~Wang, P.~Wang, T.~Cao, H.~Tian, Y.~Zhang, and L.~Guo, ``Performance
  investigation of underwater wireless optical communication system using m
  -ary oamsk modulation over oceanic turbulence,'' \emph{IEEE Photonics
  Journal}, vol.~9, no.~5, pp. 1--15, Oct 2017.

\bibitem{Boucouvalas}
A.~C. Boucouvalas, K.~P. Peppas, K.~Yiannopoulos, and Z.~Ghassemlooy,
  ``Underwater optical wireless communications with optical amplification and
  spatial diversity,'' \emph{IEEE Photonics Technology Letters}, vol.~28,
  no.~22, pp. 2613--2616, Nov. 2016.

\bibitem{JamaliMIMO}
M.~V. Jamali, J.~A. Salehi, and F.~Akhoundi, ``Performance studies of
  underwater wireless optical communication systems with spatial diversity:
  {MIMO} scheme,'' \emph{IEEE Transactions on Communications}, vol.~65, no.~3,
  pp. 1176--1192, Mar. 2017.

\bibitem{PeppasIET}
K.~P. Peppas, A.~C. Boucouvalas, and Z.~Ghassemloy, ``Performance of underwater
  optical wireless communication with multi-pulse pulse-position modulation
  receivers and spatial diversity,'' \emph{IET Optoelectronics}, vol.~11,
  no.~5, pp. 180--185, Sep. 2017.

\bibitem{JamaliMultihop}
M.~V. Jamali, A.~Chizari, and J.~A. Salehi, ``Performance analysis of multi-hop
  underwater wireless optical communication systems,'' \emph{IEEE Photonics
  Technology Letters}, vol.~29, no.~5, pp. 462--465, Mar. 2017.

\bibitem{laserbeamscin}
L.~C. Andrews, R.~L. Phillips, and C.~Y. Hopen, \emph{Laser Beam Scintillation
  with Applications}.\hskip 1em plus 0.5em minus 0.4em\relax SPIE Press, 2001.

\bibitem{Gercekcioglu}
H.~Ger\c{c}ekcio\u{g}lu, ``Bit error rate of focused {G}aussian beams in weak
  oceanic turbulence,'' \emph{J. Opt. Soc. Am. A}, vol.~31, no.~9, pp.
  1963--1968, Sep. 2014.

\bibitem{lognmunderwater}
X.~Yi, Z.~Li, and Z.~Liu, ``Underwater optical communication performance for
  laser beam propagation through weak oceanic turbulence,'' \emph{Appl. Opt.},
  vol.~54, no.~6, pp. 1273--1278, Feb. 2015.

\bibitem{Davis}
G.~E. Davis, ``{Scattering of Light by an Air Bubble in Water},'' \emph{Journal
  of the Optical Society of America}, vol.~45, no.~7, Jul. 1955.

\bibitem{ExpLN}
M.~V. Jamali, P.~Khorramshahi, A.~Tashakori, A.~Chizari, S.~Shahsavari,
  S.~AbdollahRamezani, M.~Fazelian, S.~Bahrani, and J.~A. Salehi, ``Statistical
  distribution of intensity fluctuations for underwater wireless optical
  channels in the presence of air bubbles,'' in \emph{2016 Iran Workshop on
  Communication and Information Theory (IWCIT'16)}, Tehran, Iran, May 2016, pp.
  1--6.

\bibitem{ConfUWOC}
E.~Zedini, H.~M. Oubei, A.~Kammoun, M.~Hamdi, B.~S. Ooi, and M.~S. Alouini, ``A
  new simple model for underwater wireless optical channels in the presence of
  air bubbles,'' in \emph{IEEE Global Communications Conference (GLOBECOM'17)},
  Dec. 2017, pp. 1--6.

\bibitem{Oubei:17sg}
H.~M. Oubei, E.~Zedini, R.~T. ElAfandy, A.~Kammoun, T.~K. Ng, M.-S. Alouini,
  and B.~S. Ooi, ``Efficient weibull channel model for salinity induced
  turbulent underwater wireless optical communications,'' in \emph{22nd
  OptoElectronics and Communications Conference}, ser. Oral 2-3K-2, Singapore,
  2017.

\bibitem{Oubei:17}
H.~M. Oubei, E.~Zedini, R.~T. ElAfandy, A.~Kammoun, M.~Abdallah, T.~K. Ng,
  M.~Hamdi, M.-S. Alouini, and B.~S. Ooi, ``Simple statistical channel model
  for weak temperature-induced turbulence in underwater wireless optical
  communication systems,'' \emph{Opt. Lett.}, vol.~42, no.~13, pp. 2455--2458,
  Jul 2017.

\bibitem{7879801}
H.~M. Oubei, R.~T. ElAfandy, K.~H. Park, T.~K. Ng, M.~S. Alouini, and B.~S.
  Ooi, ``Performance evaluation of underwater wireless optical communications
  links in the presence of different air bubble populations,'' \emph{IEEE
  Photonics Journal}, vol.~9, no.~2, pp. 1--9, April 2017.

\bibitem{Tableofintegrals}
I.~S. Gradshteyn and I.~M. Ryzhik, \emph{Table of Integrals, Series and
  Products}.\hskip 1em plus 0.5em minus 0.4em\relax New York: Academic Press,
  2000.

\bibitem{parameter}
A.~C. Cohen and B.~K. Whitten, \emph{Parameter Estimation in Reliability and
  Life Span Models}.\hskip 1em plus 0.5em minus 0.4em\relax New York : M.
  Dekker, c1988., 1988.

\bibitem{Rsquare}
J.~Devore, \emph{Probability and {S}tatistics for {E}ngineering and the
  {S}ciences}, 8th~ed.\hskip 1em plus 0.5em minus 0.4em\relax Cengage Learning,
  2011.

\bibitem{HTranforms}
A.~Kilbas and M.~Saigo, \emph{{H}-{T}ransforms : {T}heory and {A}pplications
  ({A}nalytical {M}ethod and {S}pecial {F}unction)}, 1st~ed.\hskip 1em plus
  0.5em minus 0.4em\relax CRC Press, 2004.

\bibitem{PrudinkovVol3}
A.~Prudnikov, Y.~Brychkov, and O.~Marichev, \emph{Integrals and Series, Volume
  3: More Special Functions}.\hskip 1em plus 0.5em minus 0.4em\relax CRC, 1999.

\bibitem{dualhopFSO}
E.~Zedini, H.~Soury, and M.~S. Alouini, ``Dual-{H}op {FSO} transmission systems
  over {G}amma-{G}amma turbulence with pointing errors,'' \emph{IEEE
  Transactions on Wireless Communications}, vol.~16, no.~2, pp. 784--796, Feb.
  2017.

\bibitem{FerkanHFox}
F.~Yilmaz and M.-S. Alouini, ``Product of the powers of generalized
  {N}akagami-$m$ variates and performance of cascaded fading channels,'' in
  \emph{IEEE Global Telecommunications Conference (GLOBECOM'09)}, Nov. 2009,
  pp. 1--8.

\bibitem{capacityFSOlinks}
A.~Lapidoth, S.~Moser, and M.~Wigger, ``On the capacity of free-space optical
  intensity channels,'' \emph{IEEE Transactions on Information Theory},
  vol.~55, no.~10, pp. 4449--4461, Oct. 2009.

\bibitem{AnasTCOM}
A.~Chaaban, J.~M. Morvan, and M.~S. Alouini, ``Free-{S}pace optical
  communications: Capacity bounds, approximations, and a new sphere-packing
  perspective,'' \emph{IEEE Transactions on Communications}, vol.~64, no.~3,
  pp. 1176--1191, Mar. 2016.

\bibitem{AFN}
F.~Yilmaz and M.-S. Alouini, ``Novel asymptotic results on the high-order
  statistics of the channel capacity over generalized fading channels,'' in
  \emph{Proceedings of IEEE International Workshop on Signal Processing
  Advances in Wireless Communications (SPAWC'12)}, 2012, pp. 389--393.

\bibitem{heterodyne1}
T.~Tsiftsis, ``Performance of heterodyne wireless optical communication systems
  over {G}amma-{G}amma atmospheric turbulence channels,'' \emph{Electronics
  Letters}, vol.~44, no.~5, pp. 372--373, Feb. 2008.

\bibitem{proakis2008digital}
J.~Proakis and M.~Salehi, \emph{Digital Communications}, ser. McGraw-Hill
  International Edition.\hskip 1em plus 0.5em minus 0.4em\relax McGraw-Hill,
  2008.

\end{thebibliography}
\end{document}